\documentclass[apj]{emulateapj}
\usepackage{graphicx}
\usepackage{subfigure}
\usepackage[usenames]{color}

\newcommand{\boldsymbol}[1]{\mbox{\boldmath{${#1}$}}}

\newcommand{\bd}{\begin{displaymath}}
\newcommand{\ed}{\end{displaymath}}
\newcommand{\be}{\begin{equation}}
\newcommand{\ee}{\end{equation}}
\newcommand{\beaa}{\begin{eqnarray*}}
\newcommand{\eeaa}{\end{eqnarray*}}
\newcommand{\bea}{\begin{eqnarray}}
\newcommand{\eea}{\end{eqnarray}}

\newcommand\HST{\textit{HST}}

\def\parlensvec{\boldsymbol{\eta}}

\def\data{\boldsymbol{d_{\rm pos}}}

\shorttitle{CLASH-VLT: Insights on mass substructures in MACS J0416.1$-$2403 through strong lensing}
\shortauthors{Grillo et al.}

\begin{document}


\title{CLASH-VLT: Insights on the mass substructures in the Frontier Fields Cluster MACS~J0416.1$-$2403 through accurate strong lens modeling$^\star$}


\author{C.~Grillo\altaffilmark{1}, S.~H.~Suyu\altaffilmark{2},
  P.~Rosati\altaffilmark{3}, A.~Mercurio\altaffilmark{4},
  I.~Balestra\altaffilmark{5}, E.~Munari\altaffilmark{6,5},
  M.~Nonino\altaffilmark{5}, G.~B.~Caminha\altaffilmark{3},
  M.~Lombardi\altaffilmark{7}, G.~De~Lucia\altaffilmark{5},
  S.~Borgani\altaffilmark{6,5}, R.~Gobat\altaffilmark{8},
  A.~Biviano\altaffilmark{5}, M.~Girardi\altaffilmark{6,5},
  K.~Umetsu\altaffilmark{2}, D.~Coe\altaffilmark{9},
  A.~M.~Koekemoer\altaffilmark{9}, 
  M.~Postman\altaffilmark{9}, A.~Zitrin\altaffilmark{10,11},
  A.~Halkola, T.~Broadhurst\altaffilmark{12},
  B.~Sartoris\altaffilmark{6}, V.~Presotto\altaffilmark{6},
  M.~Annunziatella\altaffilmark{6,5}, C. Maier\altaffilmark{13},
  A.~Fritz\altaffilmark{14}, E.~Vanzella\altaffilmark{15},
  B.~Frye\altaffilmark{16}} 
\email{grillo@dark-cosmology.dk}


\altaffiltext{$\star$}{ This work is based in large part on data
  collected at ESO VLT (prog.ID 186.A-0798) and NASA \HST.}
\altaffiltext{1}{Dark Cosmology Centre, Niels Bohr Institute,
  University of Copenhagen, Juliane Maries Vej 30, DK-2100 Copenhagen,
  Denmark} 
\altaffiltext{2}{Institute of Astronomy and Astrophysics, Academia
  Sinica, P.O. Box 23-141, Taipei 10617, Taiwan}
\altaffiltext{3}{Dipartimento di Fisica e Scienze della Terra,
  Universit\`a degli Studi di Ferrara, Via Saragat 1, I-44122 Ferrara,
  Italy} 
\altaffiltext{4}{INAF - Osservatorio Astronomico di Capodimonte, Via
  Moiariello 16, I-80131 Napoli, Italy} 
\altaffiltext{5}{INAF - Osservatorio Astronomico di Trieste, via
  G. B. Tiepolo 11, I-34143, Trieste, Italy} 
\altaffiltext{6}{Dipartimento di Fisica, Universit\`a  degli Studi di
  Trieste, via G. B. Tiepolo 11, I-34143 Trieste, Italy}
\altaffiltext{7}{Dipartimento di Fisica, Universit\`a  degli Studi di
  Milano, via Celoria 16, I-20133 Milano, Italy}
\altaffiltext{8}{Laboratoire AIM-Paris-Saclay,
  CEA/DSM-CNRS-Universit\`e Paris Diderot, Irfu/Service
  d'Astrophysique, CEA Saclay, Orme des Merisiers, F-91191 Gif sur
  Yvette, France} 
\altaffiltext{9}{Space Telescope Science Institute, 3700 San Martin
  Drive, Baltimore, MD 21208, USA} 
\altaffiltext{10}{Cahill Center for Astronomy and Astrophysics,
  California Institute of Technology, MS 249-17, Pasadena, CA 91125,
  USA} 
\altaffiltext{11}{Hubble Fellow}
\altaffiltext{12}{Ikerbasque, Basque Foundation for Science, Alameda
Urquijo, 36-5 Plaza Bizkaia, E-48011, Bilbao, Spain}
\altaffiltext{13}{University of Vienna, Department of Astrophysics,
  T\"urkenschanzstr. 17, A-1180, Wien, Austria}
\altaffiltext{14}{INAF - Istituto di Astrofisica Spaziale e Fisica
  cosmica (IASF) Milano, via Bassini 15, I-20133 Milano, Italy}
\altaffiltext{15}{INAF - Osservatorio Astronomico di Bologna, Via
  Ranzani 1, I- 40127 Bologna, Italy}
\altaffiltext{16}{Department of Astronomy/Steward Observatory,
  University of Arizona, 933 North Cherry Avenue, Tucson, AZ 85721,
  USA}


\begin{abstract}

We present a detailed mass reconstruction and a novel study on the
substructure properties in the core of the Cluster Lensing And
Supernova survey with Hubble (CLASH) and Frontier Fields galaxy
cluster MACS J0416.1$-$2403. We show and employ our extensive
spectroscopic data set taken with the VIsible Multi-Object
Spectrograph (VIMOS) 
instrument as part of our CLASH-VLT program, to confirm
spectroscopically 10 strong lensing systems and to select a sample of
175 plausible cluster members to a limiting stellar mass of $\log(M_*/M_\odot)\simeq 8.6$.
We reproduce the measured positions of a set of 30 multiple
images with a remarkable median offset of only 0.3\arcsec$\,$ by means
of a comprehensive strong lensing model comprised of 2 cluster dark-matter
halos, represented by cored
elliptical pseudo-isothermal mass distributions, and the cluster
member components, parametrized with dual pseudo-isothermal total mass
profiles. The latter have total mass-to-light ratios increasing with
the galaxy \HST/WFC3 near-IR (F160W) luminosities. 
The measurement of the total enclosed mass within the Einstein radius
is accurate to $\sim5\%$, including the systematic uncertainties estimated from six distinct mass models.
We emphasize that the use of multiple-image systems with spectroscopic
redshifts and knowledge of cluster membership based on extensive
spectroscopic information is key to constructing robust
high-resolution mass maps. 
We also produce magnification maps over the central area that is covered with
\HST\ observations. We investigate the galaxy contribution, both in
terms of total and stellar mass, to the total mass budget of the
cluster. When compared with the outcomes of cosmological $N$-body
simulations, our results point to a lack of massive subhalos in
the inner regions of simulated clusters with total masses similar
to that of MACS J0416.1$-$2403. 
Our findings of the location and shape of the cluster dark-matter halo
density profiles and on the cluster substructures provide intriguing
tests of the assumed collisionless, cold nature of dark matter and of
the role played by baryons in the process of structure formation.

\end{abstract}


\keywords{gravitational lensing $-$ galaxies: clusters: general $-$ galaxies: clusters: individuals: MACS J0416.1$-$2403 $-$ Dark matter}



\section{Introduction}
\label{sec:intro}

The currently accepted cold dark matter dominated model with the 
cosmological constant ($\Lambda$CDM) predicts that structures in our
Universe assemble hierarchically, with more massive systems forming
later through accretion and mergers of smaller, self-bound dark-matter
halos (e.g., \citealt{tor97}; \citealt{moo99}; \citealt{kly99};
\citealt{SpringelEtal01}). In $N$-body cosmological simulations, dark
matter halos of all masses converge to a roughly ``universal'' and
cuspy density profile that steepens with radius, the so-called
Navarro, Frenk and White profile (NFW;
\citealt{nav96,nav97}). Moreover, the degree of central concentration
of a halo depends on its formation epoch and 
hence on its total mass (e.g., \citealt{wec02};
\citealt{zha03}). Within this scenario, early virialized 
objects are compact when they get accreted into a larger halo. Such
objects are usually referred to as subhalos or substructures of their
host and, as they orbit within the host potential well, they are
strongly affected by tidal forces and dynamical friction, causing
mass, angular momentum, and energy loss (e.g., \citealt{ghi98};
\citealt{tor98}; \citealt{del04}; \citealt{gao04}). In the $\Lambda$CDM 
framework, more massive halos are predicted to have a larger fraction
of mass in subhalos than lower mass halos because in the former
there has been less time for tidal destruction to take place (e.g.,
\citealt{gao04}; \citealt{ContiniEtal12}).
On galaxy cluster scales, observational tests of these predictions
have been attempted in some previous works (e.g.,
\citealt{nat07,nat09}), but highly accurate analyses are becoming
possible only now, thanks to the substantially improved quality of the
available photometric and spectroscopic data.
From an observational point of view, more investigations are still
required to fully answer key questions
on the formation and evolution of subhalos.
How much mass of subhalos is
stripped as they fall into the host potential?  How many subhalos
survive as bound objects?  What are the spatial and mass distributions
of the subhalos?

Significant progress (\citealt{biv13}; \citealt{lem13};
\citealt{ume14}; \citealt{mer14}; \citealt{men14}; \citealt{don14}) in
the fields of galaxy cluster formation  
and evolution has lately been made thanks to the data collected within
the \textit{Hubble Space Telescope} (\HST) Multi-Cycle Treasury program Cluster
Lensing And Supernova survey with Hubble (CLASH; P.I.: M. Postman;
\citealt{pos12}), often complemented with the spectroscopic campaign
carried out with the Very Large Telescope (VLT)  (the CLASH-VLT Large
Programme; P.I.: P. Rosati; Rosati et al., in prep.). 
Recent new results from systematic in-depth studies of lensing
clusters with \HST\ have led to the conception of the  
Hubble Frontier Fields (HFF; P.I.: J. Lotz) that will target, 
using Director Discretionary Time, up to six massive galaxy clusters,
for a total of 140 \HST\ orbits on each cluster, in 7 broadband filters, 
achieving in all of them unprecedented depth of ~$\approx$~29 mag
(AB). Not only will this program detect the highest  
redshift galaxies and characterize for the first time this sample of
star-forming galaxies in a statistically meaningful way, the HFF data
will provide a great opportunity to study the structure of the dark
matter halos hosting these clusters. 

In this paper, we focus on the HFF cluster MACS~J0416.1$-$2403
(hereafter MACS~0416) that was first discovered in the X-rays by
\citet{MannEbeling12} as part of the Massive Cluster Survey (MACS).
MACS~0416 is an elongated cluster undergoing a merger, and such an
elongated geometry makes MACS~0416 an efficient gravitational lens for
highly magnifying background sources and forming multiple images of
each background source (the regime of ``strong'' lensing).  Given its
high magnifications and the upcoming HFF infrared observations, there
has been several recent studies of MACS~0416.  In particular,
\citet{zit13} has identified 70 multiple images and candidates that
are associated with 23 background sources, confirming the enhanced
lensing efficiency of MACS~0416 relative to other
clusters. \citet{JauzacEtal14a} further identified 51 strongly lensed
background sources, yielding 194 multiple images of lensed background
sources.
\citet{JauzacEtal14b} further complement the strong
lensing data with weak lensing and X-ray observations to study the
dynamics of MACS~0416.  Using a free-form mass modeling approach,
\citet{DiegoEtal14} found that the mass distribution in MACS~0416
overall traces its light distribution.  \citet{JohnsonEtal14} and
\citet{RichardEtal14} have also modeled MACS~0416 as
part of the HFF sample, and provided the mass and magnifications maps
of the clusters.

Building upon and extending these previous studies,
we perform a thorough strong
lensing analysis of MACS~0416 with the following new ingredients:
(1) a large number of spectroscopic redshifts of strongly lensed background
sources obtained through our CLASH-VLT program, (2) a robust approach
of selecting cluster galaxies based on multi-color information
calibrated on 113 spectroscopic members in the \HST\ field of view (FoV), and
(3) a detailed mass model that tests 
various methodological assumptions with 
our best model reproducing the
observed multiple-image positions substantially better than all previous
studies.  Using our mass model, we compare the distribution of
the cluster galaxies with those of the cluster subhalos from
$N$-body simulations to probe with
unexampled accuracy the substructure properties of a galaxy cluster.

The outline of the paper is as follows.  In Section \ref{sec:data}, we
describe the imaging and spectroscopic observations of MACS~0416.  We
detail our strong lens modeling of the cluster in Section
\ref{sec:lensmod}, including the selection of background source
galaxies and cluster galaxies.  Our resulting mass model is then
compared to those published in the literature in Section
\ref{sec:complit}.  We compare in Section \ref{sec:nbody} the mass
distribution of the cluster galaxies of MACS~0416 with those of the
cluster subhalos of analog clusters from $N$-body
simulations, before presenting conclusions in Section
\ref{sec:conclude}.

Throughout this paper, we assume a flat $\Lambda$CDM cosmology with
$H_{0}=70$ km s$^{-1}$ Mpc$^{-1}$, $\Omega_{\Lambda}=1-\Omega_{\rm m}=
0.7$.  In this cosmology, $1''$ corresponds to 5.34\,kpc at the lens
redshift of $z_{\rm lens}=0.396$. All magnitudes are given in the AB
system. 
Parameter constraints are given as the median values with
uncertainties given by the 16th and 84th percentiles (corresponding to
68\% confidence levels (CLs)) unless otherwise stated.

\section{Data}
\label{sec:data}

\subsection{\HST\ imaging}

Being a target of the CLASH program, MACS~0416 was observed in \HST\
Cycle 19, between July 24 and September 27 2012, in 16 broadband
filters, from the UV to the near-IR, to a total depth of 20 orbits
(see \citealt{pos12}). The images were processed for debias, flats,
superflats, and darks using standard techniques, and then co-aligned
and combined using drizzle algorithms to pixel scales of
0.030\arcsec$\,$ and 0.065\arcsec$\,$ (for details, see
\citealt{koe07,koe11}).

\begin{figure*}
\centering
\includegraphics[width=0.8\textwidth]{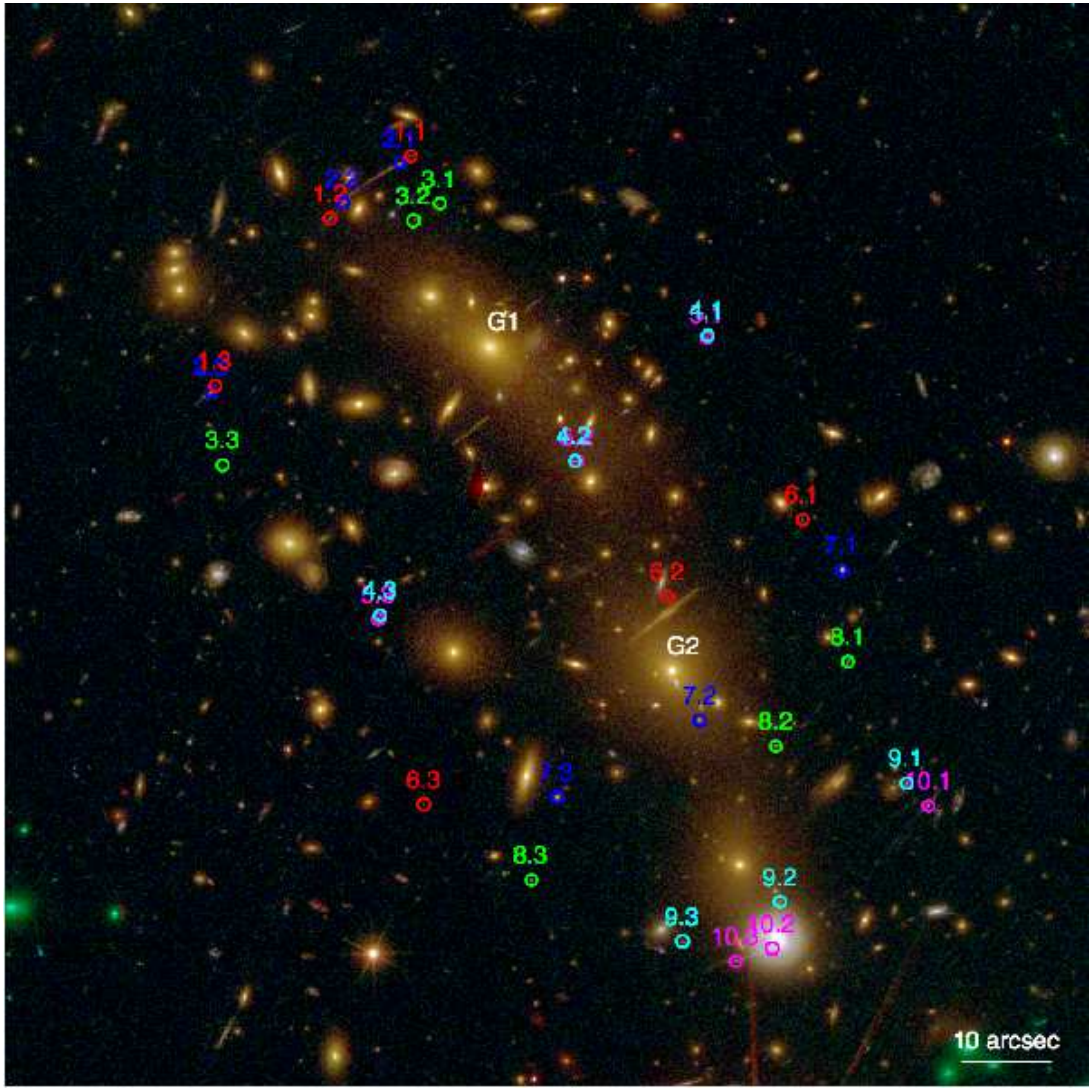}
\caption{A $2'$$\times$$2'$ color-composite image of MACS
  J0416.1$-$2403 obtained by combining the 16 filters of \HST/ACS and
  WFC3. North is top and East is left. The thirty multiple images
  modeled in this paper and the two brightest cluster galaxies, G1 and
  G2, are labeled. More information about these objects is provided
  in Tables \ref{tab5} and  \ref{tab1} and Figure \ref{fi01}.}
\label{fi06}
\end{figure*}

A color image of the inner regions of the galaxy cluster, obtained
through a combination of the \HST/ACS and WFC3 filters,
is shown in Figure
\ref{fi06}. There, we have marked the two brightest galaxies of the
cluster, G1 and G2 (see also Table \ref{tab5}), and the multiple image
systems studied in this paper (see Section \ref{sec:lensmod:multsys}).

\begin{table*}
\centering
\caption{Photometric and spectroscopic properties of the two brightest
  cluster galaxies G1 and G2.}
\begin{tabular}{ccccccc}
\hline\hline \noalign{\smallskip}
ID & R.A. & Decl. & $x^{\mathrm{a}}$ & $y^{\mathrm{a}}$ & $z_{\mathrm{sp}}$ & F160W \\
 & (J2000) & (J2000) & (\arcsec) & (\arcsec) & & (mag) \\
\noalign{\smallskip} \hline \noalign{\smallskip}
G1 & 04:16:09.154 & $-$24:04:02.90 & $\equiv$0.000 & $\equiv$0.000 & 0.400 & 17.02 \\ 
G2 & 04:16:07.671 & $-$24:04:38.75 & 20.310 & $-$35.846 & 0.396 & 17.24 \\ \noalign{\smallskip} \hline
\end{tabular}
\begin{list}{}{}
\item[$^{\mathrm{a}}$]With respect to the luminosity center of G1 and
  positive in the West and North directions.
\end{list}
\label{tab5}
\end{table*}

\subsection{VLT spectroscopy}
\label{sec:data:vlt}

MACS~0416 was first observed between December 2012 and February 2013,
as part of the ESO Large Programme 186.A-0798 ``Dark Matter Mass Distributions of Hubble 
Treasury Clusters and the Foundations of $\Lambda$CDM Structure
Formation Models'' 
(CLASH-VLT) using the VIsible Multi-Object
Spectrograph (VIMOS; \citealt{lef03}) at the ESO VLT. 
The VIMOS data were acquired using 
eight pointings with one quadrant always locked 
on the cluster core, thus allowing longer exposures on the arcs. The exposure time for 
each pointing was 60 minutes, except for the last two pointings that had shorter exposure times 
(30 minutes), because they targeted cluster member galaxies. Therefore, the final integration
times for arcs and other background galaxies varied between 30 minutes and 4 hours. 
A log of our VIMOS observations is presented in Table~\ref{obs}. 
We used the LR-blue grism, with a spectral resolution of approximately 28 \AA$\,$ with 1\arcsec-slits and a wavelength coverage of 3700-6700 \AA. 

\begin{table}
\caption{Log of VIMOS observations of MACS~0416, taken as part of our 
CLASH-VLT spectroscopic campaign.}
\begin{center}
\begin{tabular}{l c c}
\hline
\hline
OBS ID & Date & Exp. time (min.) \\
(1)     & (2)  & (3)           \\
\hline
\multicolumn{3}{l}{LR-blue masks} \\
\hline
848955  &  Dec. 2012  & $60$ \\
848957  &  Jan. 2013   & $60$ \\
848959  &  Dec. 2012  & $60$ \\
856270  &  Feb. 2013   & $60$ \\
916733  &  Feb. 2013   & $60$ \\
916723  &  Feb. 2013   & $60$ \\
915893  &  Feb. 2013   & $30$ \\
915903  &  Feb. 2013   & $30$ \\
\hline
\end{tabular}
\end{center}
\begin{list}{}{}
\item[Notes. ]Columns list the following information: (1) VIMOS mask identification number, (2) date of the observations, and (3) exposure time.
\end{list}
\label{obs}
\end{table}

We assign a quality flag (QF, indicated as Quality in Figures \ref{fi01} and \ref{fi04}) to each redshift, which indicates the 
reliability of a redshift measurement. We define four redshift quality classes: 
\texttt{SECURE} (QF=3), \texttt{LIKELY} (QF=2), \texttt{INSECURE} (QF=1), and \texttt{BASED ON A SINGLE EMISSION-LINE} (QF=9). To assess the reliability of these four quality classes we 
compared pairs of duplicate observations having at least one secure measurement. 
In this way, we could quantify the reliability of each quality class
as follows: redshifts  
with QF=3 are correct with a probability of $>99.99$\%, QF=9 with
$\sim92$\% probability,  
QF=2 with $\sim75$\% probability, and QF=1 with $<40$\%
probability. In this paper we will only consider redshifts with QF=3,
2, or 9. To date, we have 4160 reliable redshifts 
over a field $\sim$25 arcmin across, over 800 of which are cluster members. 
Full details on the spectroscopic sample observations and 
data reduction will be given in Balestra et al. (in prep.).

In the spirit of the 
open-data access 
of the HFF initiative,
we had an early release (July 2013) of a redshift catalog of 118
sources in the \HST\ FoV, including the redshifts of the
multiple image 
systems presented in the next section. This information has been used
when building recent lensing models of MACS~0416
(see Section \ref{sec:complit}; \citealt{JohnsonEtal14};
\citealt{RichardEtal14}; \citealt{JauzacEtal14a,JauzacEtal14b};
\citealt{DiegoEtal14}). We note that the redshift of system 7 (see
Table \ref{tab1}) has been revised to a more reliable value of 1.637,
thanks to spectroscopic data that became available only after our
first release. As a result of our continuing spectroscopic campaign,
which will be completed at the end of 2014, in this paper we use the
current spectroscopic redshift information for 215 galaxies in the
\HST\ FoV, 113 of
which are cluster members (see Section \ref{sec:lensmod:masscomp:gal}).

\section{Lens modeling} 
\label{sec:lensmod}

We describe the mass modeling of MACS~0416 based on the strong lensing
features.  In Section \ref{sec:lensmod:multsys}, we identify the
multiple image systems of strongly lensed background sources.  In
Section \ref{sec:lensmod:glee}, we give an overview of the method and
software used to model the lens mass distribution, with the
decomposition of cluster members and extended dark-matter halos described in
Section \ref{sec:lensmod:masscomp}.  We detail our collection of mass
models of the galaxy cluster in Section \ref{sec:lensmod:massmod}, and
present the resulting total and luminous mass distribution of MACS~0416 in Section
\ref{sec:lensmod:results}.  

\subsection{Multiple image systems}
\label{sec:lensmod:multsys}

\begin{table*}
\centering
\caption{Photometric and spectroscopic properties of the multiple image systems.}
\begin{tabular}{cccccccc}
\hline\hline \noalign{\smallskip}
ID & R.A. & Decl. & $x^{\mathrm{a}}$ & $y^{\mathrm{a}}$ & $z_{\mathrm{sp}}$ & $\delta_{x,y}$ & ID Z13$^{\mathrm{b}}$ \\
 & (J2000) & (J2000) & (\arcsec) & (\arcsec) & & (\arcsec) & \\
\noalign{\smallskip} \hline \noalign{\smallskip}
1.1 & 04:16:09.784 & $-$24:03:41.76 & $-$8.626 & 21.137 & 1.892 & 0.065 & 1.1 \\
1.2 & 04:16:10.435 & $-$24:03:48.69 & $-$17.549 & 14.214 & 1.892 & 0.065 & 1.2 \\
1.3 & 04:16:11.365 & $-$24:04:07.21 & $-$30.285 & $-$4.312 & 1.892 & 0.065 & 1.3 \\ \noalign{\smallskip} \hline \noalign{\smallskip}
2.1 & 04:16:09.871 & $-$24:03:42.59 & $-$9.823 & 20.308 & 1.892 & 0.065 & 2.1 \\
2.2 & 04:16:10.329 & $-$24:03:46.96 & $-$16.101 & 15.937 & 1.892 & 0.065 & 2.2 \\
2.3 & 04:16:11.395 & $-$24:04:07.86 & $-$30.698 & $-$4.962 & 1.892 & 0.065 & 2.3 \\ \noalign{\smallskip} \hline \noalign{\smallskip}
3.1 & 04:16:09.549 & $-$24:03:47.08 & $-$5.419 & 15.819 & 2.087 & 0.065 & c7.1 \\
3.2 & 04:16:09.758 & $-$24:03:48.90 & $-$8.282 & 14.001 & 2.087 & 0.065 & c7.2 \\
3.3 & 04:16:11.304 & $-$24:04:15.94 & $-$29.451 & $-$13.040 & 2.087 & 0.065 & c7.3 \\ \noalign{\smallskip} \hline \noalign{\smallskip}
4.1 & 04:16:07.385 & $-$24:04:01.62 & 24.221 & 1.280 & 1.990 & 0.065 & 3.1 \\
4.2 & 04:16:08.461 & $-$24:04:15.53 & 9.492 & $-$12.630 & 1.990 & 0.065 & 3.2 \\
4.3 & 04:16:10.031 & $-$24:04:32.62 & $-$12.019 & $-$29.719 & 1.990 & 0.065 & 3.3 \\ \noalign{\smallskip} \hline \noalign{\smallskip}
5.1 & 04:16:07.390 & $-$24:04:02.01 & 24.157 & 0.890 & 1.990 & 0.065 & 4.1 \\
5.2 & 04:16:08.440 & $-$24:04:15.57 & 9.776 & $-$12.671 & 1.990 & 0.065 & 4.2 \\
5.3 & 04:16:10.045 & $-$24:04:33.03 & $-$12.206 & $-$30.134 & 1.990 &
0.065 & 4.3 \\ \noalign{\smallskip} \hline \noalign{\smallskip}
6.1 & 04:16:06.618 & $-$24:04:21.99 & 34.731 & $-$19.086 & 3.223 & 0.065 & 13.1 \\
6.2 & 04:16:07.709 & $-$24:04:30.56 & 19.788 & $-$27.661 & 3.223 & 0.065 & 13.2 \\
6.3 & 04:16:09.681 & $-$24:04:53.53 & $-$7.219 & $-$50.632 & 3.223 & 0.065 & 13.3 \\\noalign{\smallskip} \hline \noalign{\smallskip}
7.1 & 04:16:06.297 & $-$24:04:27.60 & 39.130 & $-$24.700 & 1.637 & 0.065 & 14.1 \\
7.2 & 04:16:07.450 & $-$24:04:44.23 & 23.334 & $-$41.334 & 1.637 & 0.065 & 14.2\\
7.3 & 04:16:08.600 & $-$24:04:52.76 & 7.580 & $-$49.860 & 1.637 & 0.065 & 14.3\\ \noalign{\smallskip} \hline \noalign{\smallskip}
8.1 & 04:16:06.246 & $-$24:04:37.76 & 39.818 & $-$34.861 & 2.302 & 0.065 & 10.1 \\
8.2 & 04:16:06.832 & $-$24:04:47.10 & 31.799 & $-$44.204 & 2.302 & 0.065 & 10.2 \\
8.3 & 04:16:08.810 & $-$24:05:01.93 & 4.707 & $-$59.028 & 2.302 & 0.065 & c10.3 \\ \noalign{\smallskip} \hline \noalign{\smallskip}
9.1 & 04:16:05.779 & $-$24:04:51.22 & 46.217 & $-$48.320 & 1.964 & 0.065 & 16.1 \\
9.2 & 04:16:06.799 & $-$24:05:04.35 & 32.249 & $-$61.452 & 1.964 & 0.065 & 16.2 \\
9.3 & 04:16:07.586 & $-$24:05:08.72 & 21.465 & $-$65.822 & 1.964 & 0.065 & 16.3 \\ \noalign{\smallskip} \hline \noalign{\smallskip}
10.1 & 04:16:05.603 & $-$24:04:53.70 & 48.625 & $-$50.798 & 2.218 & 0.065 & c17.3 \\
10.2 & 04:16:06.866 & $-$24:05:09.50 & 31.331 & $-$66.598 & 2.218 & 0.065 & c17.2 \\
10.3 & 04:16:07.157 & $-$24:05:10.91 & 27.344 & $-$68.010 & 2.218 & 0.065 & c17.1 \\ \noalign{\smallskip} \hline
\end{tabular}
\begin{list}{}{}
\item[$^{\mathrm{a}}$]With respect to the luminosity center of G1 and
  positive in the West and North directions.
\item[$^{\mathrm{b}}$]Corresponding image identifier in \citet{zit13}.
\end{list}
\label{tab1}
\end{table*}

\begin{figure*}
\centering
\includegraphics[width=0.63\textwidth]{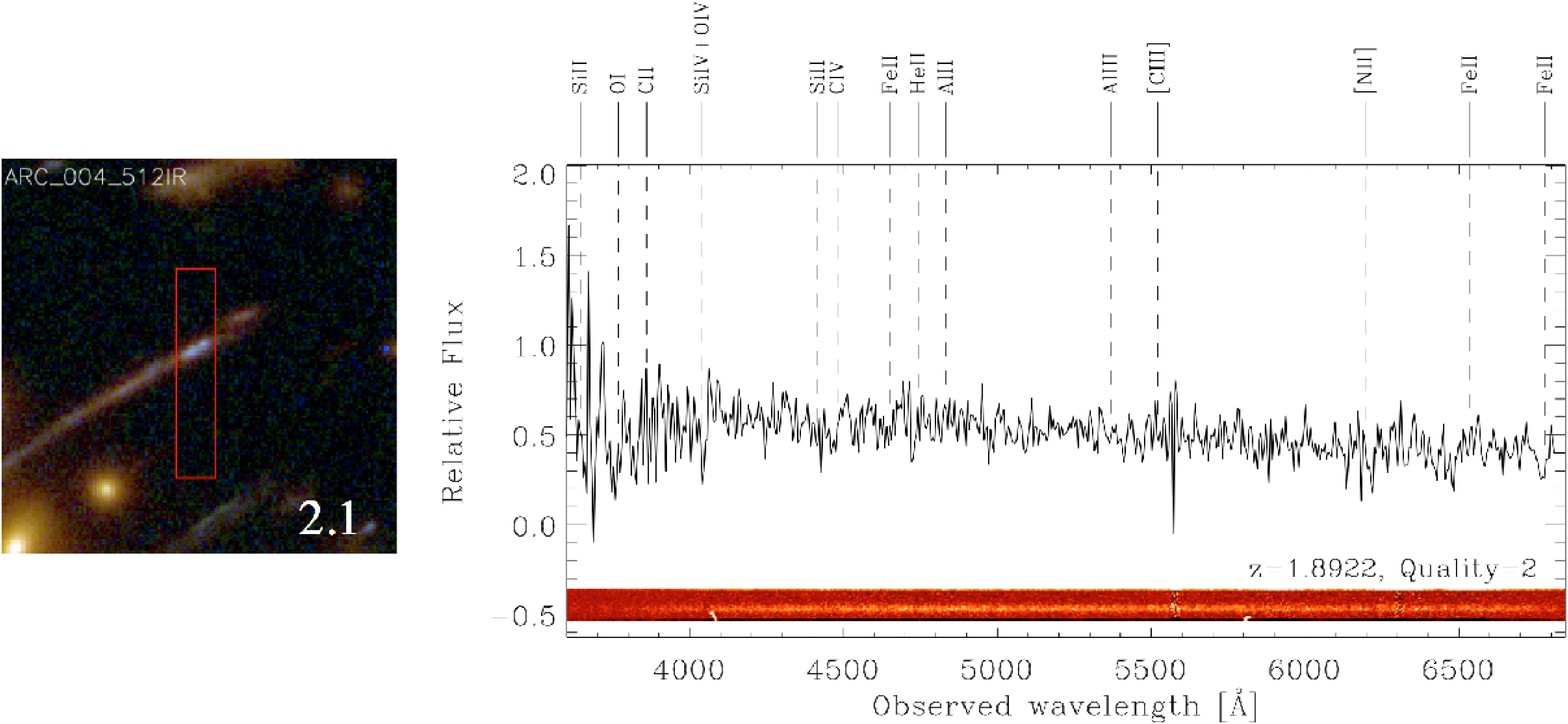}
\includegraphics[width=0.63\textwidth]{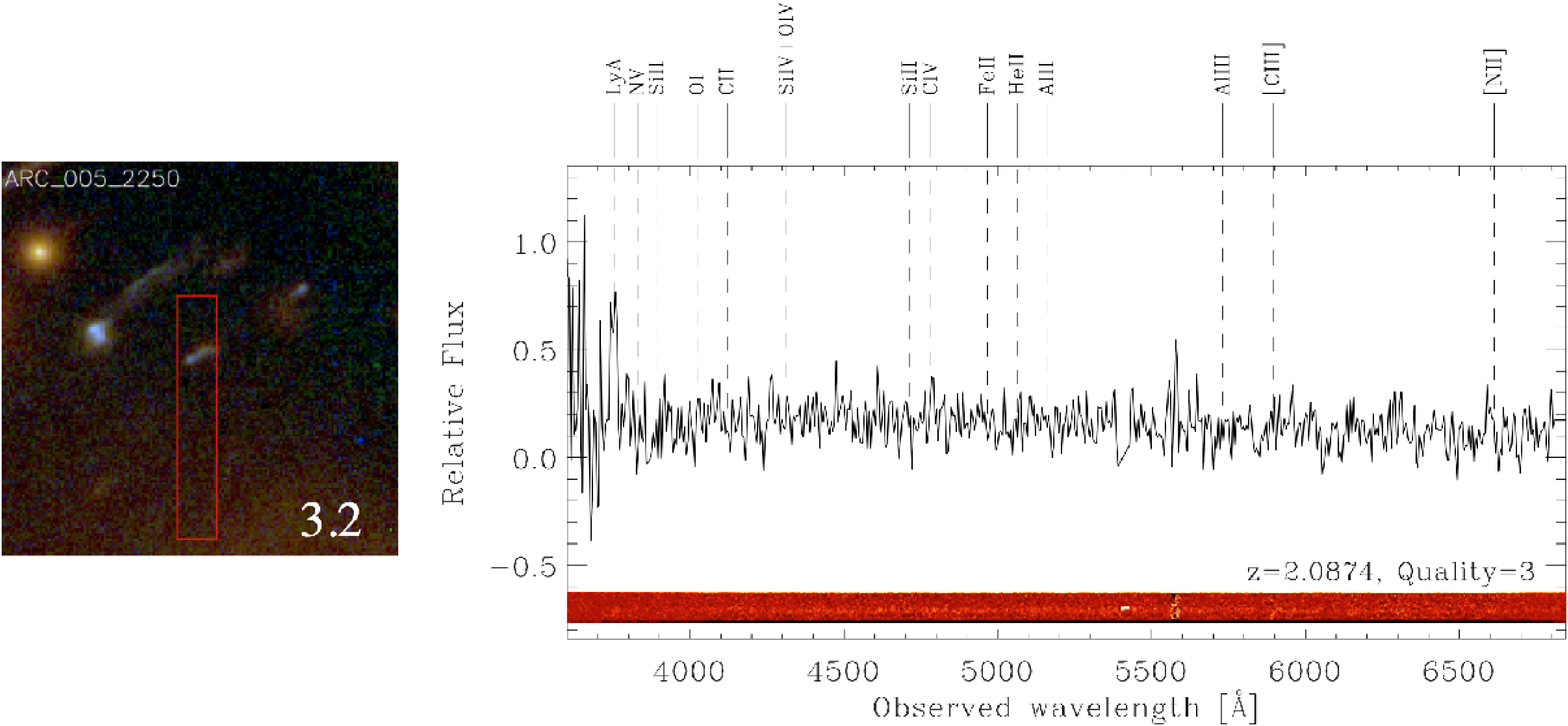}
\includegraphics[width=0.63\textwidth]{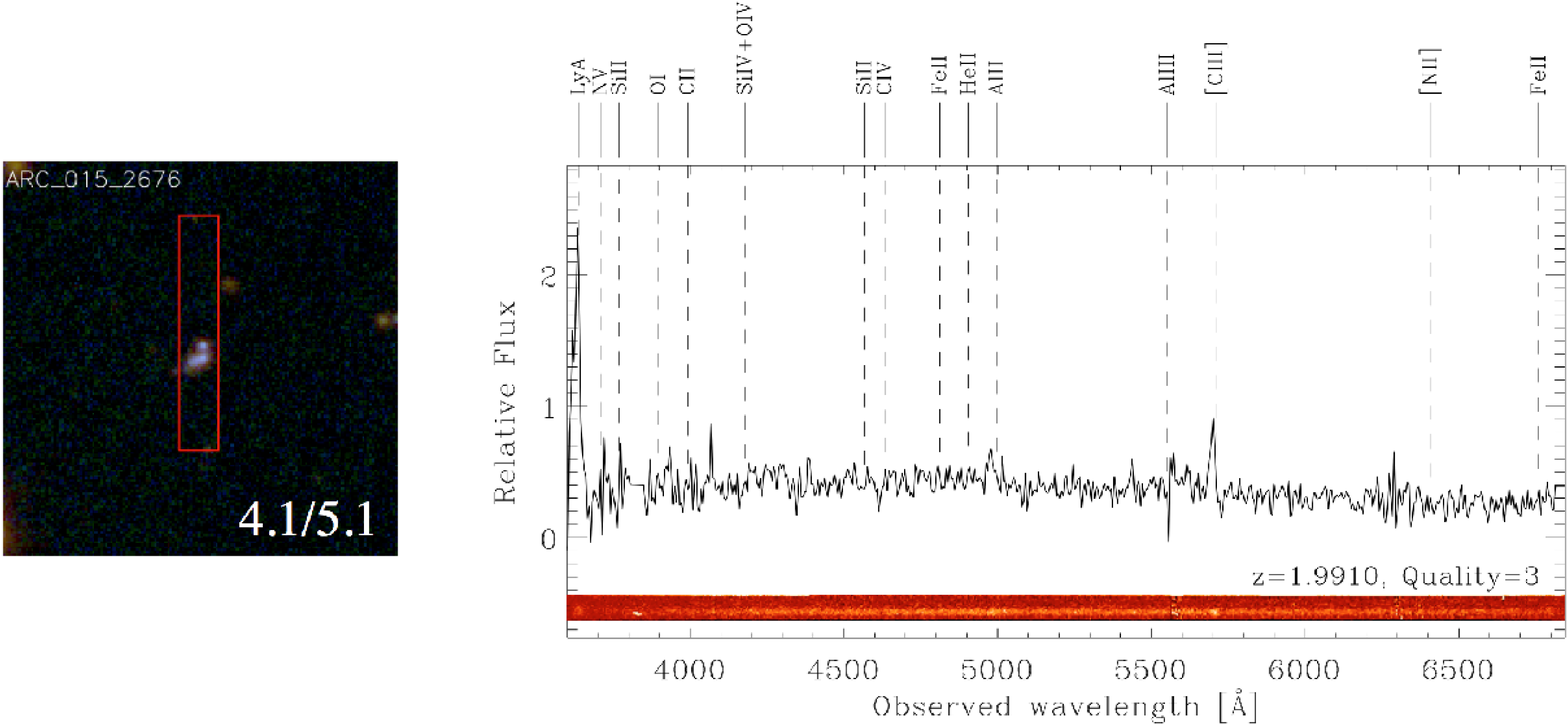}
\includegraphics[width=0.63\textwidth]{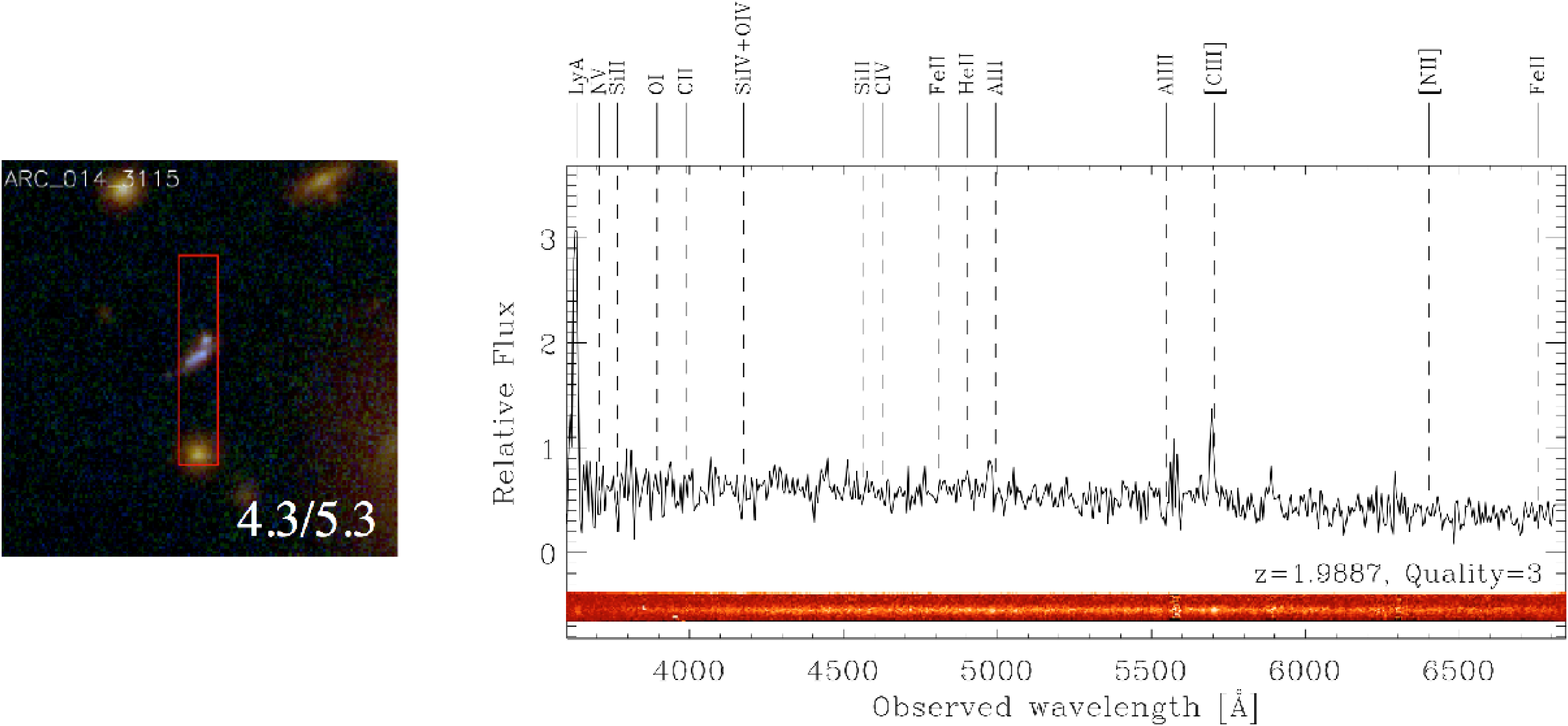}
\caption{VLT/VIMOS slits and spectra of the multiple image systems. For each lensed image, we show, on the left, a multi-color \HST\ snapshot with the VIMOS 1\arcsec-wide slit position and orientation (in red) and the associated ID from Table \ref{tab1} and, on the right, the 1D and 2D spectra with the estimated redshift value and spectroscopic quality flag (see Section \ref{sec:data:vlt}). The main emission and absorption lines of a template shifted to the measured redshift value are also indicated.}
\label{fi01}
\end{figure*}

\begin{figure*}
\figurenum{\ref{fi01}}
\centering
\includegraphics[width=0.63\textwidth]{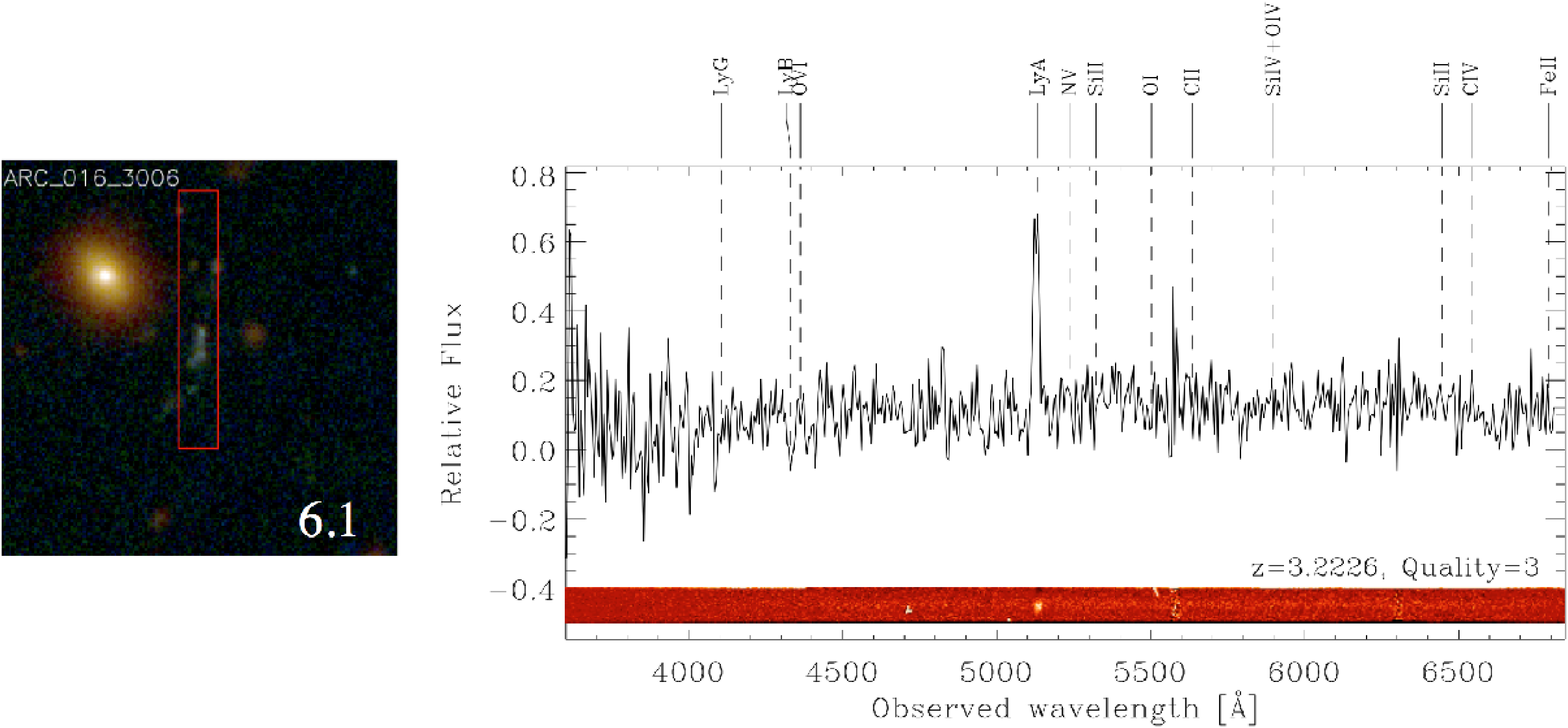}
\includegraphics[width=0.63\textwidth]{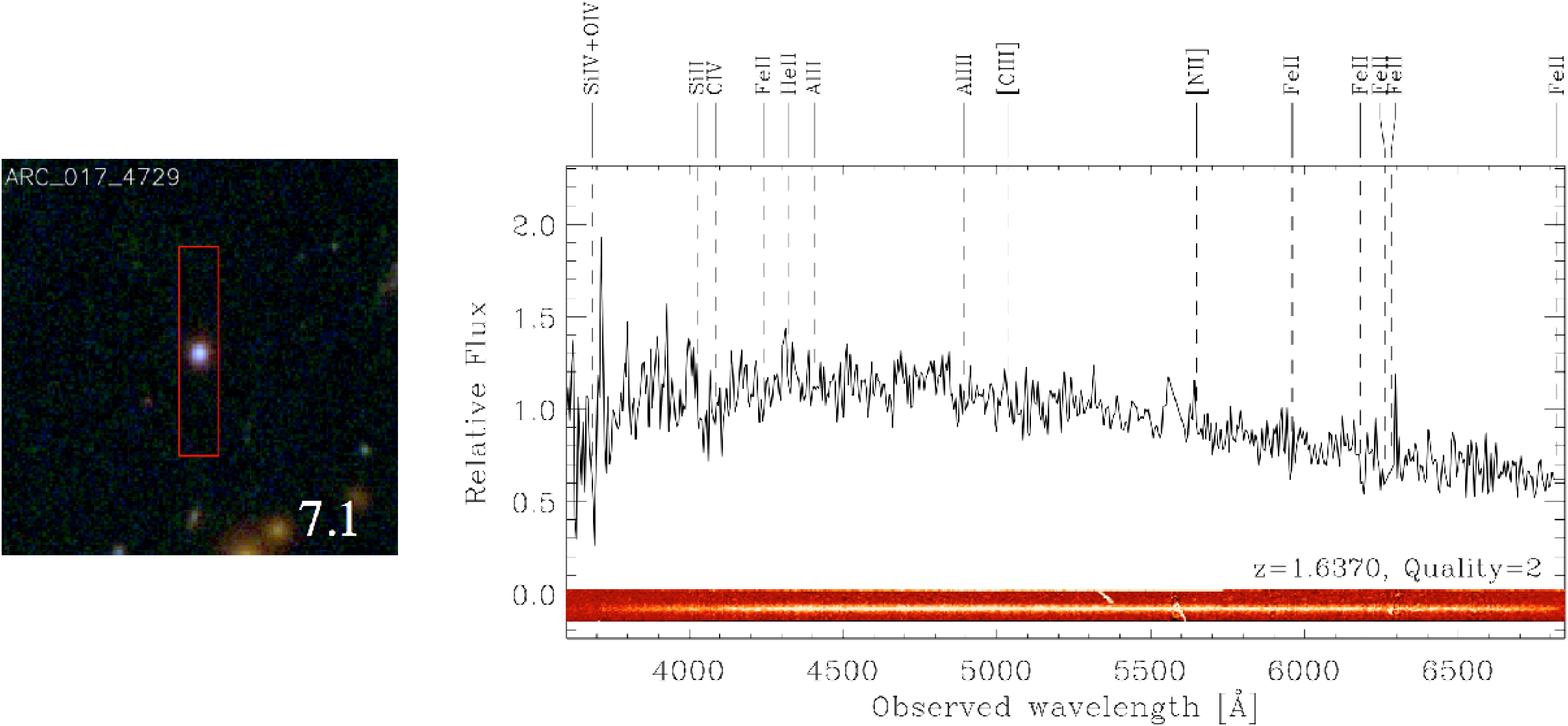}
\includegraphics[width=0.63\textwidth]{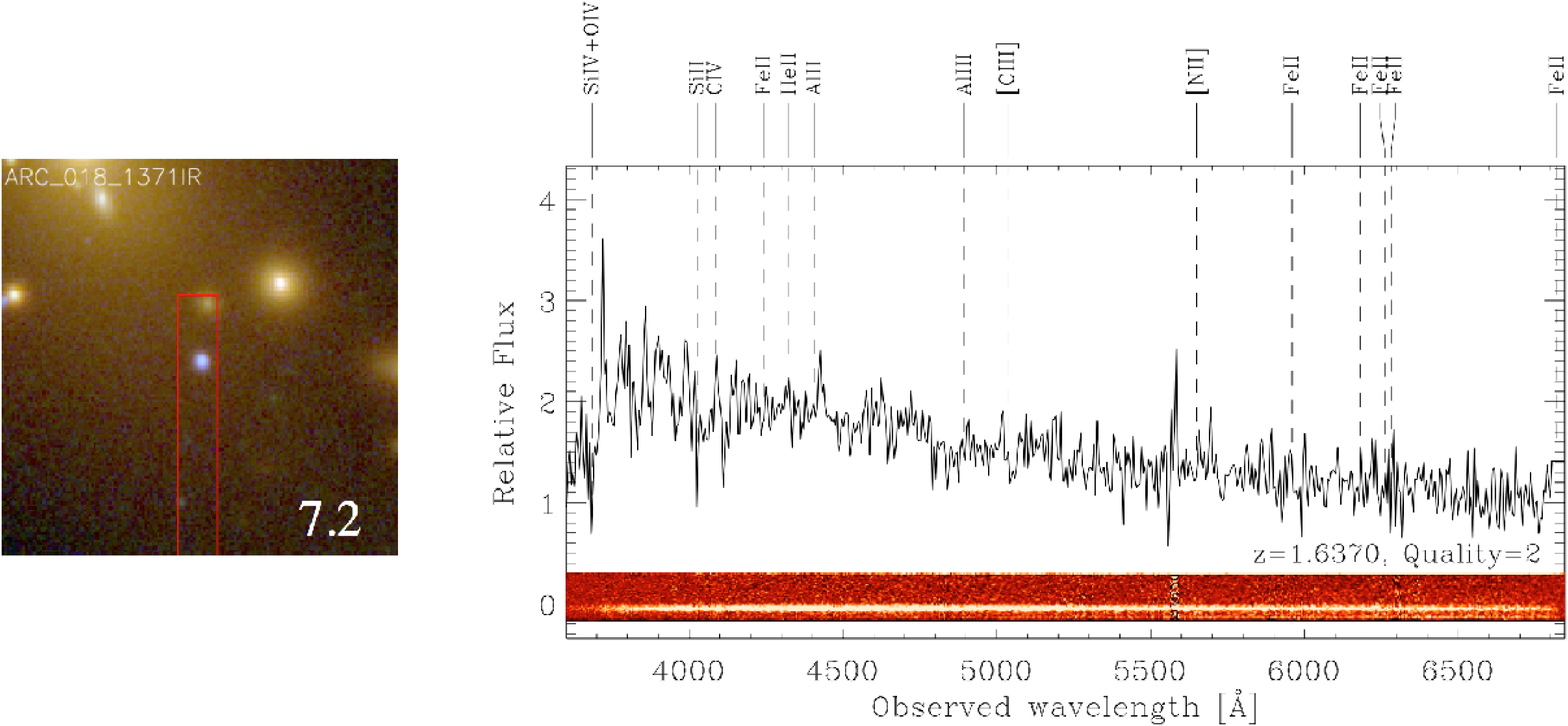}
\includegraphics[width=0.63\textwidth]{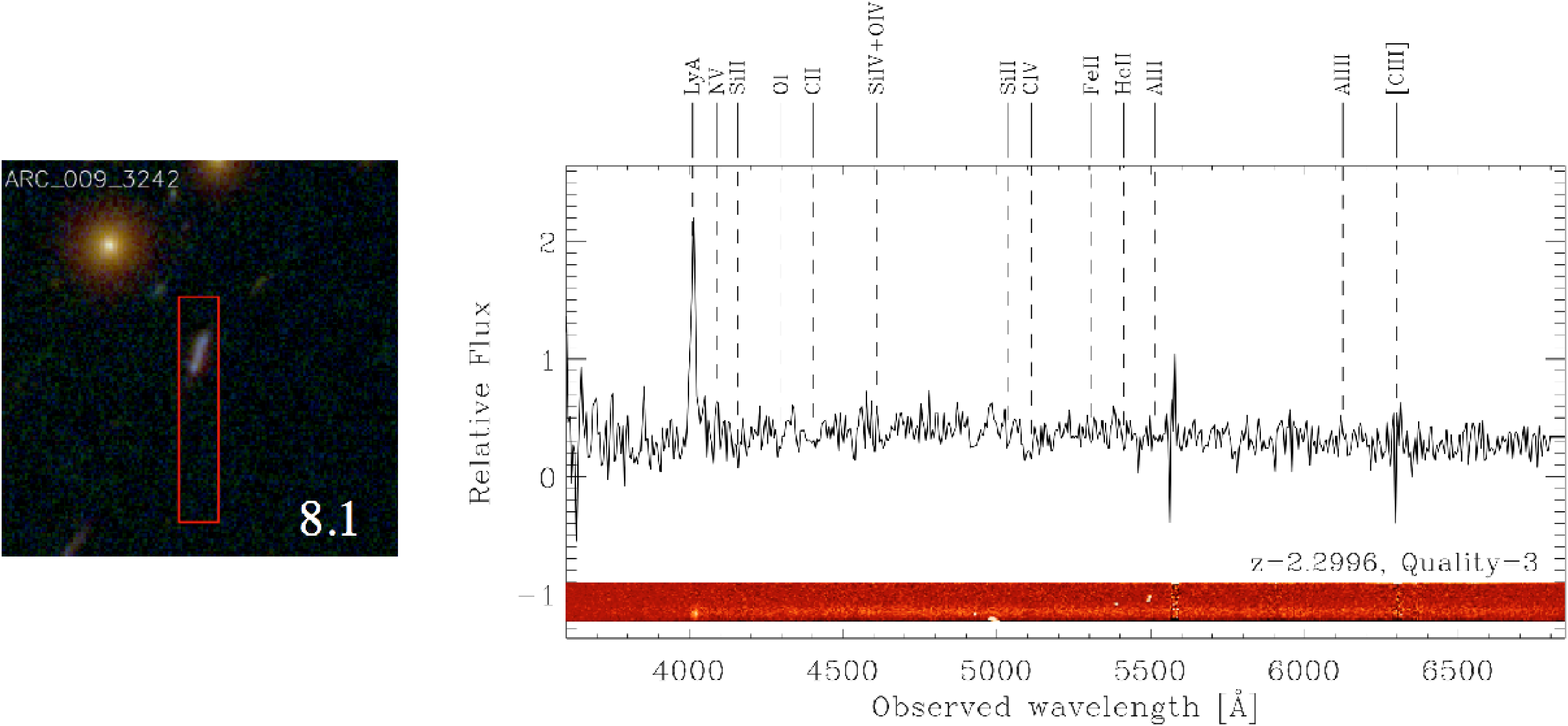}
\caption{{\it continued.}}
\end{figure*}

\begin{figure*}
\figurenum{\ref{fi01}}
\centering
\includegraphics[width=0.63\textwidth]{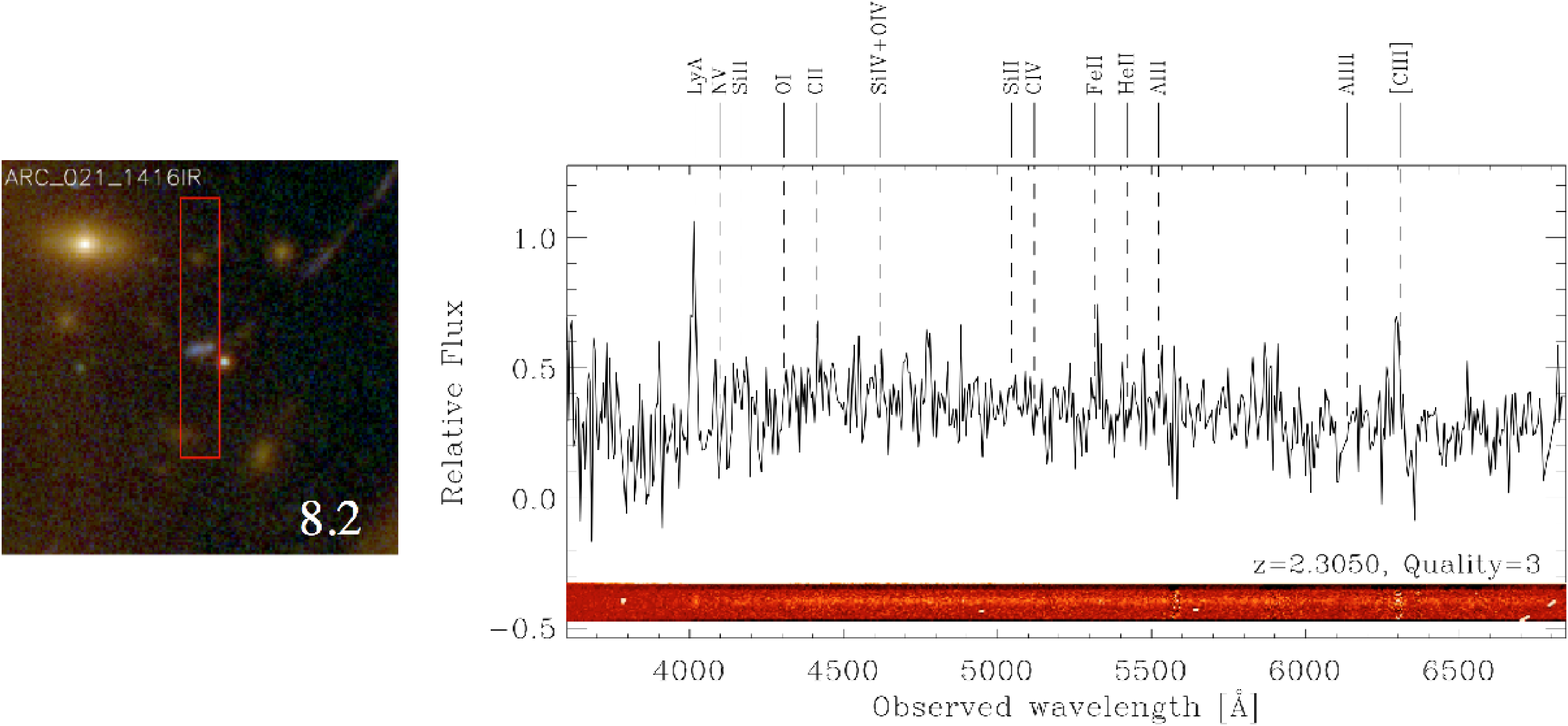}
\includegraphics[width=0.63\textwidth]{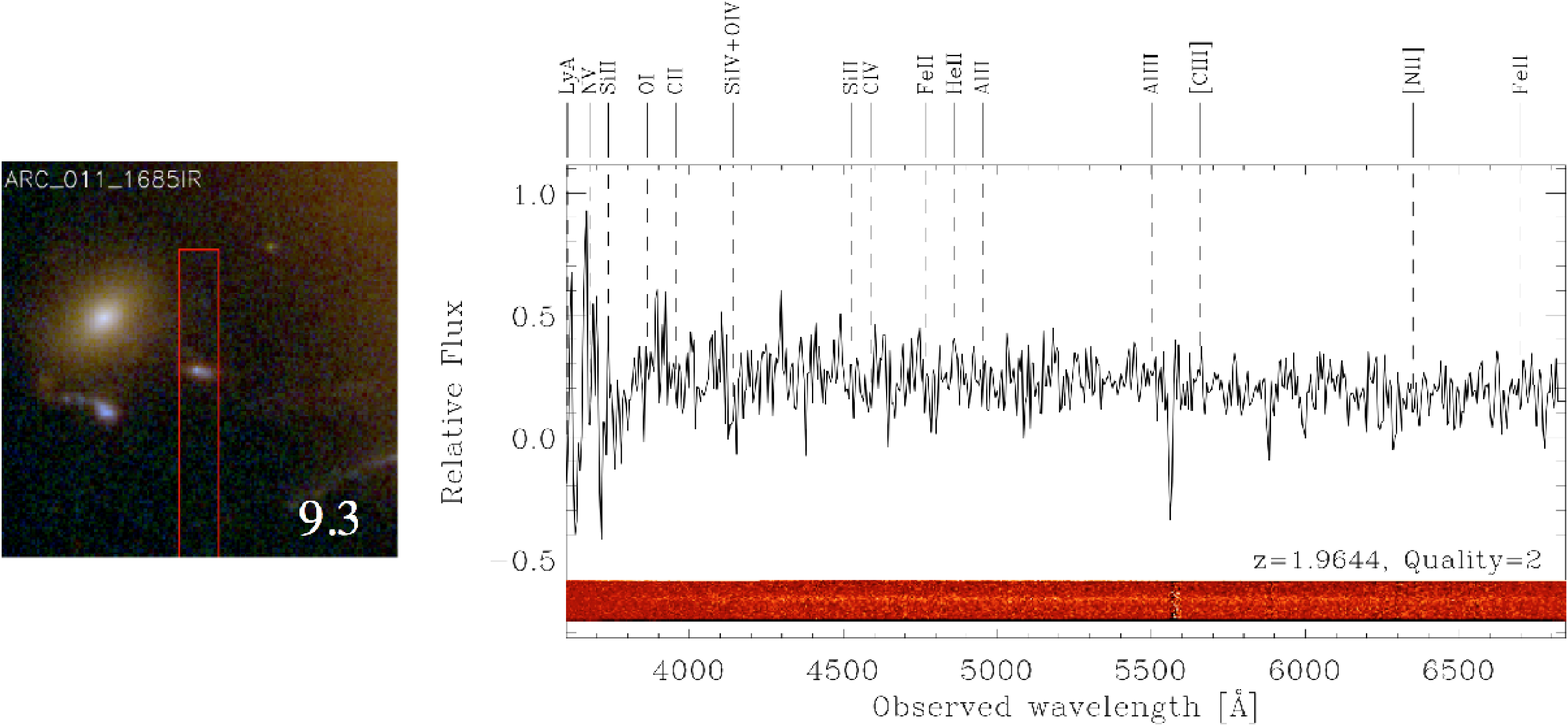}
\includegraphics[width=0.63\textwidth]{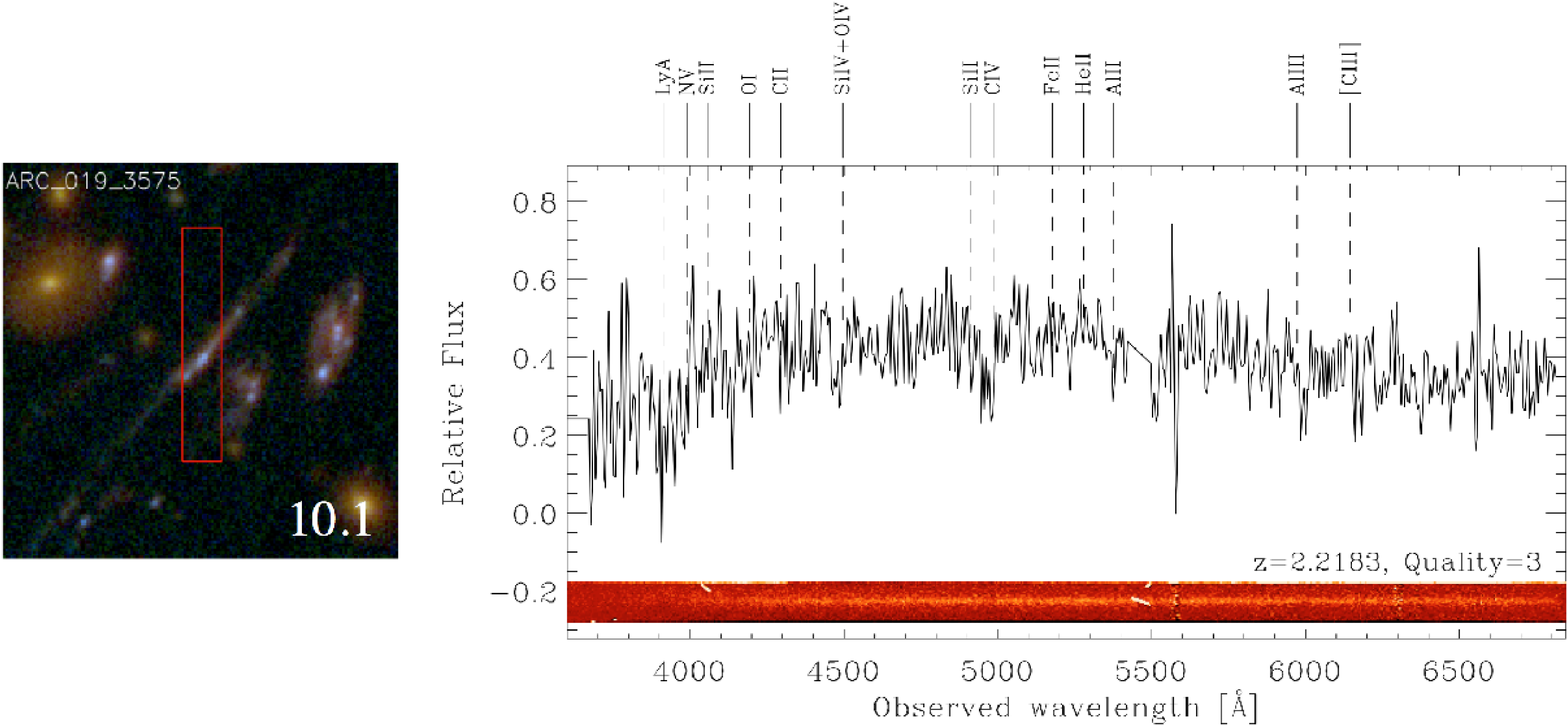}
\caption{{\it continued}}
\end{figure*}

\begin{table}
\centering
\caption{Photometric and spectroscopic properties of the images
  magnified but not multiply-imaged.}
\begin{tabular}{cccc}
\hline\hline \noalign{\smallskip}
ID & R.A. & Decl. & $z_{\mathrm{sp}}$ \\
 & (J2000) & (J2000) &  \\
\noalign{\smallskip} \hline \noalign{\smallskip}
s\_1 & 04:16:08.205 & $-$24:02:33.29 & 2.814 \\ 
s\_2 & 04:16:10.781 & $-$24:03:27.94 & 2.807 \\
s\_3 & 04:16:15.262 & $-$24:05:30.90 & 2.207 \\ \noalign{\smallskip} \hline
\end{tabular}
\label{tab4}
\end{table}

\begin{figure*}
\centering
\includegraphics[width=0.63\textwidth]{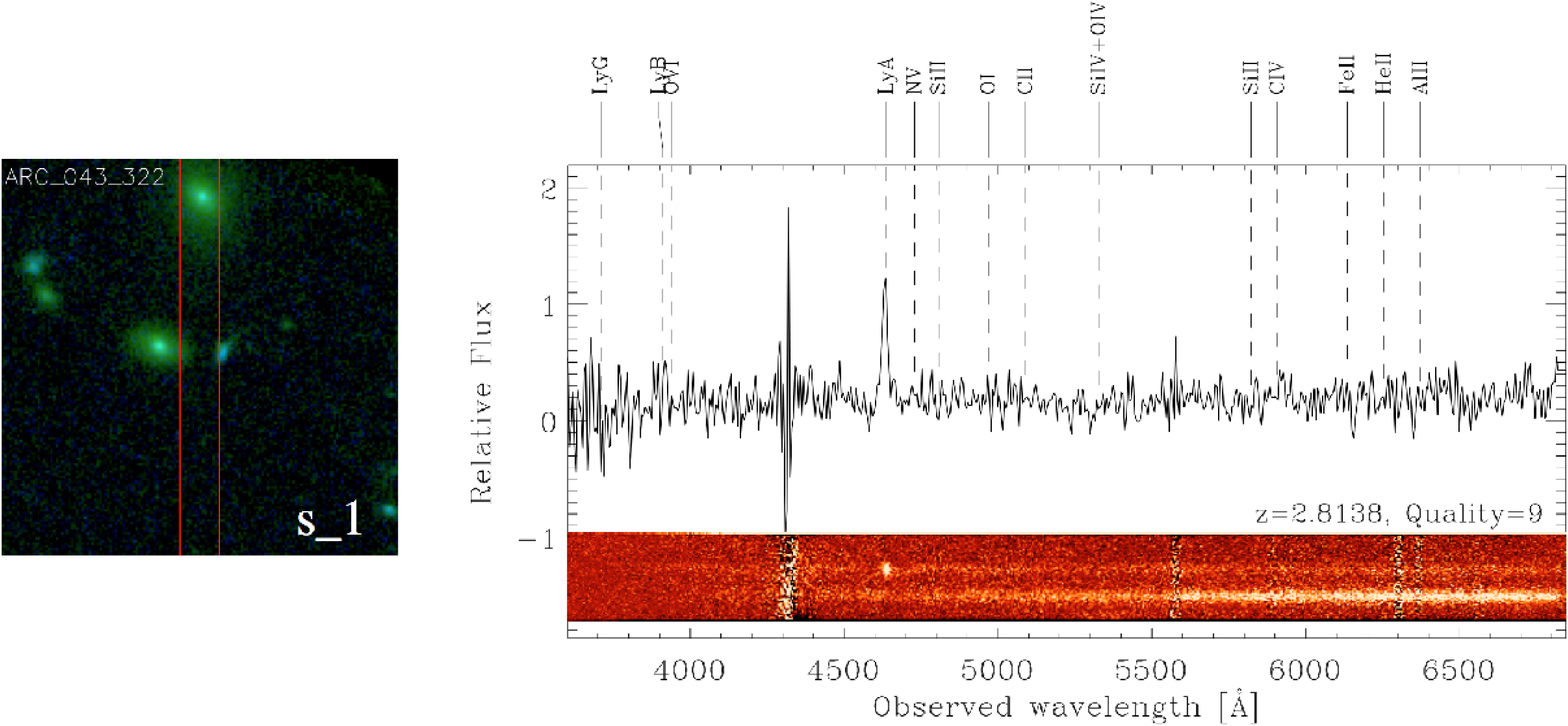}
\includegraphics[width=0.63\textwidth]{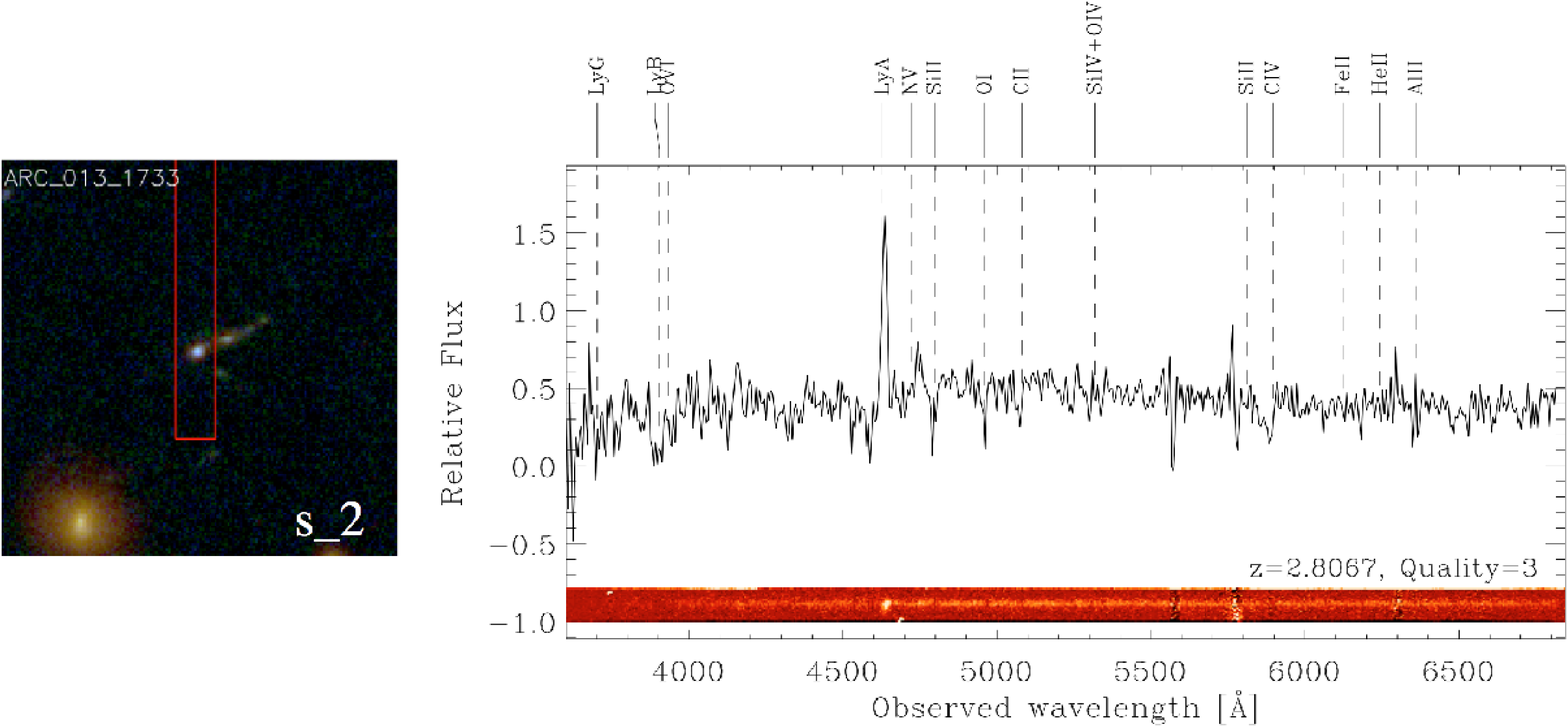}
\includegraphics[width=0.63\textwidth]{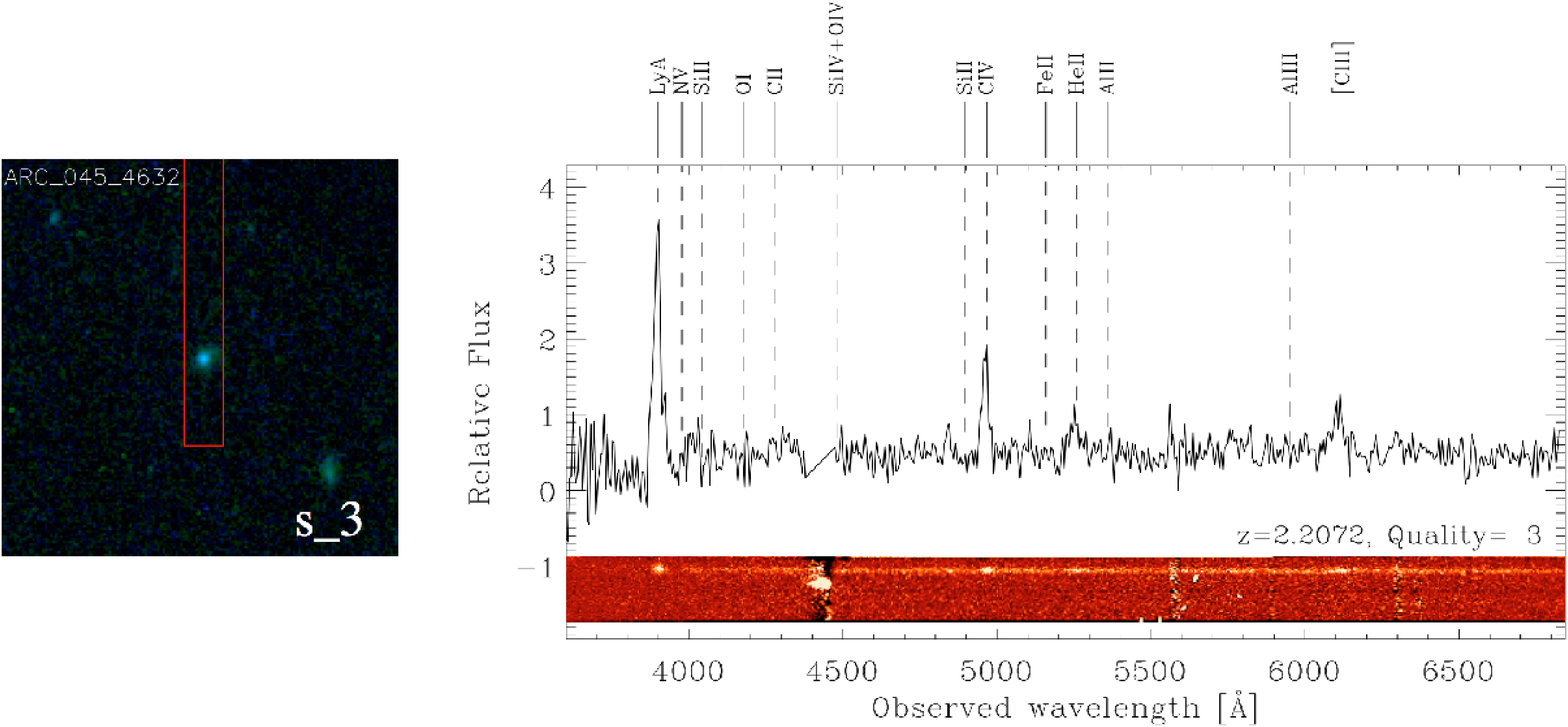}
\caption{VLT/VIMOS slits and spectra of the images that are magnified but not
  multiply-imaged. For each magnified  
image, we show, on the left, a multi-color \HST\ snapshot with the VIMOS
1\arcsec-wide slit position and orientation (in red) and the
associated ID from Table \ref{tab4} and, on the right, the 1D and 2D
spectra with the estimated redshift value and spectroscopic quality
flag (see Section \ref{sec:data:vlt}). The main emission and absorption lines of a template shifted to the measured redshift value are also indicated.}
\label{fi04}
\end{figure*}

The physical observables that we want to reconstruct are the positions
of ten multiple image systems, each of which is composed of three
images associated to one background source. We choose our multiple
image systems among the reliable or candidate systems selected by
\citet{zit13} and now spectroscopically confirmed by our VLT/VIMOS
observations. Every system has at least one multiple image with a
spectroscopic redshift value classified as either \texttt{SECURE}
(i.e., Quality = 3 in Figure \ref{fi01}) or \texttt{LIKELY} (i.e.,
Quality = 2 in Figure \ref{fi01}), according to the criteria defined
in Section 2.2. If a system has spectroscopic observations for two
images, both of which with \texttt{SECURE} estimates, we adopt for
that system the average of the \texttt{SECURE} redshift values. If a
system has spectroscopic observations for two images, one with a
\texttt{SECURE} and the other with a \texttt{LIKELY} estimate, we
adopt for that system the \texttt{SECURE} redshift value. In Figure
\ref{fi06}, we show that the multiple image systems cover a relatively
large area of the cluster central region and are distributed in a
fairly uniform way around the two brightest cluster members G1 and
G2. We remark that all sources are rather compact and well
approximated by point-like objects. Nonetheless, to exploit better the
information contained in the surface brightness distribution of two
sources, we split each of these sources into two systems (see in
Figure \ref{fi06}, systems 1 and 2 and systems 4 and 5). 

The observed angular positions, $x$ and $y$ (measured with respect to
the luminosity center of the galaxy G1 and positive in the
West and North directions), and spectroscopic redshifts,
$z_{\mathrm{sp}}$, of the thirty multiple images are listed in Table
\ref{tab1}.  The positional uncertainty for each image,
$\delta_{x,y}$, is one pixel of the chosen \HST\ images (i.e.,
0.065\arcsec). In 
Figure \ref{fi01}, we show \HST\ color-composite snapshots, with
the VIMOS 1\arcsec-wide slits marked, and the reduced 2D and 1D spectra with
the estimated redshift values of the targeted objects. We notice that
the background lensed sources span a relatively large redshift range
extending from 1.637 to 3.223. In total, the multiple images provide
sixty observables to be reproduced by a strong lensing model.  

In Table \ref{tab4} and Figure \ref{fi04}, we present three additional
background sources with reliable high-redshift measurements. They are
located outside the strong lensing region, where multiple images of a
source are created. Therefore, these objects are distorted and
magnified, but not multiply imaged, by the cluster lensing effect.

\subsection{GLEE}
\label{sec:lensmod:glee}

We model the mass distribution of MACS~0416 with {\sc Glee}, a software
developed by A.~Halkola and S.~H.~Suyu \citep{SuyuHalkola10,
  SuyuEtal12}.  We use simply-parametrized mass profiles to describe
the cluster galaxies and dark matter halo, and we denote the lens
parameters collectively as $\parlensvec$.  The image positions of the 10
multiple image systems in Table \ref{tab1} are then used to constrain
the parameters $\parlensvec$.

We use Bayesian analysis to infer the mass model parameters.  In
particular, we sample the posterior probability distribution function
(PDF) of the lens mass parameters $\parlensvec$ given the data of
observed image positions $\data$,
\be
\label{eq:bayes}
P(\parlensvec|\data) \propto P(\data|\parlensvec)\,P(\parlensvec).
\ee
The proportionality in the above equation follows from Bayes' Theorem,
and the first term to the right of the proportionality is the
likelihood whereas the second term is the prior.  The likelihood of
the lensing data is
\be
\label{eq:lenslike}
P(\data |\parlensvec) = \frac{1}{Z_{\rm {pos}}} \exp
  {\left[-\frac{1}{2}\displaystyle\sum_{j=1}^{N_{\rm
          sys}}\displaystyle\sum_{i=1}^{N_{\rm im}^j}
      \frac{\vert\boldsymbol{R}_{i,j}^{\rm obs}-\boldsymbol{R}_{i,j}^{\rm
          pred}(\parlensvec)\vert^2}{\sigma_{i,j}^2} \right]},
\ee
where $N_{\rm sys}$ is the number of multiply imaged systems (=10, as
listed in Table \ref{tab1}), 
$N_{\rm im}^j$ is the number of multiple images in system
$j$, $\boldsymbol{R}_{i,j}^{\rm obs}=(x_{i,j}^{\rm obs}, y_{i,j}^{\rm obs})$ is the observed image position, 
$\boldsymbol{R}_{i,j}^{\rm pred}(\parlensvec)$ is the predicted/modeled image
position (given the lens parameters $\parlensvec$), $\sigma_{i,j}$ is
the uncertainty in the observed image 
position, and $Z_{\rm pos}$ is the normalization given by
\be
\label{eq:lenslike_norm}
Z_{\rm pos}={(2\pi)^{N_{\rm pos}}
  \displaystyle\prod_{j=1}^{N_{\rm sys}} \displaystyle\prod_{i=1}^{N_{\rm im}^j} \sigma_{i,j}^2}
\ee
with
\be
N_{\rm pos}=\displaystyle\sum_{j=1}^{N_{\rm sys}} {N_{\rm im}^j} = 30.
\ee  
We adopt a uniform distribution as the prior $P(\parlensvec)$ on the
parameters. 

The source position for each system of multiple images is needed to
predict the image positions.  For each system, we use the deflection
angles of the lens mass model to map the observed image positions to
the source plane and take the weighted average of these mapped
positions as our source position. 
Specifically, we weight the mapped source position
$\boldsymbol{\beta}_k$ by $\sqrt{\mu_{k}}/\sigma_{k}$, where $\mu_{k}$
and $\sigma_{k}$ are the modeled magnification and the positional
uncertainty of image $k$, respectively.  In other words, we
approximate the lensing likelihood as having a delta function at the
weighted source position for each image system, thus effectively
marginalizing the source position parameters.  This approximation
works well and is computationally efficient compared to optimizing the
source position \citep[e.g.,][]{SuyuEtal12}.

We can either optimize or sample the lens parameters $\parlensvec$ in
{\sc Glee}.  
To sample the posterior PDF of $\parlensvec$, we use Markov chain
Monte Carlo (MCMC) methods that are based on \citet{DunkleyEtal05} for
efficient MCMC sampling and for assessing chain convergence.

\subsection{Mass components}
\label{sec:lensmod:masscomp}

\subsubsection{Cluster members}
\label{sec:lensmod:masscomp:gal}

\begin{figure*}
\centering
\includegraphics[width=0.8\textwidth]{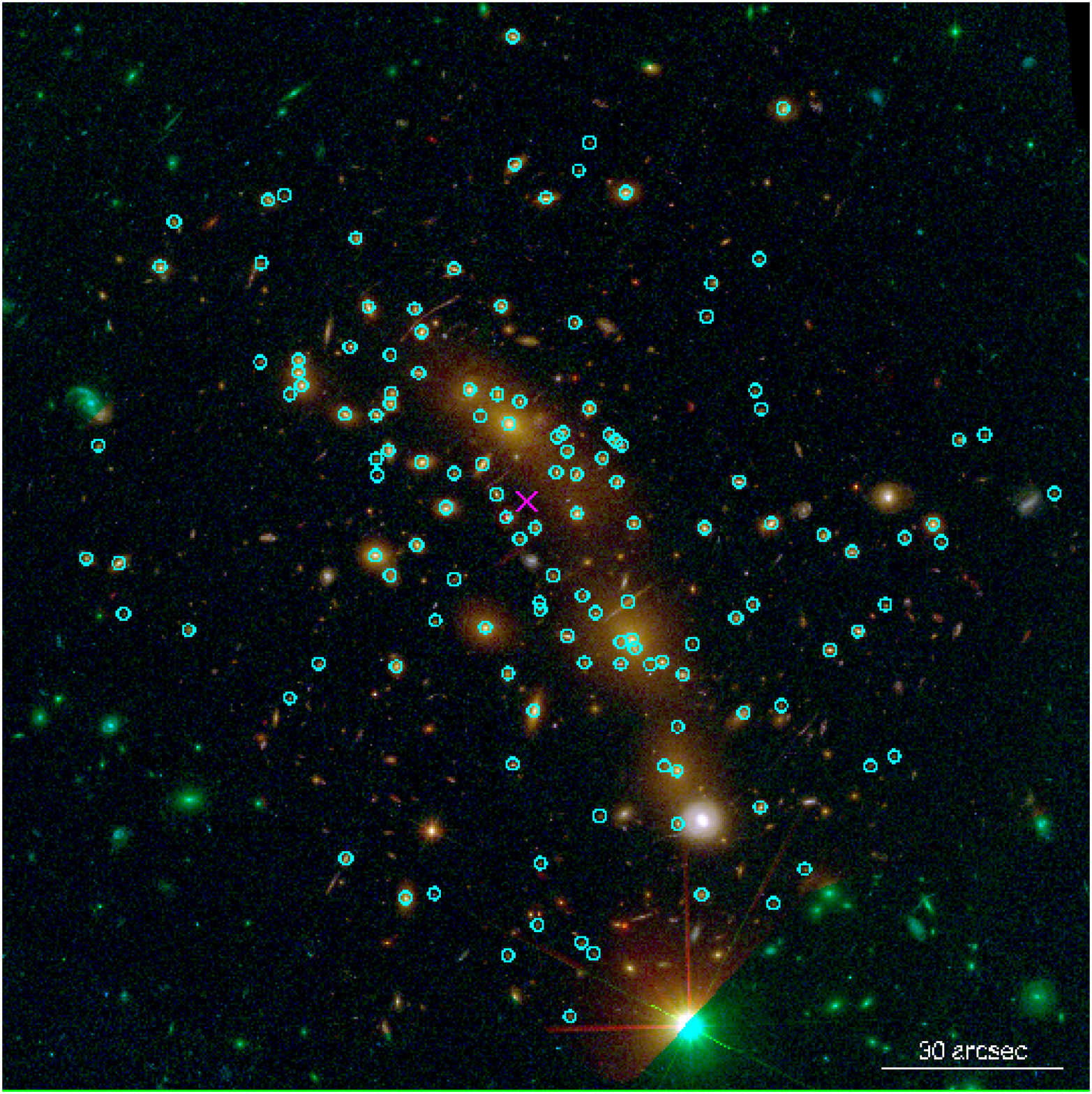}
\caption{A $3'$$\times$$3'$ color-composite image of MACS~0416 showing
  in cyan the 175 cluster members, with near-IR F160W
  magnitudes, selected with the method described in
  Section \ref{sec:lensmod:masscomp:gal}, based on spectroscopic and
  multi-color (12 bands) data, and used in the cluster strong 
  lensing models presented in Section \ref{sec:lensmod:massmod}. The
  magenta cross locates the cluster luminosity center, estimated
  by weighting the positions of the candidate cluster members with
  their F160W magnitudes. North is top and East is left.}
\label{fi05}
\end{figure*}

\begin{figure}
\centering
\includegraphics[width=0.4\textwidth]{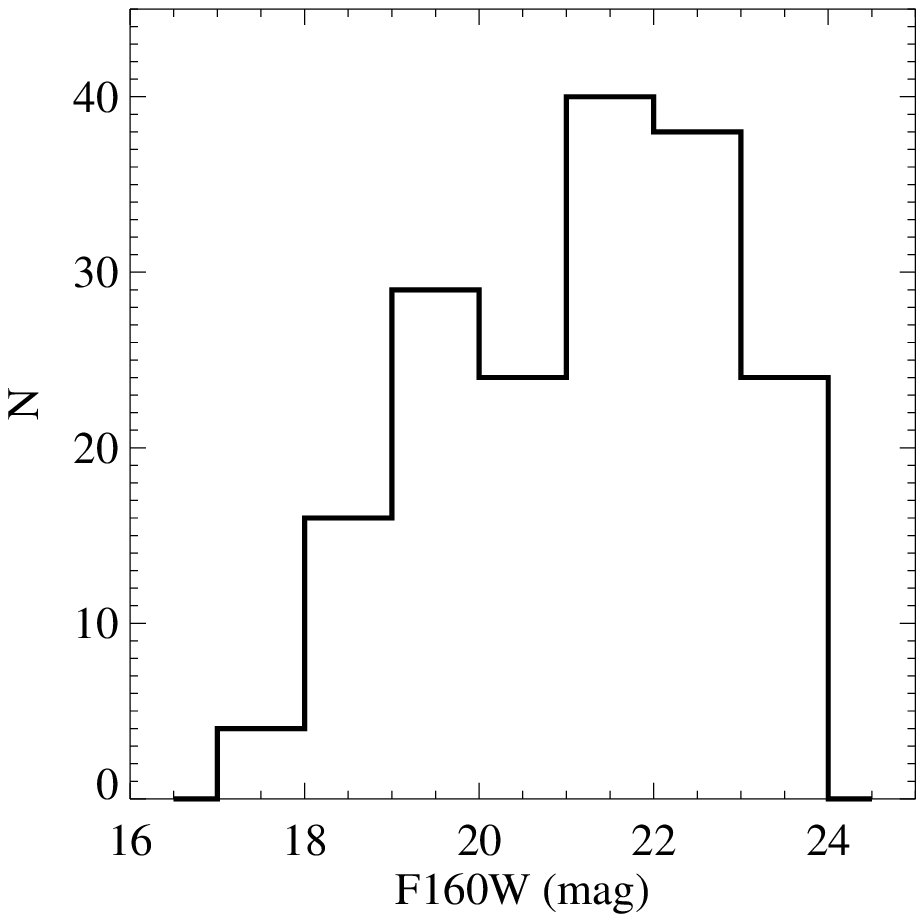}
\caption{Histogram of the near-IR F160W magnitudes of the 175 candidate cluster members shown in Figure \ref{fi05}.}
\label{fi30}
\end{figure}

The selection of cluster members is a critical step for a reliable
gravitational lensing model. In MACS~0416 we have at our disposal
already a large set of more than 800
spectroscopically confirmed cluster members, 113 of which are in the \HST\
FoV (63 with WFC3 photometry). This sample is used to
identify the locus of the member galaxies, within the \HST/WFC3 FoV, in a
multi-dimensional color 
space from 12 CLASH bands. Based on the $n$-dimensional distance
of a given galaxy from the color distribution of spectroscopic
members, we can assign a membership probability to each galaxy.

Specifically, we first select all galaxies (113) with spectroscopic redshift
in the range $0.396 \pm 0.014$, corresponding to $\pm 3000$ km s$^{-1}$
rest-frame, and with good photometric data. We exclude the F225W,
F275W, F336W, and F390W bands from the CLASH photometric data set due to the low
signal-to-noise of these data for the faint member galaxies. We then
compute the average colors and the covariance matrix from the color
distribution of spectroscopic members using a Minimum Covariance
Determinant method (MCD, \citealt{rou84a}).

Similarly, we select a representative set of field galaxies 
with redshifts \textit{outside} the range associated with cluster
members (102) and compute the mean and the covariance matrix of the colors.
We assume that the population of cluster members and of field
galaxies can each be well described by a multivariate normal
distribution with the previously determined averages and
covariances. Despite this approximation, we verify a-posteriori that
it produces catalogs of 
cluster members with good purity and completeness.

We tune the member probability threshold in order to maximize the
purity of the cluster members, particularly at the bright-end of the
luminosity function, where the most massive galaxies (those that
provide the most important 
contribution to the mass model of the cluster) reside.

With this method we select 109 members. We find that this represents
a pure sample of cluster members, at the expense of some moderate
incompleteness. The latter can be significantly alleviated by studying
the color magnitude relation (CMR) of spectroscopic and
photometric members. Note that the method outlined above does not use
any a priori knowledge of the color-magnitude distribution of
galaxies, particularly for specific colors which straddle the H+K
break and hence produce well distinguished color sequences for cluster
galaxies. We therefore supplement the photometric sample obtained from
the galaxy distribution in color space with galaxies, fainter than the
brightest cluster galaxies, lying on the cluster sequence of the color-magnitude diagram of
F606W$-$F814W vs F606W. We 
define the mean CMR by using the biweight estimator on
spectroscopically confirmed members and select 66 more galaxies lying
within a scatter of 0.15 mag from the mean. We verify that the
extended sample of cluster galaxies is $\ge 95\%$ complete down to
F160W(AB)=21. We fix our F160W magnitude limit at 24 mag,
corresponding to approximately $m^{\star}+4.5$ and one magnitude 
lower than the value of our faintest spectroscopically confirmed
cluster member (see Figure \ref{fi21}), beyond which it becomes very
difficult to have a reliable estimate of the purity and completeness of
the sample. We obtain a final catalogue of
candidate cluster members containing 175 
objects (see Figures \ref{fi05} and \ref{fi30}). Further details on
the selection and statistical analysis of the photometric sample of
cluster galaxies will be included in Balestra et al. (in prep.).

\begin{figure}
\centering
\includegraphics[width=0.46\textwidth]{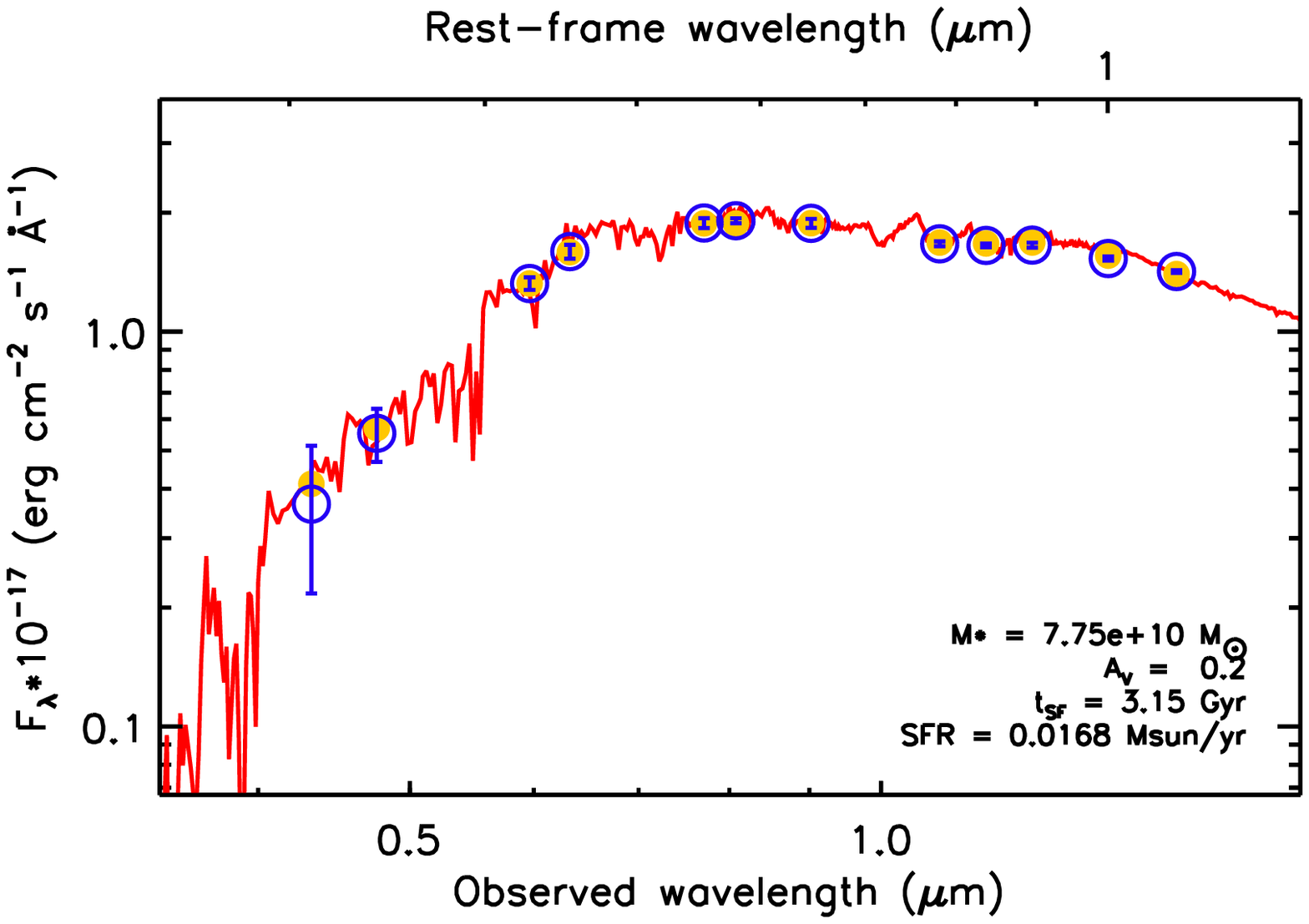}
\caption{Example of the composite stellar population modeling of the
  12 reddest \HST\ bands of a spectroscopically confirmed galaxy cluster
  member. We use \citet{bru03} templates (the best-fitting one is
  shown here in red) at solar metallicity, with dust and a Salpeter
  stellar IMF. Observed fluxes with 1$\sigma$ errors are represented
  with blue empty circles and bars, model-predicted fluxes are shown
  as orange filled circles.}
\label{fi20}
\end{figure}

\begin{figure}
\centering
\includegraphics[width=0.46\textwidth]{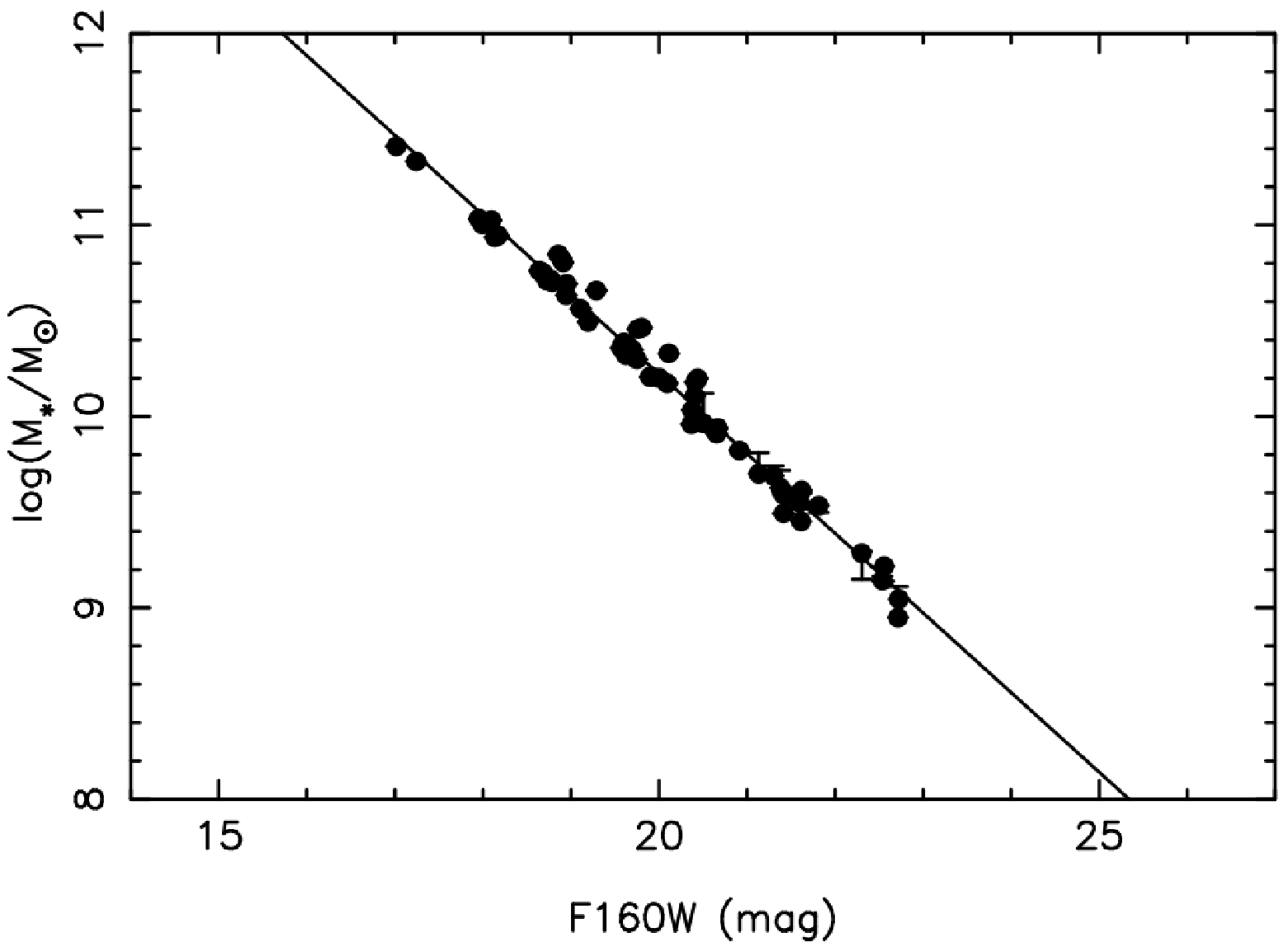}
\caption{Best-fitting stellar mass values, obtained from the SED modeling of the multicolor \HST\ photometry, as a function of the magnitude values measured in the reddest \HST/WFC3 broadband (F160W). The points, with 1$\sigma$ error bars, represent 63 spectroscopically confirmed galaxy cluster members. The solid line shows the best-fitting line.}
\label{fi21}
\end{figure}

To determine the mass in the form of stars present in the
spectroscopically 
confirmed cluster members, we fit their spectral energy distributions
(SEDs), composed of the 12 reddest \HST\ bands, through composite stellar
population (CSP) models based on \citet{bru03} templates at solar
metallicity and with a \citet{sal55} stellar IMF. We consider delayed
exponential star formation histories (SFHs) and allow for the presence
of dust, according to \citet{cal00} (see also \citealt{gri09,gri14}). We
show a representative example in Figure \ref{fi20}. From Figure
\ref{fi21}, we remark that the values of the cluster member stellar
masses and F160W magnitudes are very tightly correlated. We find that
the best-fitting line is $\log(M_{*}/M_{\odot}) = 18.541-0.416 \times
{\rm F160W}$. According to this relation and to the 
F160W galaxy luminosities, we assign a stellar mass value to each
candidate cluster member. Thus, our F160W magnitude limit corresponds
to $\log(M_*/M_\odot)\simeq 8.6$.

For each cluster galaxy selected, we model its projected dimensionless
surface mass density, a.k.a.~convergence, as 
a dual pseudoisothermal elliptical mass distribution 
\citep[dPIE;\ ][]{EliasdottirEtal07, SuyuHalkola10} with vanishing
ellipticity and core radius,
\be
\label{eq:dpie}
\kappa_{\rm g}(x,y) = \frac{\vartheta_{\rm
    E}}{2}\left(\frac{1}{R} -
  \frac{1}{\sqrt{R^2+r_{\rm t}^2}} \right),
\ee
where $(x,y)$ are coordinates on the image/lens
plane, $\vartheta_{\rm E}$ is the cluster galaxy lens strength
(a.k.a.~Einstein radius),
$R$ (=$\sqrt{x^2+y^2})$ is the 
radial coordinate, and $r_{\rm t}$ is the
``truncation radius''.  The truncated isothermal distribution is
suitably translated by the centroid position of the cluster galaxy
luminosity ($x_{\rm g}$, $y_{\rm g}$).  

The lensing convergence depends in general on the lens and source
redshifts.  In MACS~0416, there is one lens redshift (that of the
galaxy cluster), and various source redshifts of the multiple image
systems.  The convergence defined in Equation (\ref{eq:dpie}) is
relative to a background source at redshift $\infty$.  For a source at a
redshift $z_{\rm s}$, the convergence associated with that particular
source is
\be
\label{eq:dpie:sscl}
\kappa_{\rm g}|_{z=z_{\rm s}} = \frac{D_{\rm ds}}{D_{\rm s}} \kappa_{\rm g}|_{z=\infty},
\ee
where $D_{\rm ds}$ is the angular diameter distance of the source as
viewed from the lens, and $D_{\rm s}$ is the angular diameter distance
to the source from us.  Therefore, the factor ${D_{\rm ds}}/{D_{\rm
    s}}$ is used to relate the deflection angles for the background
sources at different redshifts.

For an isothermal profile, we can relate the velocity dispersion of
the cluster galaxy to its Einstein radius $\vartheta_{\rm E}$ via
\be
\label{eq:vdisp_thE}
\sigma/c = \sqrt{\frac{\vartheta_{\rm E}}{4\pi}},
\ee
where $c$ is the speed of light.  Furthermore, the circular velocity of
the galaxy, $v_{\rm c}$, is related to its velocity dispersion
$\sigma$ via 
\be
\label{eq:vcirc_vdisp}
v_{\rm c} = \sqrt{2} \sigma.
\ee

The three-dimensional mass density
distribution corresponding to Equation (\ref{eq:dpie}) is
\be
\label{eq:dpie_rho}
\rho(r) \propto \frac{1}{r^2(r^2+r_{\rm t}^2)},
\ee
where $r$ is the three dimensional radius ($r=\sqrt{x^2+y^2+z^2}$).
Note that for $r \gg r_{\rm t}$ the mass density distribution scales
as $r^{-4}$ and is thus ``truncated'', and $r_{\rm t}$ is roughly
the half-mass radius \citep[e.g.,][]{EliasdottirEtal07}.

\subsubsection{Cluster dark-matter halos}
\label{sec:lensmod:masscomp:halo}

To complete the total mass modeling of the cluster on radial scales
larger than those typical of the cluster members, we include two
additional mass components. We use two components because the
cluster luminosity distribution shows two main peaks.
These components are intended to represent
the contribution to the total mass budget of all remaining mass (intra
cluster light, hot gas and, mainly, dark matter) not associated to the
galaxy luminous and dark matter mass distributions. We consider two
forms of mass distributions for the cluster mass components: (1)  
two-dimensional, pseudo-isothermal, elliptical (\citealt{kas93};
hereafter PIEMD), and (2) three-dimensional, prolate, Navarro, Frenk and
White (\citealt{ogu03}; hereafter PNFW) mass profiles.  Below we
describe the convergence relative to a background source at
redshift $\infty$; for a source at $z_{\rm s}$, the convergence is
scaled analogously to Equation (\ref{eq:dpie:sscl}).
In Section \ref{sec:lensmod:results}, we compare the performances of
the two mass models.

The dimensionless surface mass density of PIEMD
is of the form 
\be
\label{eq:piemd}
\kappa_{\rm h}(x,y) = \frac{\vartheta_{\rm
    E,h}}{2\sqrt{R_{\rm \epsilon}^2 + r_{\rm c,h}^2}},
\ee
where 
\be
\label{eq:rem}
R_{\rm \epsilon}^2 =\frac{{x^2}}{(1+\epsilon)^2}+\frac{y^2}{(1-\epsilon)^2}, 
\ee
$\epsilon$ is the ellipticity defined as $\epsilon\equiv(1-q_{\rm
  h})/(1+q_{\rm h})$ with $q_{\rm h}$ being the axis ratio.  The
strength of the halo is $\vartheta_{\rm E,h}$, and the distribution has a
central core with radius $r_{\rm c,h}$ that marks the transition in the radial
dependence in the convergence from $R^0$ to $R^{-1}$.  The distribution
is appropriately translated by the centroid position of the cluster
halo ($x_{\rm h}$, $y_{\rm h}$) and rotated by the
position angle $\phi_{\rm h}$.  Each PIEMD thus requires 6 parameters to
characterize ($x_{\rm h}$, $y_{\rm h}$, $q_{\rm h}$, $\phi_{\rm
  h}$, $\vartheta_{\rm E,h}$, $r_{\rm c,h}$).

The three-dimensional density distribution of PNFW is given by
\begin{equation}
\label{eq:pnfw}
\rho_{\rm h}(r) = \frac{\rho_{\rm 0,h}}{(r/r_{\rm
    s,h})(1+r/r_{\rm s,h})^2}
\end{equation}
where
\begin{equation}
\label{eq:pnfw_r}
  r^2 = c^2\left(\frac{x^2+y^2}{a^2} + 
    \frac{z^2}{c^2}\right), \quad a\leq c.
\end{equation}
The parameter $a/c$ describes the prolateness of the halo: $a/c=1$
corresponds to a spherical halo, whereas $a/c \ll 0$ corresponds to a
highly elongated halo. 
The orientation of the dark matter halo as seen by a distant observer
can be described by two angles: (1) $\varphi_{\mathrm{h}}$, the
viewing angle responsible for the level of ellipticity of the
two-dimensional projection where $\varphi_{\rm h}=0$ yield a projected
axis ratio of 1 and $\varphi_{\rm h}=90\degr$ yields the projected
axis ratio of $a/c$, and (2) $\phi_{\mathrm{h}}$, the projected major axis
position angle.  We use the Einstein radius of the cluster halo,
$\vartheta_{\rm E,h}$, instead of $\rho_{\rm 0,h}$ to characterize the
strength/mass of the halo since strong lensing allows us to measure
robustly $\vartheta_{\rm E,h}$.  We refer to \citet{SuyuEtal12} for
the relation between $\vartheta_{\rm E,h}$ and $\rho_{\rm 0,h}$.  In
summary, the PNFW is described by 7 parameters: $x_{\rm h}$, $y_{\rm
  h}$, $a/c$, $\varphi_{\mathrm{h}}$, $\phi_{\mathrm{h}}$,
$\vartheta_{\rm E,h}$, $r_{\rm s,h}$.

\subsection{Mass models}
\label{sec:lensmod:massmod}

In our analysis, we explore different mass models for the galaxy cluster, varying the mass weighting of the cluster members and the mass parametrization of the cluster dark-matter halos.

We start with a model (labeled as 2PIEMD) with only two PIEMD mass
profiles (see Section \ref{sec:lensmod:masscomp:halo}), describing the
extended and smooth total mass distribution of the cluster. Then, we
add to the two PIEMDs the mass contribution on smaller scales of the
175 cluster members selected in Section \ref{sec:lensmod:masscomp:gal}. We decide to use their luminosity values, $L$, in the reddest WFC3 band (i.e., the F160W) to
assign the relative total mass weights to their dPIE profiles (see
Section \ref{sec:lensmod:masscomp:gal}).
In detail, we choose the following scaling relations for the values of the Einstein radius, $\vartheta_{\mathrm{E},i}$, and truncation radius, $r_{\mathrm{t},i}$, of the $i$-th cluster member:
\begin{equation}
\label{eq:sr1}
\vartheta_{\mathrm{E},i}=\vartheta_{\mathrm{E,g}}
\left(\frac{L_{i}}{L_{\rm g}}\right)^{0.5} \quad \textrm{and} \quad
r_{\mathrm{t},i}=r_{\mathrm{t,g}} \left(\frac{L_{i}}{L_{\rm g}}\right)^{0.5} \, ,
\end{equation}
where $\vartheta_{\mathrm{E},g}$ and $r_{\mathrm{t},g}$ are two reference values, corresponding, in particular, to those of the brightest cluster galaxy G1. Recalling that for a dPIE profile the total mass, $M_{\mathrm{T}}$, is proportional to the product of the squared value of the effective velocity dispersion, $\sigma$ (where $\sigma \sim \vartheta_{\mathrm{E}}^{0.5}$), and the truncation radius, the relations adopted in Equation (\ref{eq:sr1}) imply that
\begin{equation}
\label{eq:mr1}
\frac{M_{\mathrm{T},i}}{L_{i}} \sim \frac{\sigma_{i}^{2}r_{\mathrm{t},i}}{L_{i}} \sim \frac{L_{i}^{0.5}L_{i}^{0.5}}{L_{i}} \sim L_{i}^{0}.
\end{equation}
This is therefore equivalent to having cluster members with constant
total mass-to-light ratios. We identify this model with 2PIEMD +
175(+1)dPIE ($M_{\mathrm{T}}L^{-1}=k$). Note that in this model, as
well as in the following ones, we include an extra (+1) dPIE mass
component to take into account the lensing contribution of the bright
foreground galaxy (R.A.: 04:16:06.820; Decl.: $-$24:05:08.45;
$z_{\mathrm{sp}}$ = 0.114, Quality = 3) that is in projection very close to image
10.2 (see Figure \ref{fi06}). We postpone to the future a more complex
and rigorous multi-plane lensing analysis and take here into account the
different redshift of this particular galaxy through optimizing its
\textit{effective} values of $\vartheta_{\mathrm{E}}$ and
$r_{\mathrm{t}}$ without any constraints.

Next, we investigate whether we can find a better lensing model by changing our assumption on the constant total mass-to-light ratio for the candidate cluster members. In particular, we test the following two relations:
\begin{equation}
\label{eq:sr2}
\vartheta_{\mathrm{E},i}=\vartheta_{\mathrm{E,g}}
\left(\frac{L_{i}}{L_{\rm g}}\right)^{0.7} \quad \textrm{and} \quad
r_{\mathrm{t},i}=r_{\mathrm{t,g}} \left(\frac{L_{i}}{L_{\rm g}}\right)^{0.5} \, ,
\end{equation}
\begin{equation}
\label{eq:sr3}
\vartheta_{\mathrm{E},i}=\vartheta_{\mathrm{E,g}}
\left(\frac{L_{i}}{L_{\rm g}}\right)^{0.5} \quad \textrm{and} \quad
r_{\mathrm{t},i}=r_{\mathrm{t,g}} \left(\frac{L_{i}}{L_{\rm g}}\right)^{0.25} \, .
\end{equation}
The first one corresponds to values of the total mass-to-light ratio that increase with the luminosity. In particular, we have that
\begin{equation}
\label{eq:mr2}
\frac{M_{\mathrm{T},i}}{L_{i}} \sim \frac{\sigma_{i}^{2}r_{\mathrm{t},i}}{L_{i}} \sim \frac{L_{i}^{0.7}L_{i}^{0.5}}{L_{i}} \sim L_{i}^{0.2}.
\end{equation}
This relation between $M_{\mathrm{T}}/L$ and $L$ is particularly
interesting because it has been used to interpret the systematic
increase of galaxy effective mass-to-light ratio with effective mass
(also known as the tilt of the Fundamental Plane; e.g., \citealt{fab87};
\citealt{ben92}) observed 
in early-type galaxies. The second one instead is motivated by
theoretical studies (e.g., \citealt{mer83}) that predict a linear
relation between 
truncation radius and velocity dispersion for galaxies residing in a
cluster environment. We refer to these two models as 2PIEMD +
175(+1)dPIE ($M_{\mathrm{T}}L^{-1}\sim L^{0.2}$) and 2PIEMD +
175(+1)dPIE ($r_{\mathrm{t}} \sim \sigma$), respectively. 

Finally, we try four additional models, analogous to the previous ones, in which only the mass distribution of the two extended cluster dark-matter halos is substituted with PNFW profiles (see Section \ref{sec:lensmod:masscomp:halo}). We label these models as 2PNFW, 2PNFW + 175(+1)dPIE ($M_{\mathrm{T}}L^{-1}=k$), 2PNFW + 175(+1)dPIE ($M_{\mathrm{T}}L^{-1}\sim L^{0.2}$), and 2PNFW + 175(+1)dPIE ($r_{\mathrm{t}} \sim \sigma$), respectively.

\subsection{Results}
\label{sec:lensmod:results}

\begin{table}
\centering
\caption{The investigated strong lensing models and their best-fitting, minimum-$\chi^{2}$ values.}
\begin{tabular}{cc}
\hline\hline \noalign{\smallskip}
Model & $\chi^{2}$ \\
\noalign{\smallskip} \hline \noalign{\smallskip}
2PIEMD & 6032 \\
2PIEMD + 175(+1)dPIE ($M_{\mathrm{T}}L^{-1}=k$) & 1169 \\
2PIEMD + 175(+1)dPIE ($M_{\mathrm{T}}L^{-1}\sim L^{0.2}$) & 915 \\ 
2PIEMD + 175(+1)dPIE ($r_{\mathrm{t}} \sim \sigma$) & 1262 \\
2PNFW & 6973 \\
2PNFW + 175(+1)dPIE ($M_{\mathrm{T}}L^{-1}=k$) & 1767 \\
2PNFW + 175(+1)dPIE ($M_{\mathrm{T}}L^{-1}\sim L^{0.2}$) & 1529 \\ 
2PNFW + 175(+1)dPIE ($r_{\mathrm{t}} \sim \sigma$) & 1901 \\
\noalign{\smallskip} \hline
\end{tabular}
\label{models}
\end{table}

We show the best-fitting, minimum-$\chi^{2}$ values of the eight different mass models in Table
\ref{models}. First, we notice that the inclusion of the cluster
members results in $\chi^2$ values that are always more than a factor
of 3 lower
than those obtained with only the two extended cluster dark-matter halos. Then, we 
find that the models with scaling of the
galaxy total mass-to-light ratio increasing with the luminosity are slightly
better in reproducing the observed multiple image systems. Interestingly,
we also see that there is significant evidence that cored elliptical
pseudo-isothermal profiles are better-suited than three-dimensional,
prolate, Navarro, Frenk and White profiles to represent the extended
total mass distribution of MACS~0416; considering the models with the cluster member contribution, 
the $\chi^2$ values of the prolate NFW
halos are more than 50\% higher than those of the PIEMD halos, despite the
prolate NFW halos having more parameters. 

In summary, we conclude that the mass model of MACS~0416 that best
fits the strong lensing observables is composed of 2 cored elliptical
pseudo-isothermal mass distributions and numerous (175) dual
pseudo-isothermal mass distributions, scaled with total
mass-to-light ratios increasing with the near-IR luminosities of the
candidate cluster members. We confirm that detailed modeling on the
small mass/radial scales of the many cluster galaxies is fundamental to
a precise multiple image reconstruction. 

\subsubsection{The best-fitting model}

\begin{figure}
\centering
\includegraphics[width=0.36\textwidth]{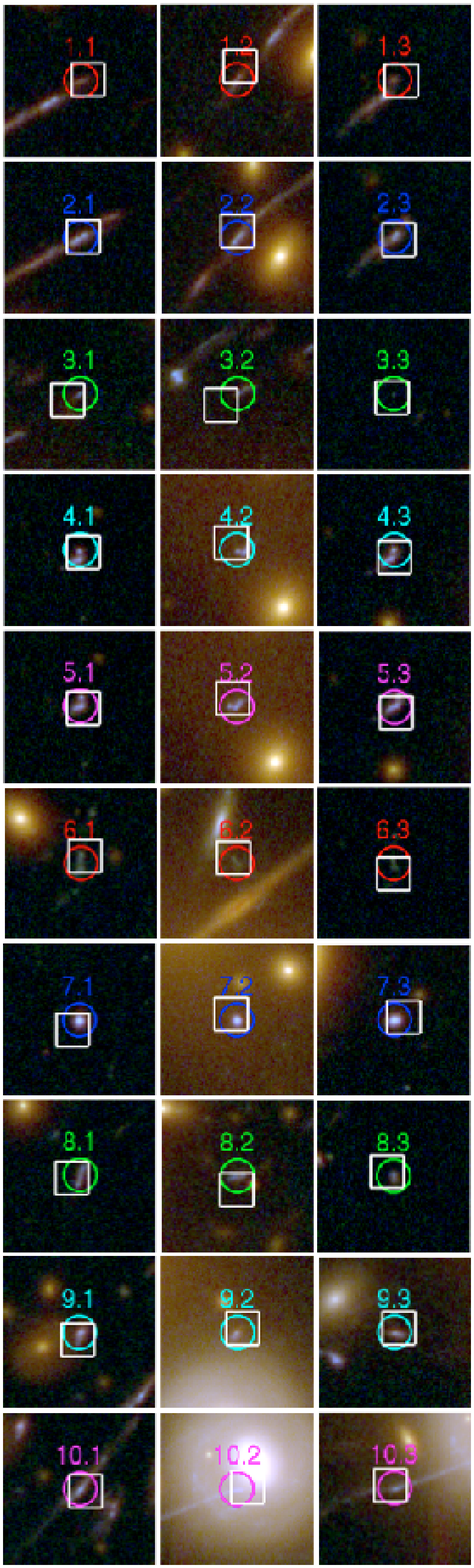}
\caption{Snapshots (6 arcsec across) of the 10 strong lensing systems
  with spectroscopic redshifts. The observed and predicted (by the best-fitting model, see Table \ref{models}) positions of the multiple images are marked, respectively, with colored circles and white squares.}
\label{fi11}
\end{figure}

\begin{figure*}
\centering
\includegraphics[width=0.85\textwidth]{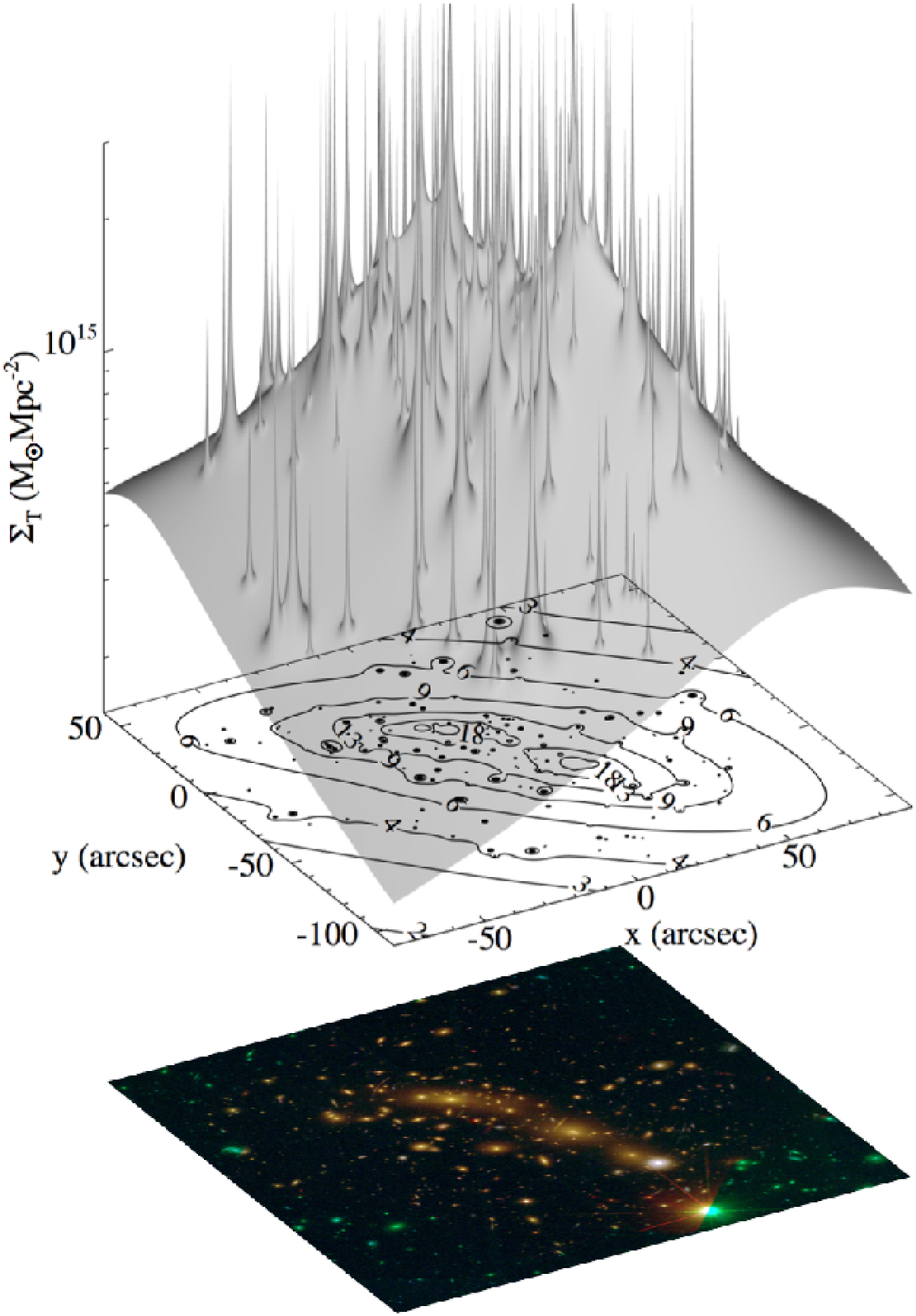}
\caption{The total surface mass density $\Sigma_{\mathrm{T}}$ in the inner regions of MACS~0416 reconstructed from the best-fitting strong lensing model (see Table \ref{models}). The different contributions of the two extended dark-matter halo and many candidate cluster member components are visible. The contour levels on the lens plane are in units of $10^{14}$ M$_{\odot}$Mpc$^{-2}$.}
\label{fi07}
\end{figure*}

The best-fitting model, with a minimum $\chi^{2}$ value of 915 (see
Table \ref{models}), can reproduce the multiple images of the 10
strong lensing systems very accurately, with a median (rms) offset
between the observed and model-predicted positions of only 0.31\arcsec
$\,$ (0.36\arcsec), i.e., approximately 5 (6) pixels. In Figure
\ref{fi11}, we compare the positions of the multiple images measured
and listed in Table \ref{tab1} (indicated by circles) with those
reconstructed according to our model (squares). We notice that every
system is almost perfectly reconstructed, without any systematic
offset in the predicted image positions. This implies that the
expected complex total mass distribution of the cluster is globally
described very well by our simple parametrized mass profiles. 

\begin{figure}
\centering
\includegraphics[width=0.4\textwidth]{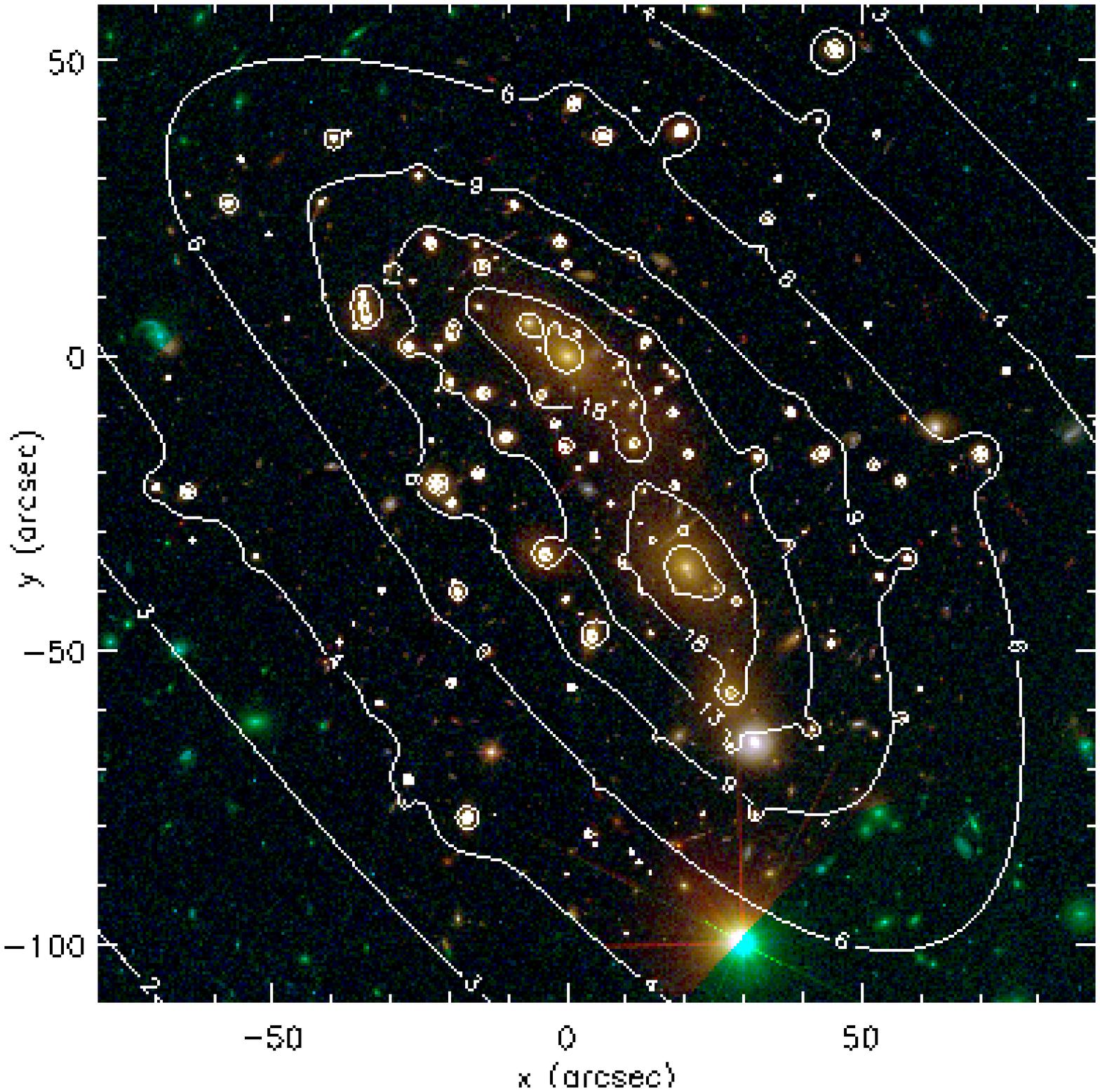}
\includegraphics[width=0.4\textwidth]{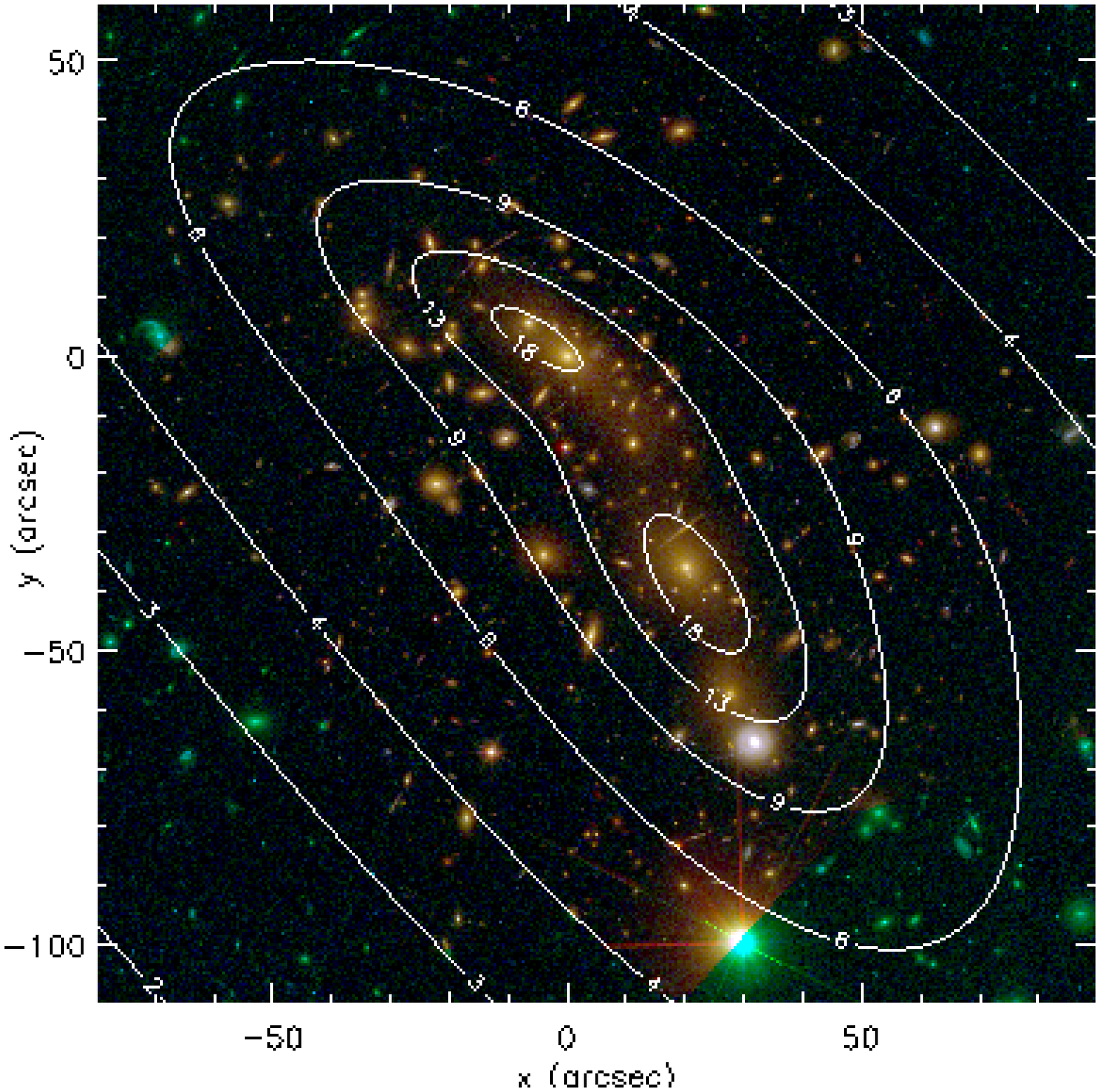}
\includegraphics[width=0.4\textwidth]{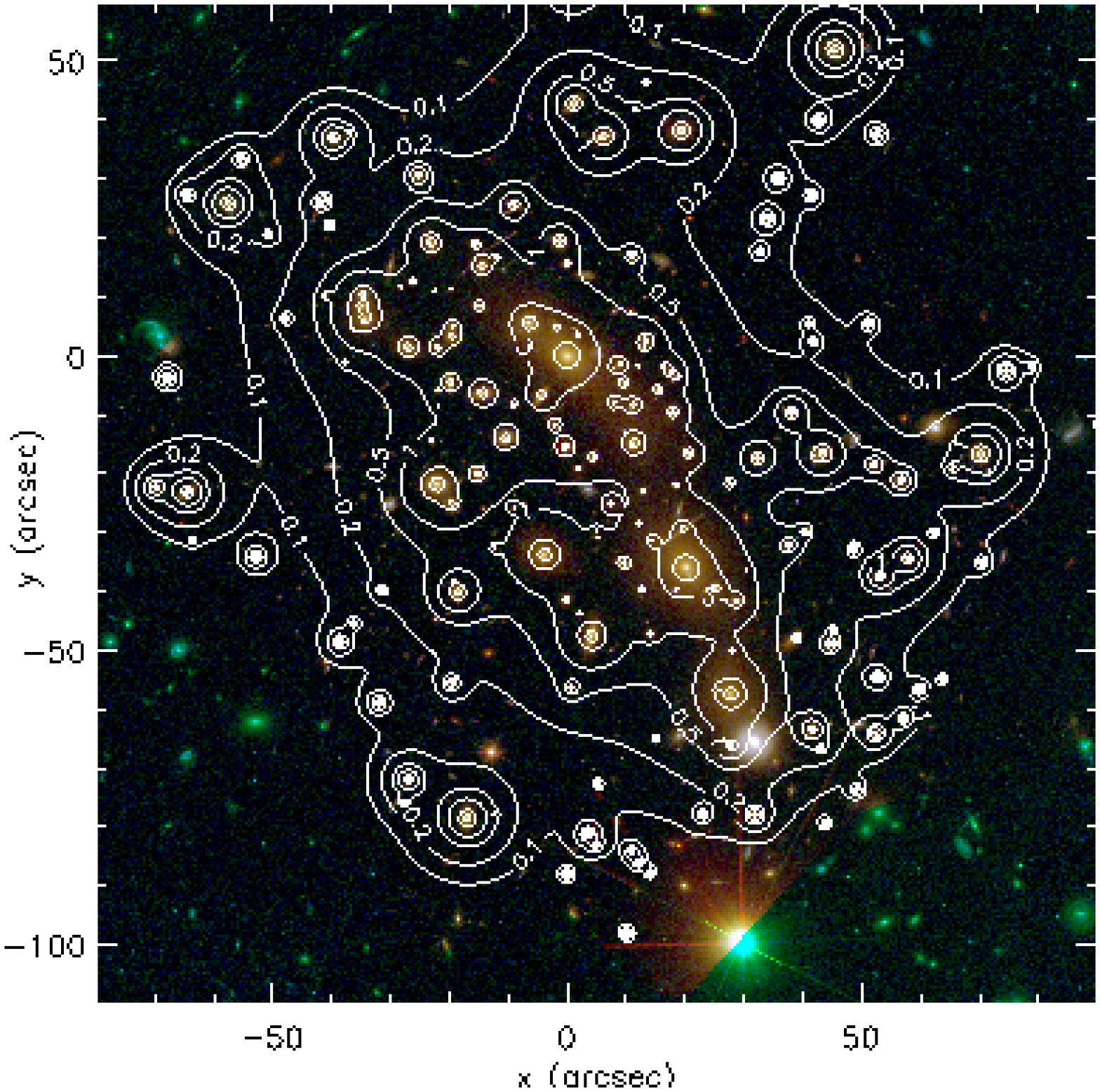}
\caption{Decomposition of the reconstructed total surface mass density, $\Sigma_{\mathrm{T}}$, (\emph{on the top}) into the surface mass densities of the two extended cluster dark-matter halos, $\Sigma_{\mathrm{H}}$, (\emph{in the middle}) and many candidate cluster members, $\Sigma_{\mathrm{G}}$ (\emph{on the bottom}). The contour levels on the lens plane are in units of $10^{14}$ M$_{\odot}$Mpc$^{-2}$.}
\label{fi08}
\end{figure}

In Figures \ref{fi07} and \ref{fi08}, we show the reconstructed
surface mass density of MACS~0416. We illustrate the total mass
density, $\Sigma_{\mathrm{T}}$, the smooth and extended contribution
of the cluster dark-matter halos, $\Sigma_{\mathrm{H}}$, and the more
concentrated and localized mass density of the cluster members,
$\Sigma_{\mathrm{G}}$. We remark that the cluster dark-matter halo
components are traced reasonably well by the total light distribution of the cluster.

\begin{figure*}
\centering
\includegraphics[width=0.99\textwidth]{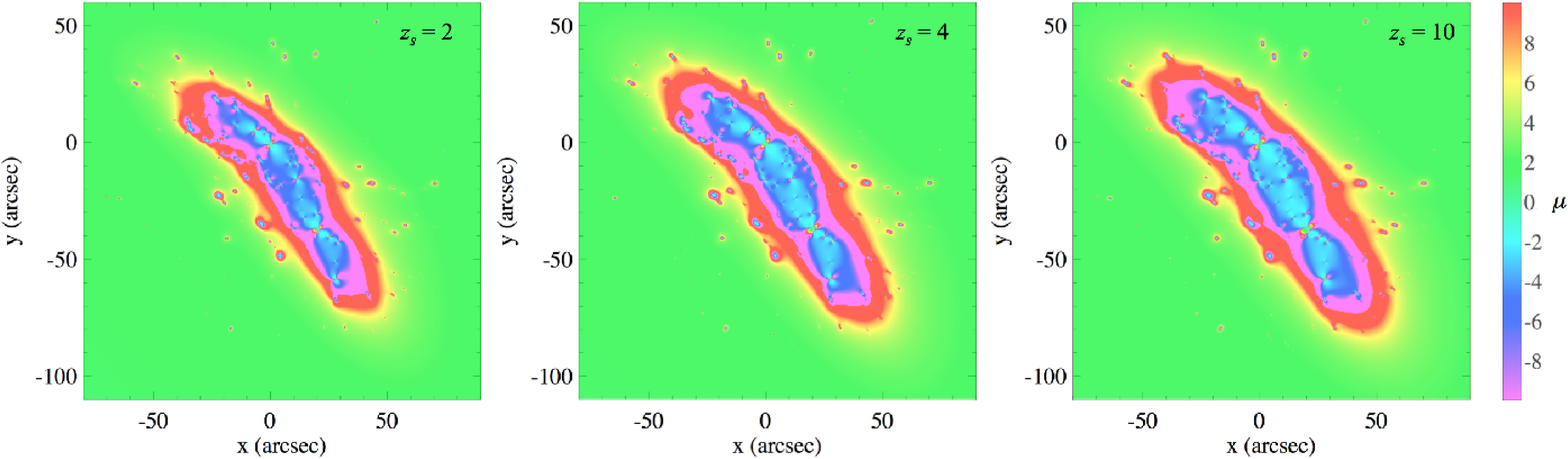}
\caption{Magnification maps in the inner regions of MACS~0416 reconstructed from the best-fitting strong lensing model (see Table \ref{models}) for sources at redshifts $z_{\rm s}$ equal to 2 (\emph{on the left}), 4 (\emph{in the middle}), and 10 (\emph{on the right}). The different colors, as indicated by the color bar on the right side, represent the different values of the magnification factor $\mu$ on a linear scale extending from $-10$ to $10$.}
\label{fi09}
\end{figure*}

\begin{table*}
\centering
\caption{Values of the area $A$ on the lens plane where the
  magnification factor $\mu$ is included in the specified ranges for
  different source redshifts $z_{\rm s}$.}
\begin{tabular}{cccccc}
\hline\hline \noalign{\smallskip}
 & $A(\mu  < 0)$ & $A(3 \le | \mu | < 5)$ & $A(5 \le | \mu | < 10)$  & $A(10 \le | \mu | < 30)$  & $A(| \mu | \ge 30)$  \\
 & (arcmin$^2$) & (arcmin$^2$) & (arcmin$^2$) & (arcmin$^2$) & (arcmin$^2$) \\
\noalign{\smallskip} \hline \noalign{\smallskip}
$z_{\rm s} = 2$ & 0.57 & 0.88 & 0.55 & 0.34 & 0.17 \\
$z_{\rm s} = 3$ & 0.70 & 1.01 & 0.62 & 0.40 & 0.20 \\
$z_{\rm s} = 4$ & 0.77 & 1.07 & 0.67 & 0.42 & 0.22 \\
$z_{\rm s} = 6$ & 0.84 & 1.15 & 0.71 & 0.46 & 0.23 \\ 
$z_{\rm s} = 8$ & 0.88 & 1.19 & 0.73 & 0.48 & 0.23 \\
$z_{\rm s} = 10$ & 0.90 & 1.22 & 0.74 & 0.50 & 0.23 \\
\noalign{\smallskip} \hline
\end{tabular}
\label{tab2}
\end{table*}

\begin{figure}
\centering
\includegraphics[width=0.48\textwidth]{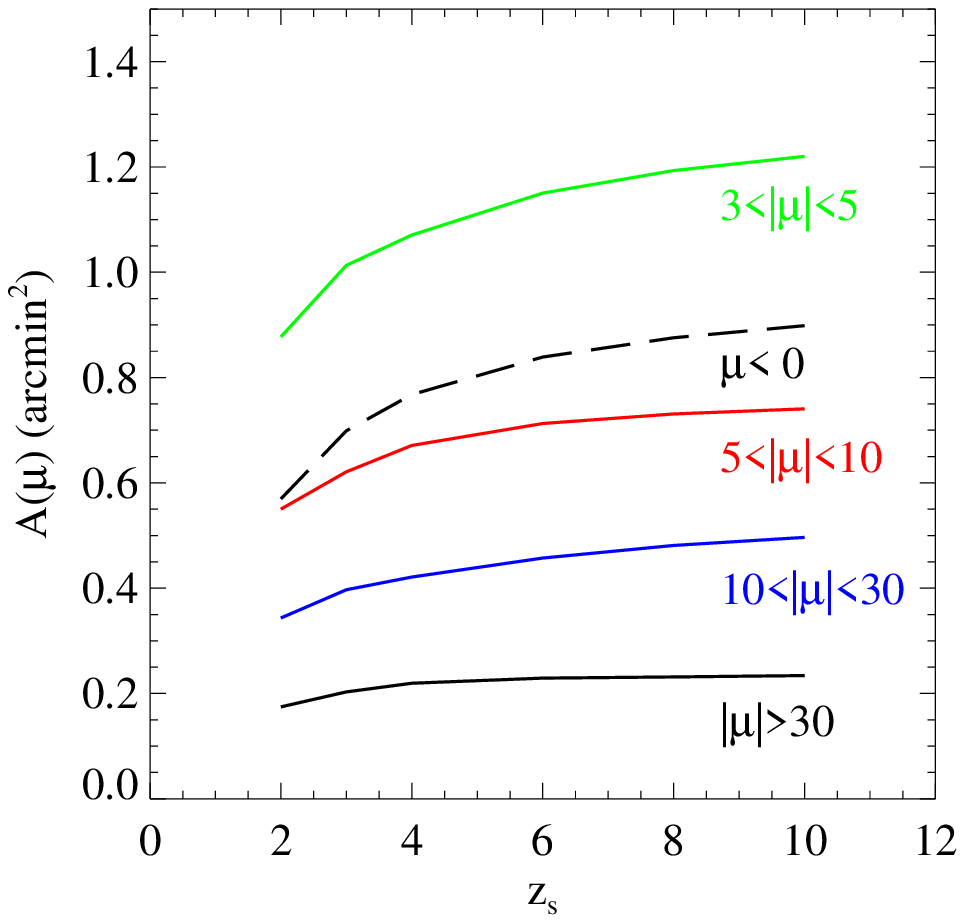}
\caption{Values of the area $A$ on the lens plane where the
  magnification factor $\mu$ is included in the specified ranges for
  different source redshifts $z_{\rm s}$ (see also Table \ref{tab2}).}
\label{fi14}
\end{figure}

Finally, we calculate the magnification factor values, $\mu$, in the
central regions of the cluster for several redshifts $z_{\rm s}$ of a
possible background source and show them in Figure \ref{fi09}. In
Table \ref{tab2} and Figure \ref{fi14}, we also show the sizes of the area $A$ inside
which the magnification factor is within the tabulated ranges.
We remark that MACS~0416 is an efficient deflector with extended
regions of high magnification. More quantitatively, by looking at the
values of $A(\mu  < 0)$, corresponding approximately to the surface
enclosed within the tangential critical curve, we notice that the size
of this area is enlarged by a factor of 1.5 when the source redshift is
increased from 2 to 10. Varying the source redshift within the same
range, the area with the largest magnification factors, $A(|\mu | \ge
30)$, grows by a factor of 1.3. Moreover, raising the value of
$z_{\rm s}$ from 2 to 10 also increases the percentage of surface of the
lens plane with medium magnification values.

\subsubsection{MCMC analysis}
\label{sec:MCMCanalysis}

\begin{table}
\centering
\caption{Median values and intervals at 1$\sigma$, 2$\sigma$, and 3$\sigma$ confidence level of the parameters $\parlensvec$ of the 2PIEMD + 175(+1)dPIE ($M_{\mathrm{T}}L^{-1}\sim L^{0.2}$) model.}
\begin{tabular}{ccccc}
\hline\hline \noalign{\smallskip}
 & Median & 1$\sigma$ CL & 2$\sigma$ CL & 3$\sigma$ CL \\
\noalign{\smallskip} \hline \noalign{\smallskip}
$x_{\mathrm{h1}}$ (\arcsec) & $-7.8$ & $^{+1.5}_{-1.8}$ & $^{+2.8}_{-4.4}$ & $^{+3.9}_{-7.5}$ \\
$y_{\mathrm{h1}}$ (\arcsec) & $5.0$ & $^{+1.7}_{-1.5}$ & $^{+3.6}_{-2.9}$ & $^{+5.7}_{-4.2}$ \\
$q_{\mathrm{h1}}$ & $0.35$ & $^{+0.04}_{-0.04}$ & $^{+0.08}_{-0.08}$ & $^{+0.13}_{-0.12}$ \\
$\phi_{\mathrm{h1}}$ (rad) & $2.61$ & $^{+0.04}_{-0.04}$ & $^{+0.09}_{-0.07}$ & $^{+0.14}_{-0.10}$ \\
$\vartheta_{\mathrm{E,h1}}$ (\arcsec) & $21.0$ & $^{+2.7}_{-2.5}$ & $^{+5.4}_{-4.8}$ & $^{+8.6}_{-7.1}$ \\
$r_{\mathrm{c,h1}}$ (\arcsec) & $12.9$ & $^{+2.1}_{-1.9}$ & $^{+4.5}_{-3.8}$ & $^{+7.5}_{-5.7}$ \\
$x_{\mathrm{h2}}$ (\arcsec) & $23.0$ & $^{+0.9}_{-0.9}$ & $^{+1.8}_{-1.7}$ & $^{+2.7}_{-2.4}$ \\
$y_{\mathrm{h2}}$ (\arcsec) & $-40.9$ & $^{+1.2}_{-1.2}$ & $^{+2.4}_{-2.3}$ & $^{+3.5}_{-3.4}$ \\
$q_{\mathrm{h2}}$ & $0.44$ & $^{+0.03}_{-0.02}$ & $^{+0.06}_{-0.05}$ & $^{+0.09}_{-0.07}$ \\
$\phi_{\mathrm{h2}}$ (rad) & $2.19$ & $^{+0.01}_{-0.01}$ & $^{+0.03}_{-0.03}$ & $^{+0.04}_{-0.05}$ \\
$\vartheta_{\mathrm{E,h2}}$ (\arcsec) & $32.8$ & $^{+3.1}_{-2.8}$ & $^{+6.3}_{-5.2}$ & $^{+9.5}_{-7.4}$ \\
$r_{\mathrm{c,h2}}$ (\arcsec) & $14.0$ & $^{+1.5}_{-1.4}$ & $^{+3.0}_{-2.7}$ & $^{+4.8}_{-4.0}$ \\
$\vartheta_{\mathrm{E,g}}$ (\arcsec) & $2.3$ & $^{+0.6}_{-0.4}$ & $^{+1.8}_{-0.7}$ & $^{+2.6}_{-1.1}$ \\
$r_{\mathrm{t,g}}$ (\arcsec) & $21$ & $^{+13}_{-10}$ & $^{+27}_{-16}$ & $^{+39}_{-17}$ \\
\noalign{\smallskip} \hline
\end{tabular}
\begin{list}{}{}
\item[Notes. ] The parameters $q_{\rm h1}$ and $q_{\rm h2}$ are the axis ratios of the two cluster dark-matter halos (introduced in Section \ref{sec:lensmod:masscomp:halo}). The angles $\phi_{\mathrm{h1}}$ and $\phi_{\mathrm{h2}}$ are measured counterclockwise from the positive $x$-axis (West).
\end{list}
\label{tab3}
\end{table}

\begin{figure*}
\centering
\includegraphics[width=0.99\textwidth]{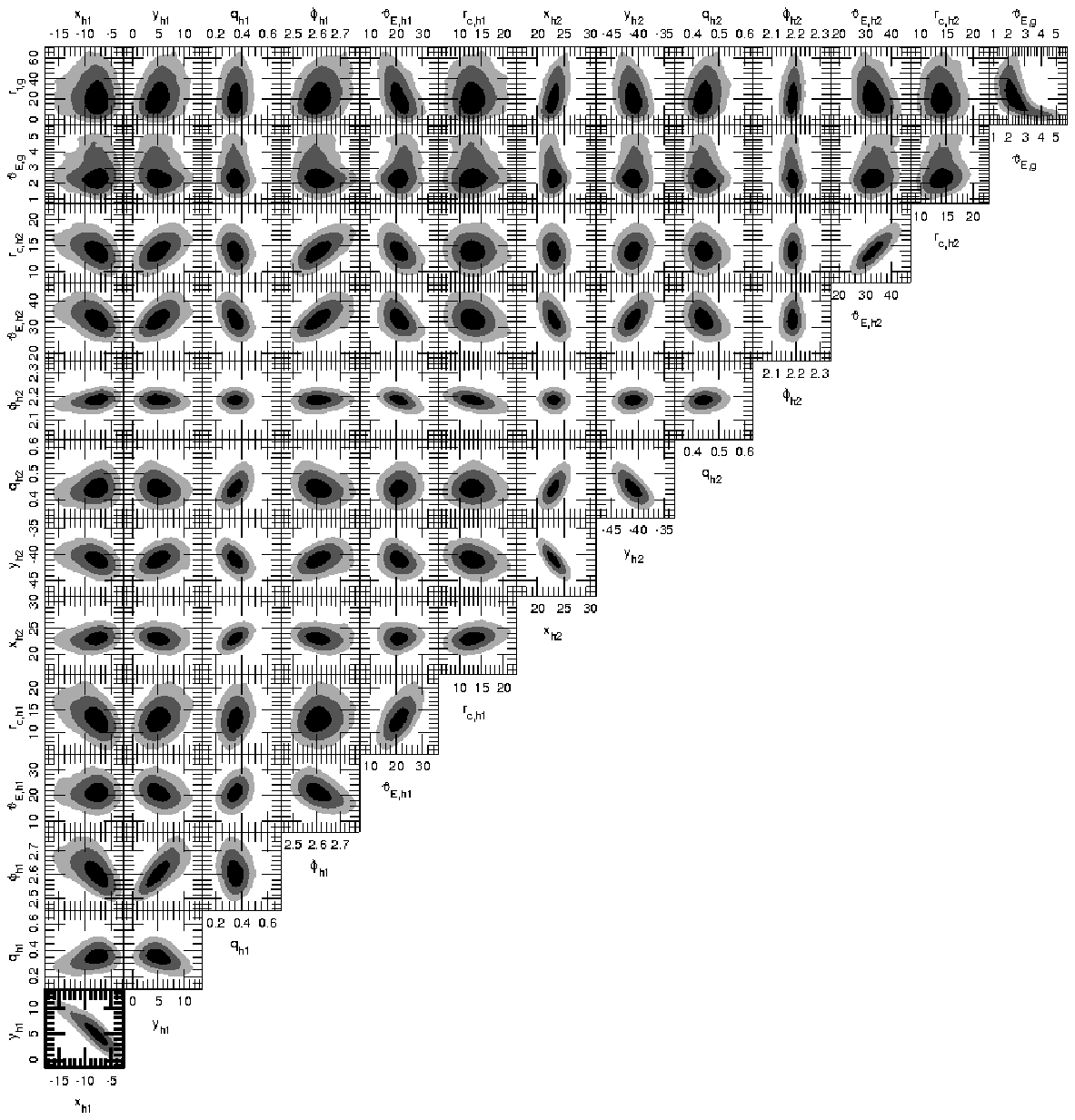}
\caption{Estimates of the uncertainties and correlations of the parameters
  $\parlensvec$ of the 2PIEMD + 175(+1)dPIE ($M_{\mathrm{T}}L^{-1}\sim L^{0.2}$) model. The gray contour levels on the
  planes represent the 1$\sigma$, 2$\sigma$, and 3$\sigma$ confidence
  regions and are obtained from a MCMC chain with $2 \times 10^{6}$ samples.}
\label{fi10}
\end{figure*}

\begin{figure}
\centering
\includegraphics[width=0.48\textwidth]{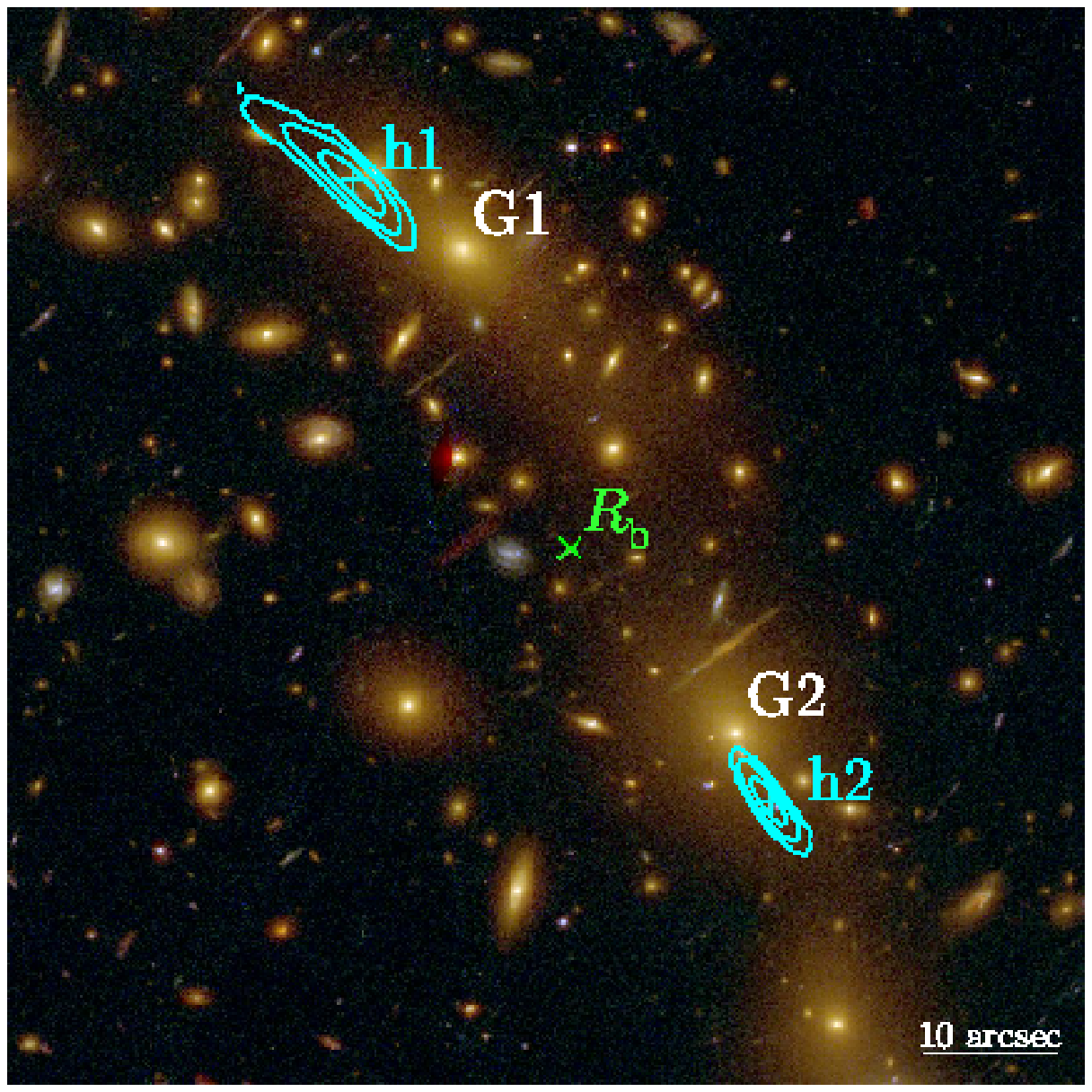}
\caption{An $80''$$\times$$80''$ cluster-core image showing the positions
  of the centers of the two extended dark-matter halos ($\mathrm{h1}$
and $\mathrm{h2}$, cyan plus symbols, see also Table \ref{tab3}) and of the cluster barycenter
($\boldsymbol{R}_\mathrm{b}$, green cross, defined in Eq. (\ref{eq:bary})) of the best-fitting lens
model. Given the uncertainties shown in Figure \ref{fi10}, reproduced
here with cyan contours, the dark-matter-halo centers are offset from
the luminosity centers of the closest brightest cluster galaxies
(G1 and G2) at more than $3\sigma$ CL. North is up and East is left.}  
\label{fi22}
\end{figure}

\begin{figure*}
\centering
\includegraphics[width=0.48\textwidth]{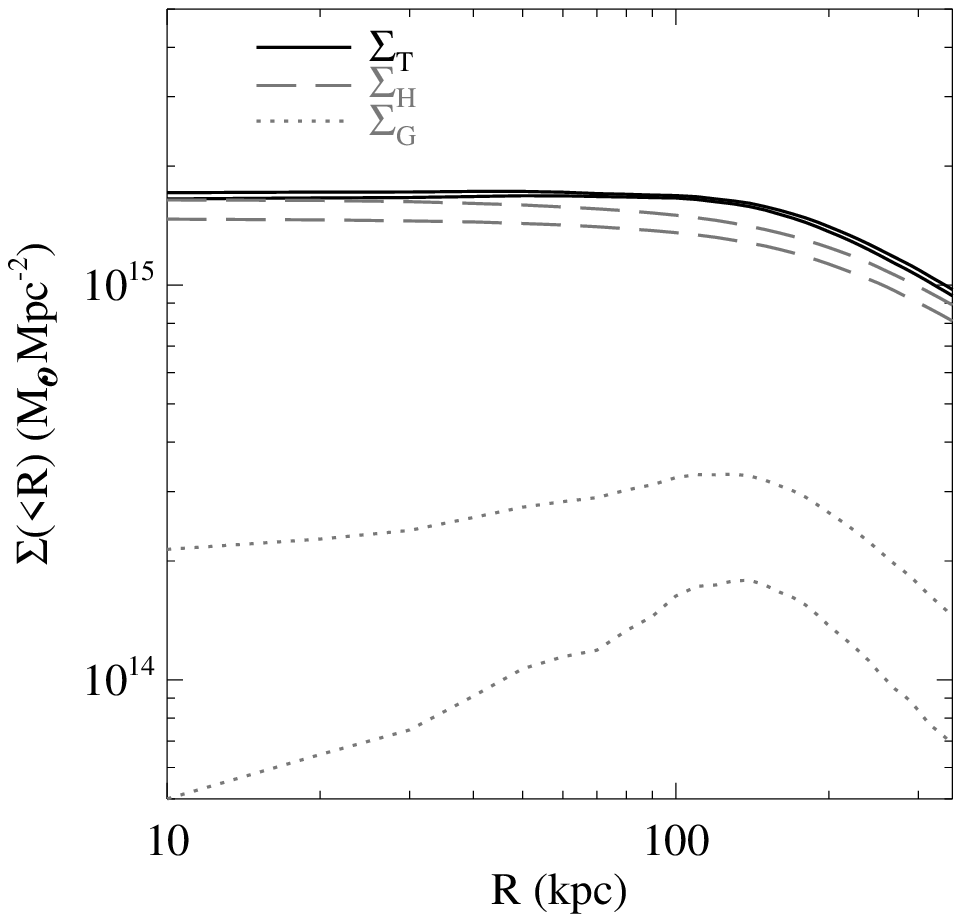}
\includegraphics[width=0.48\textwidth]{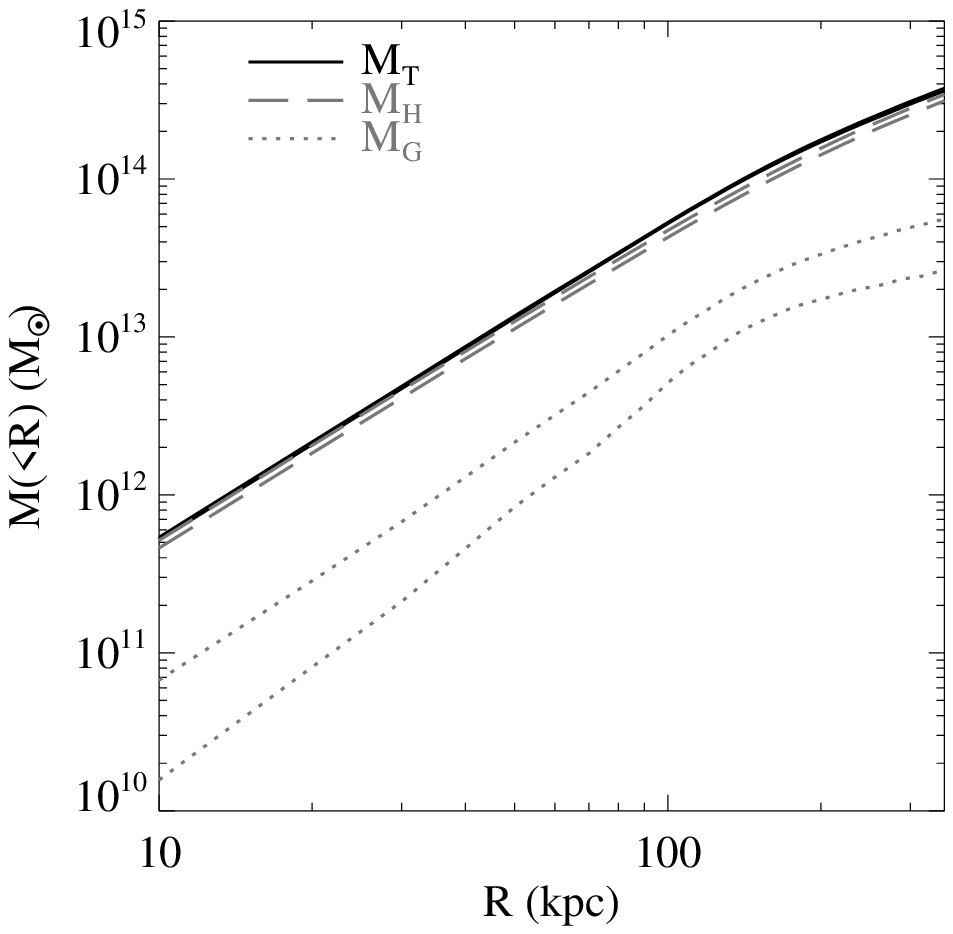}
\caption{Average surface mass density, $\Sigma(<R)$, and cumulative
  projected mass, $M(<R)$, profiles of the 2PIEMD + 175(+1)dPIE ($M_{\mathrm{T}}L^{-1}\sim L^{0.2}$) model. The solid, long-dashed, and dotted
  lines represent, respectively, the total, cluster dark-matter halo,
  and cluster member profiles at 1$\sigma$ confidence level.
}
\label{fi12}
\end{figure*}

As mentioned in Section \ref{sec:lensmod:glee}, we apply a MCMC
technique to sample the posterior PDF of the parameters $\parlensvec$
of the best-fitting 2PIEMD + 175(+1)dPIE ($M_{\mathrm{T}}L^{-1}\sim L^{0.2}$)
model. The uncertainties obtained from a MCMC analysis for the parameters of a model, and thus for the quantities derived from them, correlate with the uncertainties assigned to the observables. To get realistic uncertainties for the model parameters, the uncertainties of the observables have to be scaled so that the value of the best-fitting $\chi^{2}$ is comparable to the number of degrees of freedom of the investigated system (in other words, the reduced $\chi^{2}$ value should be approximately equal to 1). For this reason, in our best-fitting strong lensing model we increase the positional error of the observed multiple images $\delta_{x,y}$ by a factor of approximately 6, 
i.e., to 0.4\arcsec. This can account for, e.g., line-of-sight
structures and small dark-matter clumps that are not
encapsulated in our mass model. 
In this way, the value (24) of the $\chi^{2}$ is comparable to the number (24) of the degrees of freedom. The latter is given by the number of lensing observables ($x$ and $y$ of each multiple image) minus the number of parameters $\parlensvec$ of the model ($x$ and $y$ of each source, $x_{\mathrm{h}}$, $y_{\mathrm{h}}$, $q_{\mathrm{h}}$, $\phi_{\mathrm{h}}$, $\vartheta_{\mathrm{E,h}}$, and $r_{\mathrm{c,h}}$ of each cluster dark-matter halo, $\vartheta_{\mathrm{E,g}}$ and $r_{\mathrm{t,g}}$ of the scaling relations of the cluster members, and $\vartheta_{\mathrm{E}}$ and $r_{\mathrm{t}}$ of the foreground galaxy). We show the results, derived from a chain with $2 \times 10^{6}$ samples (with an acceptance rate of approximately 0.22), in Figure \ref{fi10} and Table \ref{tab3}.

Looking at Figure \ref{fi10}, we observe that the values of
$x_{\mathrm{h}}$ and $y_{\mathrm{h}}$ of each of the two cluster
dark-matter halos are anticorrelated. This means that, to preserve the
goodness of the fit, shifts of the mass centers of these two
components are only allowed in the North-East (or South-West)
direction. As expected, for the same mass components, we also find
that the values of $\vartheta_{\mathrm{E,h}}$, and $r_{\mathrm{c,h}}$
are correlated. From Equation (\ref{eq:piemd}), it is clear that in order
to obtain a fixed amount of projected mass within a given small ($R
\lesssim r_{\mathrm{c,h}}$) circle, an increase in the value of
$\vartheta_{\mathrm{E,h}}$ must be counterbalanced by a suitable
increase in the value of $r_{\mathrm{c,h}}$. Moreover, the values of
the strength of the dark-matter halos, $\vartheta_{\mathrm{E,h1}}$
and $\vartheta_{\mathrm{E,h2}}$, are anticorrelated. This follows from
the fact that the contributions of the two halos to the total mass in
the central regions of the cluster (tightly constrained by the
multiple image systems) compensate each other. Not surprisingly, we do
not have much information about the values of the truncation radius
$r_{\mathrm{t,g}}$ of the cluster members. In Figure \ref{fi06}, we
see indeed that none of the lensing systems has two or more multiple
images located close to and around a single cluster member. This would
have enabled a determination of the values of both
$\vartheta_{\mathrm{E,g}}$ and $r_{\mathrm{t,g}}$. In our model, the
total mass profiles of the cluster members are thus well approximated
by simple isothermal mass distributions (see Equation (\ref{eq:dpie}) for
large values of $r_{\mathrm{t,g}}$). 

From Tables \ref{tab5} and \ref{tab3} and Figure \ref{fi22}, we remark
that the reconstructed centers of two cluster dark-matter halos are
significantly separated, more than 3$\sigma$ away, from the centers of
the two brightest cluster galaxies. In detail, the northern halo is at
a projected distance of approximately 50 kpc from G1 and a smaller
projected distance of approximately 30 kpc separates the centers of
the southern halo and G2. Interestingly, the northern halo is
preferentially displaced towards the North-East direction, whereas the
southern halo towards the South-West direction, resulting in a
projected distance between the two cluster dark-matter halos of
approximately 300 kpc. We have checked that (1) fixing the halo
centers to those of the brightest cluster galaxies results in
$\chi^{2}$ values that are approximately a factor of 4 higher than
those obtained with the halo centers free to vary, and (2) increasing
the F160W magnitude value of G1 by 0.3 mag, i.e. mimicking an
overestimate of the galaxy luminosity due to a possible
contamination from the intra cluster light, reduces the offset between
the centers of the northern halo and G1 by only about
10\%. As a result of the superposition of two components, we
mention that the density peaks of the 
two superposed dark-matter clumps are less
distant than the individual dark-matter halo centers from the centers of the
brightest cluster galaxies (see Figure~\ref{fi08}). Furthermore, we
notice that within the adopted mass parametrization, the two
dark-matter components require appreciably large core radii, at more
than 3$\sigma$ CL. The median values of the two cores are of 69 and 75
kpc for the northern and southern halos, respectively.

Next, we extract from the MCMC chain 100 different models to quantify the statistical uncertainty on the derived circular quantities of average surface mass density
\begin{equation}
\label{eq:asmd}
\Sigma(<R) \equiv \frac{\int^{R}_{0} \Sigma(\tilde{R})2\pi\tilde{R}\,\mathrm{d}\tilde{R}}{\pi \tilde{R}^{2}}
\end{equation}
and cumulative projected mass
\begin{equation}
\label{eq:cpm}
M(<R) \equiv \int^{R}_{0} \Sigma(\tilde{R})2\pi\tilde{R}\,\mathrm{d}\tilde{R} \, ,
\end{equation}
where $\boldsymbol{\tilde{R}} = \tilde{R} \,\boldsymbol{e_{\tilde{R}}} = (x,y)$ and $ \boldsymbol{e_{\tilde{R}}} = \boldsymbol{\tilde{R}} / \tilde{R}$. As done in the previous section, we decompose the total $\Sigma(<R)$ and $M(<R)$ into their cluster dark-matter halo and cluster member components. We calculate the distances $R$ on the lens plane from the barycenter, or center of mass, of the cluster
\begin{equation}
\label{eq:bary}
\boldsymbol{R}_\mathrm{b} \equiv \frac{ \int \Sigma_{\mathrm{T}}(\boldsymbol{\tilde{R}})\boldsymbol{\tilde{R}}\,\mathrm{d}\boldsymbol{\tilde{R}}}{ \int \Sigma_{\mathrm{T}}(\boldsymbol{\tilde{R}})\,\mathrm{d}\boldsymbol{\tilde{R}} } \, .
\end{equation}
According to our best-fitting model, we find that the
coordinates of the barycenter, with respect to the
luminosity center of G1, in arcsec are $(8.14,-22.22)$. Therefore,
considering the line that connects the luminosity centers of G1 and
G2, the center of mass of MACS~0416 lies on the eastern side (see
Figure \ref{fi22}), where
more luminous cluster members are observed (see Figure \ref{fi06}). Figure \ref{fi12} illustrates the radial dependence of the functions $\Sigma(<R)$ and $M(<R)$.

We notice that the cluster member and dark-matter halo components have
remarkably similar distributions. Both $\Sigma_{\rm G}(<R)$ and
$\Sigma_{\rm H}(<R)$ show very flat inner profiles, with core radii of
approximately 100 kpc. The small bump in the cluster member component,
visible at about 130 kpc, is due to the presence of the two brightest cluster galaxies G1
and G2 at such a projected
distance from the barycenter. It is also interesting to remark that the 
cluster-galaxy and cluster-halo 
mass components
must be anticorrelated. In fact, they clearly have relative
uncertainties that are larger than those of the total quantities. At
1$\sigma$ CL, the cumulative projected total mass profile exhibits,
surprisingly, nearly constant uncertainty of a few per cent over the
investigated radial range extending from 10 to 350 kpc. At more than
100 kpc in projection from the barycenter, we measure a cluster member
over total mass ratio, $M_{\rm{G}}/M_{\rm{T}}(<R)$, of $13^{+5}_{-4}$\%.

\begin{figure}
\centering
\includegraphics[width=0.48\textwidth]{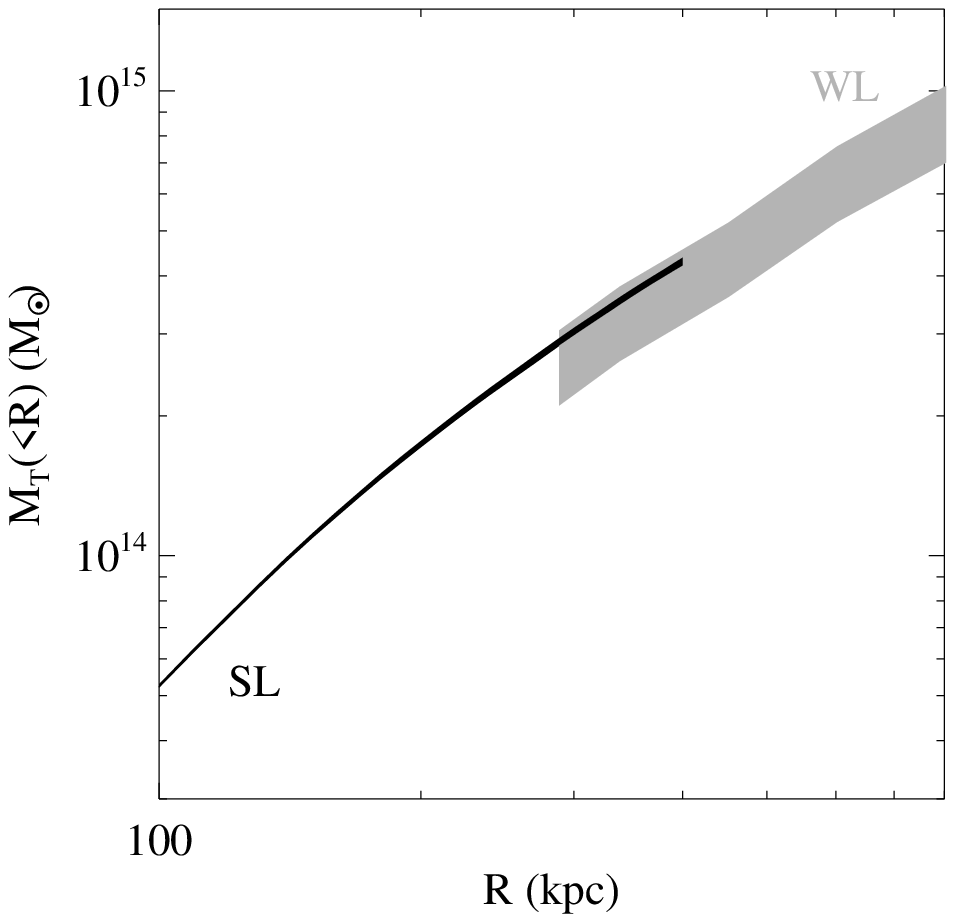}
\caption{The cumulative projected total mass, $M_{\mathrm{T}}(<R)$,
  profiles obtained from the strong lensing (SL) analysis presented in
  this work, i.e., from the 2PIEMD + 175(+1)dPIE
  ($M_{\mathrm{T}}L^{-1}\sim L^{0.2}$) model, and from the weak
  lensing (WL) analysis presented in \citet{ume14}. Intervals are at
  1$\sigma$ confidence level.} 
\label{fi15}
\end{figure}

In Figure \ref{fi15}, we plot the cumulative projected total mass
profile resulting from our best-fitting strong lensing model and that
from the independent weak lensing analysis by \citet{ume14}. It is
well known that the strong and weak lensing effects allow one to map
the projected total mass of a cluster on different radial scales. We
show here that the estimates of the two total mass diagnostics overlap
between approximately 300 and 400 kpc from the center of MACS~0416 and
over this radial range they are consistent, given the 1$\sigma$
uncertainties, despite the slightly different definition of the
cluster center in the two studies. We remark that such a good
agreement between the strong and weak lensing mass estimates has been
observed only in a few galaxy clusters 
(e.g., \citealt{ume11,ume12}; \citealt{coe12};
\citealt{eic13}; \citealt{med13}), partly due to the different
systematic uncertainties affecting the two methods and to the not
always optimal quality of the data available.    

\begin{figure*}
\centering
\includegraphics[width=0.48\textwidth]{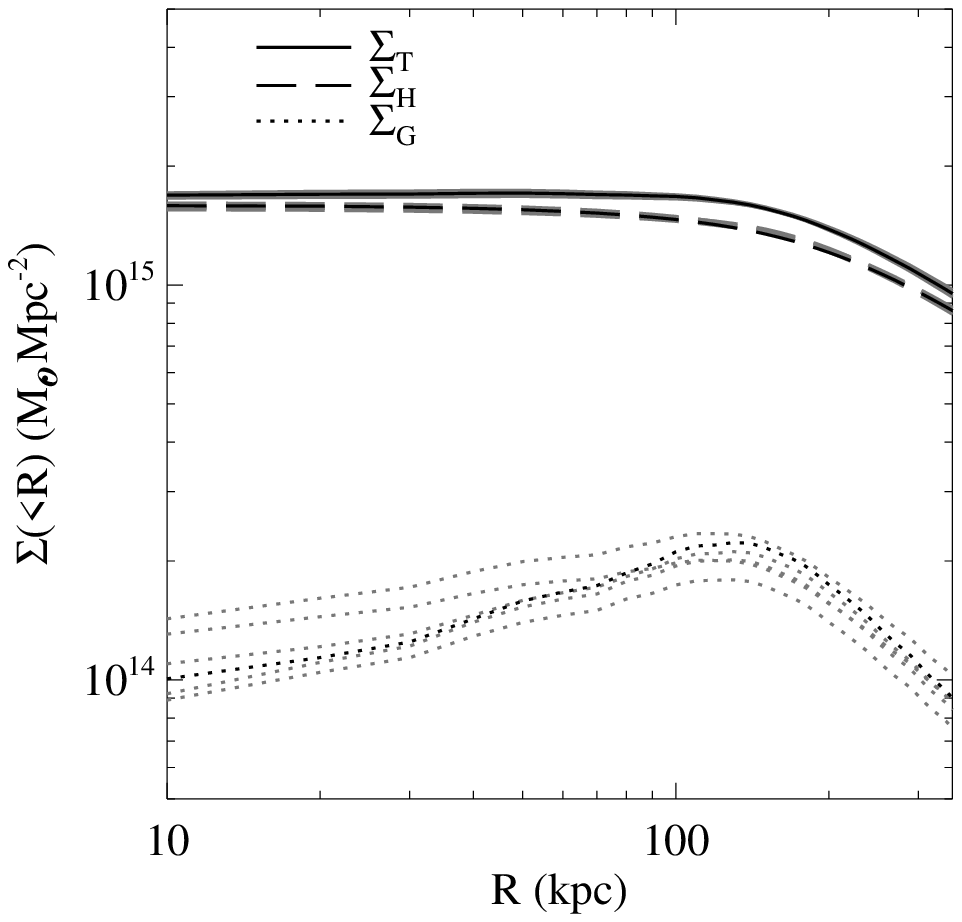}
\includegraphics[width=0.48\textwidth]{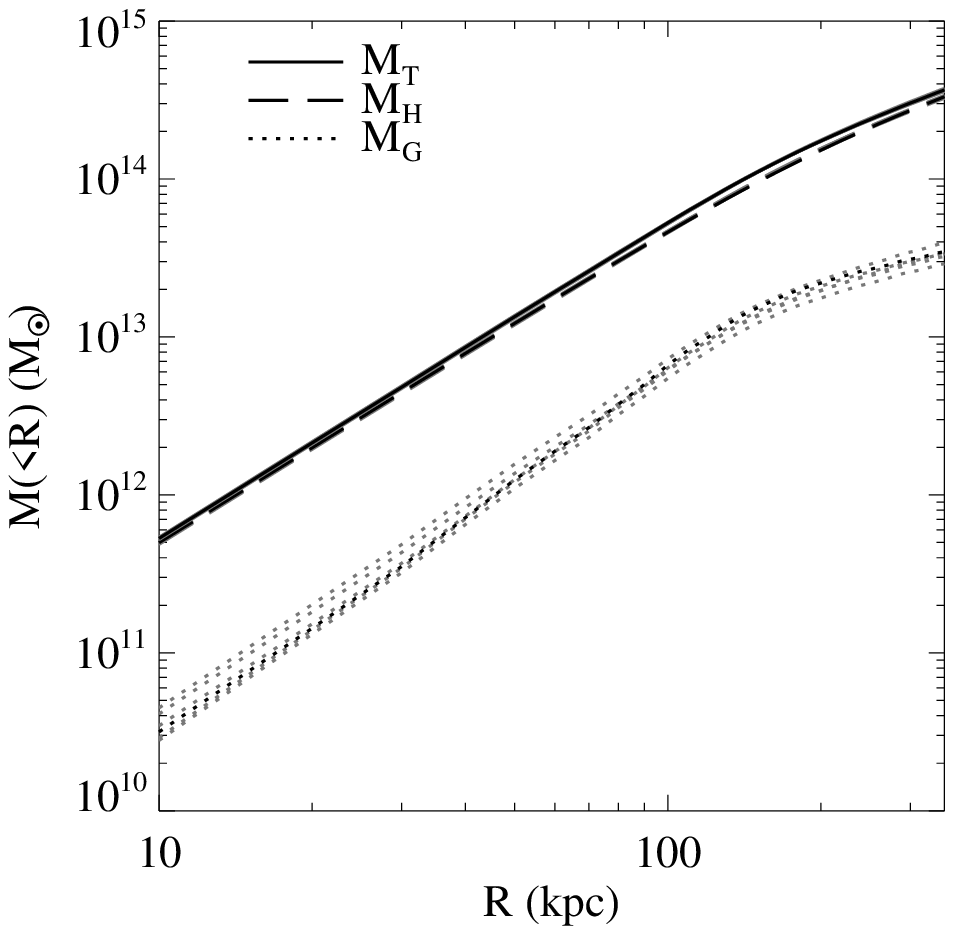}
\caption{Surface mass density, $\Sigma(<R)$, and cumulative
  projected mass, $M(<R)$, profiles of the six optimized strong lensing models that include the 175 candidate cluster members (see Table \ref{models}). The 2PIEMD + 175(+1)dPIE ($M_{\mathrm{T}}L^{-1}\sim L^{0.2}$) model is shown in black, all the others in gray. The solid, long-dashed, and dotted
  lines represent, respectively, the best-fitting total, cluster dark-matter halo,
  and cluster member profiles of the different models.
}
\label{fi16}
\end{figure*}

We give an estimate of the systematic errors in our surface mass
density and cumulative projected mass profiles by considering the
best-fitting results of the six strong lensing models, shown in Table
\ref{models}, that include the mass contribution of the galaxy cluster
members. We compare the model-reconstructed quantities in Figure
\ref{fi16}. Surprisingly, we find that all the models produce very
similar results, almost independently of the adopted mass
parametrization details, with only relatively larger variations in the
cluster member component. The profiles of the extended cluster
dark-matter halos are barely distinguishable, the different inner
radial dependence of the PIEMD and PNFW mass models
notwithstanding. This unexpected and interesting strong degeneracy can
be ascribed to three effects: the measurement of the mass quantities
a) in projection, b) within circular apertures, and c) superposing two
dark-matter halos in each lens model. We remark though that the
differing central slope values of the PIEMD and PNFW mass density
profiles are clearly visible in the reconstructed two-dimensional mass
density maps, explaining partly the variance in the
minimum-$\chi^{2}$ values of Table \ref{models}. In summary, from
these tests we can state that (1) the total mass measurements of a
galaxy cluster from accurate strong lensing modeling are robust, even
if different mass density profiles are adopted, (2) disparate mass
density profiles produce detectable differences in the multiple image
reconstruction, and (3) our values of the statistical and systematic
uncertainties on the total mass of MACS~0416 are comparable and of the
order of 5 per cent.

\begin{figure}
\centering
\includegraphics[width=0.48\textwidth]{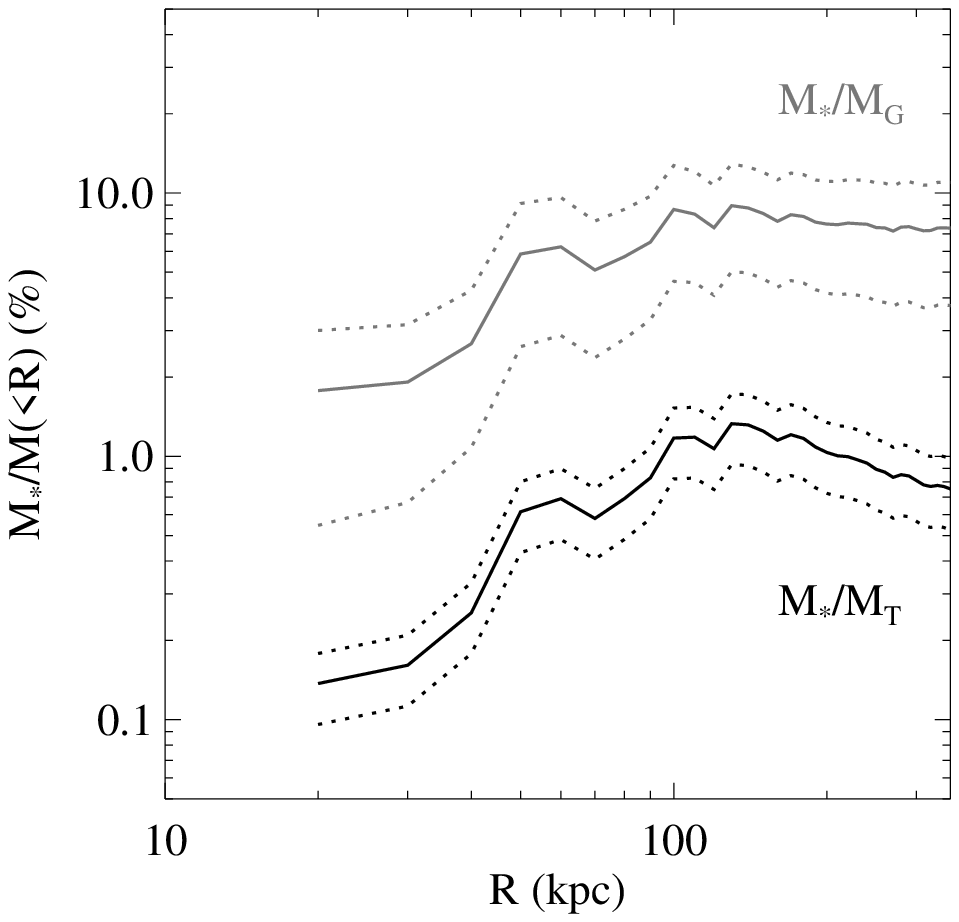}
\caption{Cumulative projected profiles of the stellar over total cluster, $M_{*}/M_{\mathrm{T}}$, and stellar over total galaxy, $M_{*}/M_{\mathrm{G}}$, mass fractions. Solid and dotted lines show the median values and the 1$\sigma$ confidence levels, respectively.}
\label{fi18}
\end{figure}

Finally, we estimate the profiles of the cumulative mass in the form
of stars as a fraction of the cumulative total mass of MACS~0416,
$M_{*}/M_{\rm{T}}(<R)$, as well as a fraction of the cluster member
mass, $M_{*}/M_{\rm{G}}(<R)$. For a given radial (two-dimensional)
aperture, the stellar mass budget is obtained as the sum of the
luminous mass values of the galaxy cluster members which have their
luminosity centers enclosed within that aperture. The method used to
measure the galaxy luminous mass values is described in Section
\ref{sec:lensmod:masscomp:gal} and the $M_{\rm{T}}(<R)$ and
$M_{\rm{G}}(<R)$ profiles are those presented above and shown in
Figure \ref{fi12}. We plot our results in Figure \ref{fi18}. We find
that at projected distances from the cluster total mass center between
100 and 350 kpc the stellar-to-total-cluster-mass ratio is slightly
decreasing, with an average value of approximately $(1.0\pm0.3)$\%, and
the stellar-to-total-galaxy-mass ratio has a fairly constant value
of approximately $(7.7\pm3.6)$\%.

\section{Comparison with the literature}
\label{sec:complit}

\begin{figure}
\centering
\includegraphics[width=0.48\textwidth]{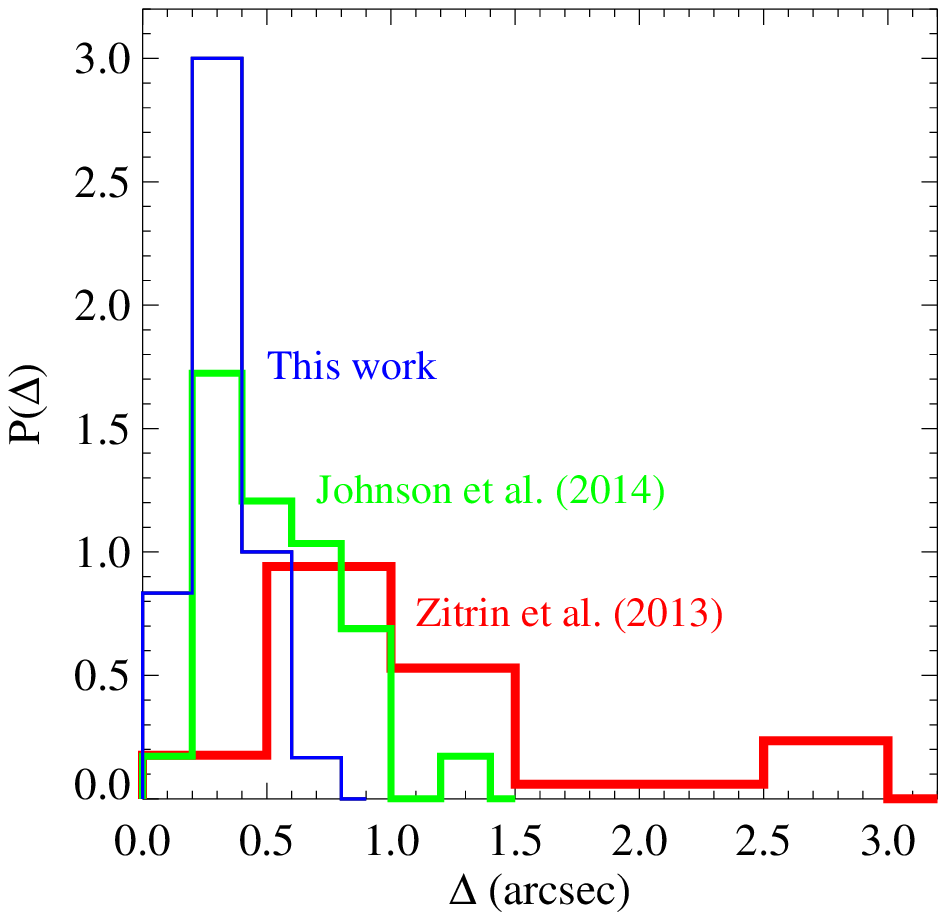}
\caption{Probability distribution functions of the absolute values of the differences ($\Delta$, in arcsec) between the observed and best-fitting model-predicted positions of the multiple-image systems considered in the different strong lensing studies of MACS~0416.}
\label{fi13}
\end{figure}

As mentioned in Section \ref{sec:intro}, MACS~0416 has been the
subject of several recent strong lensing studies (\citealt{zit13};
\citealt{JohnsonEtal14}; \citealt{RichardEtal14};
\citealt{JauzacEtal14a,JauzacEtal14b}; \citealt{DiegoEtal14}). We discuss
here the main differences and results of the previous analyses, in
particular contrasting them with ours.

The total number of images of the modeled candidate strong lensing
systems in MACS~0416 has more than doubled over the last year,
increasing from 70 (\citealt{zit13}) to 194
(\citealt{JauzacEtal14a}). The combination of the shallow 
\HST\ imaging from the CLASH survey with the more recent and
deeper observations in 3 \HST/ACS filters from the HFF program has made
such an improvement possible. More multiply-imaged systems will likely be
identified in the cluster core, thanks to the upcoming HFF data in the
remaining 4 \HST/WFC3 bands.  The various aforementioned studies have considered different 
combinations of spectroscopic and photometric redshifts for the
background lensed sources. The spectroscopic values adopted in all
studies are those obtained from our CLASH-VLT survey and presented in
Section \ref{sec:lensmod:multsys}. These measurements were shared with
several international strong lensing groups to enable the distribution
of preliminary strong lensing models (and
magnification maps) to the community, before the acquisition of the HFF observations
and the publication of this paper. The photometric redshifts were
estimated from the multicolor \HST\ photometry of CLASH (for more
details, see \citealt{jou14}) and have been used as reference values
or priors in the strong lensing modeling of previous studies.  We
restate here that we have purposely restricted our analysis to the
positions, measured from the CLASH data, of the 30 multiple images
associated with our first 10 spectroscopically confirmed strong
lensing systems.

To reconstruct the total mass distribution of MACS~0416, most of the
cited studies (\citealt{zit13}; \citealt{JohnsonEtal14}; 
\citealt{JauzacEtal14a}; \citealt{DiegoEtal14}) have focused on strong
lensing only models, while some others have considered joint strong
and weak lensing analyses (\citealt{RichardEtal14};
\citealt{JauzacEtal14b}). Various lensing codes (e.g., {\sc LTM,
  Lenstool, WSLAP+}) have been used to assign mass, by means of
physically-motivated, parametrized profiles or pixelized grids, to the
extended cluster dark-matter halos and candidate cluster members. The
latter have essentially been selected from the cluster red sequence
and their mass weights have been scaled according to their luminosity
values in the reddest \HST\ optical bands (i.e., the F775W or the
F814W). In our analysis, we have concentrated on the strong lensing
modeling of MACS~0416 with {\sc Glee} (software) that makes use
of parametrized mass profiles. We have determined the 175 candidate
cluster members to include in our models by considering the full
multicolor information, in 12 \HST\ broadbands, of our more than 60
CLASH-VLT spectroscopically confirmed cluster members in the \HST/WFC3
FoV. We have varied the cluster member mass contribution
depending on the galaxy luminosity values in the reddest \HST\ near-IR
band (i.e., the F160W). 

We confirm that the total mass distribution in the central regions of
MACS~0416 is dominated by two highly-elongated and close in projection
components, representative of two massive and extended dark-matter
halos and responsible for the large area on the lens plane with high
magnification factors (as also observed in Figure 2 in
\citealt{gri11}, on galaxy group scales). Consistent with previous
results, we find that the inner mass density profile of the cluster is
flat, with a core radius of the order of 100 kpc. The good agreement
on these last points with the outcomes of the very different strong
lensing models presented in \citet{DiegoEtal14} is particularly
remarkable. As far as aperture
total mass measurements is concerned, we estimate that (1)
$M_{\rm{T}}(<200 \,{\rm kpc})$ is between 1.72 and 1.77 $\times 
10^{14}M_{\odot}$, somewhat higher than the measurements of
$(1.63\pm0.03)\times 10^{14}M_{\odot}$ and $(1.60\pm0.01)\times
10^{14}M_{\odot}$ presented in \citet{RichardEtal14} and
\citet{JauzacEtal14a}, respectively; (2) $M_{\rm{T}}(<250 \,{\rm kpc})$
ranges between 2.35 and 2.43 $\times 10^{14}M_{\odot}$, consistent
with the value of $2.46^{+0.04}_{-0.08}\times 10^{14}M_{\odot}$
reported in \citet{JohnsonEtal14} (we also concur with the last
authors in finding that the mass measurements 
of \citealt{zit13} are noticeably higher than ours); (3)
$M_{\rm{T}}(<320 \,{\rm kpc})$ is between 3.23 and 3.35 $\times
10^{14}M_{\odot}$, in agreement with the estimates of $(3.26\pm0.03)\times
10^{14}M_{\odot}$ and $(3.15\pm0.13)\times 10^{14}M_{\odot}$ from the
strong and strong plus weak lensing analyses of, respectively,
\citet{JauzacEtal14a} and \citet{JauzacEtal14b}. The differences
in the total mass measurements quoted here might partly be connected
to the slight displacement in the adopted cluster mass centers, to the
details of the lensing models, and, most likely, to the well-known
(and, with photometric redshift values, only partially broken)
degeneracy between the mass of a lens and the redshift of a
multiply-imaged source. We have explicitly checked that our
total mass measurements do not depend appreciably on the redshift value 
of an individual multiple-image system, and, in particular, that the significant 
discrepancies listed above cannot be ascribed to the different redshift 
value of system 7 adopted in the previous analyses.
 
Despite the relatively good agreement obtained by the various groups
on the cluster total mass reconstruction, we mention that the
best-fitting values of some model parameters (see Table \ref{tab3} in
Section \ref{sec:MCMCanalysis}, Table 13 in \citealt{JohnsonEtal14},
Table 8 in \citealt{RichardEtal14}, and Table 1 in
\citealt{JauzacEtal14a}) are significantly inconsistent. 
These discrepancies are likely due to the use of different multiple image
systems as constraints, modeling assumptions and strategies.  Rather
than discussing in length on the intricate modeling details, we focus
instead on the 
goodness of the different models in reproducing the positions of the
observed multiple images, which can be compared irrespectively of 
model assumptions.  To quantify the goodness of fit, we consider the values of
$\Delta$, the modulus of the difference (in arcsec) between the
observed and best-fitting model-predicted positions of an image. In
Figure \ref{fi13}, we compare the probability distribution functions
of $\Delta$ estimated from our best-fitting, minimum-$\chi^{2}$ strong
lensing model and those of \citet{zit13} and \citet{JohnsonEtal14}
(the only two previous studies that present the numbers necessary for this
comparison). We note that for the distribution of $\Delta$ for \citet{JohnsonEtal14} we consider
exactly our 10 multiple image systems, whereas for \citet{zit13} we
use the positions of 34 multiple images from 13 different sources (23
images of 8 sources are in common with ours). From Figure \ref{fi13},
it is evident that our model reproduces the observables with extremely
good accuracy. More quantitatively, the median values of $\Delta$
are 0.31\arcsec$\,$ in our case, 0.44\arcsec$\,$ in
\citet{JohnsonEtal14} and 0.90\arcsec$\,$ in \citet{zit13}. Specifically, the rms $\Delta$ is
0.36\arcsec$\,$ in our study, 0.51\arcsec$\,$ in
\citet{JohnsonEtal14}, 0.68\arcsec$\,$ in 
\citet{JauzacEtal14a}, approximately 0.8\arcsec$\,$ in
\citet{RichardEtal14}, and 1.37\arcsec$\,$ in
\citet{zit13}\footnote{we note though that different analyses use
  different sets of multiple image systems}.

From the discussion above, we caution that strong lensing models that
reproduce the positions of observed multiple images with an accuracy
worse than ours should hardly enable total mass measurements of
MACS~0416 with a precision better than ours (approximately 5\% in
both statistical and systematical uncertainties). Moreover, we remark
that the knowledge of the spectroscopic redshifts of several cluster
members and multiply-imaged sources are essential to build detailed
and reliable strong lensing models, and thus to obtain accurate total
mass estimates of a galaxy cluster. The spectra collected within our
CLASH-VLT survey will also allow us to detect the presence of possible
mass structures along the line of sight, usually not accounted for in the
strong lensing modeling, and to quantify their effects on the offset
between the observed and model-predicted image positions. Our values
of $\Delta$ of less than 0.4\arcsec$\,$ suggests that the compound
lensing effect is weaker than previously thought and usually assumed
until now (approximately 1\arcsec; e.g. \citealt{hos12}).  

\section{Comparison with cosmological simulations}
\label{sec:nbody}

The high-quality mass reconstruction of MACS~0416, enabled by the
combination of superb \HST\ imaging and extensive VLT spectroscopy,
allows a detailed investigation of the inner structure of the
dark-matter mass distribution in this 
system and a
meaningful comparison with cosmological simulations.
We juxtapose the measured cluster mass profile and the amount of
substructures (cluster galaxy members) identified in MACS~0416 with
theoretical predictions based on a set of $N$-body simulations in
order to probe the formation history of the galaxy cluster and its
substructures.  In Section \ref{sec:nbody:sim}, we outline the
$N$-body simulations.  The comparison of the cluster mass profile is
presented in Section \ref{sec:nbody:clusterprofile} whereas the
comparison of substructure properties is in Section
\ref{sec:nbody:substruct}. An extension of this kind of analyses to a
sample of galaxy clusters, presenting different properties in terms of
total mass and dynamical state, will be possible through
accurate modeling of further CLASH-VLT targets.

\subsection{High-resolution simulation of galaxy clusters}
\label{sec:nbody:sim}

The set of $N$-body simulations consists of 29 Lagrangian regions,
extracted around massive clusters from a cosmological box, and
resimulated at high resolution using the `zoom--in' technique 
\citep{TormenEtal97}. We refer the reader to \citet{BonafedeEtal11}
 and \citet{ContiniEtal12} for a
detailed description of this set of simulations. The adopted
cosmological model has $\Omega_{\rm m}=0.24$ for the matter
density parameter, $\Omega_{\rm bar}=0.04$ for the contribution of
baryons, $H_0=72\,{\rm km\,s^{-1}Mpc^{-1}}$ for the present-day Hubble
constant, $n_s=0.96$ for the primordial spectral index, and
$\sigma_8=0.8$ for the normalization of the power spectrum.

The particle mass is $10^8 M_\odot \, h^{-1}$, with a
Plummer--equivalent softening length for the computation of the
gravitational force fixed to $\epsilon=2.3\,h^{-1}$~kpc in physical
units at redshift $z<2$, and in comoving units at higher
redshift. Each output of the simulation (93 in total between $z\sim
60$ and $z=0$) has been postprocessed to identify dark matter halos
and subhalos. Dark matter halos have been identified using a standard
friends-of-friends (FOF) algorithm, with a linking length of 0.16 in
units of the mean inter-particle separation. Each FOF group was then
decomposed into a set of disjoint subhalos using the algorithm
{\small SUBFIND} \citep{SpringelEtal01}.

As in previous work, we consider as genuine subhalos only those that
retain at least 20 bound particles after a gravitational unbinding
procedure.  Our simulation thus contain all suhalos that are more
massive than $2 \times 10^9 M_\odot \, h^{-1}$ (corresponding to the
20 particle limit).  We further impose that these subhalos have
circular velocities greater than 50\,km\,s$^{-1}$.  The circular
velocity is 
computed as the maximum value of $\sqrt{GM(<r)/r}$ where $M(<r)$ is
the mass within a distance $r$ from the center of the
subhalos. These limits on mass and circular velocity are meant to
select the largest number of well defined subhalos. We have
checked that for the adopted velocity threshold we obtain a tight
relation between the mass and the velocity dispersion of
subhalos, making the circular velocity a good proxy for
subhalo mass.

\subsection{Cluster mass profile}
\label{sec:nbody:clusterprofile}

\begin{figure}
\centering
\includegraphics[width=0.48\textwidth]{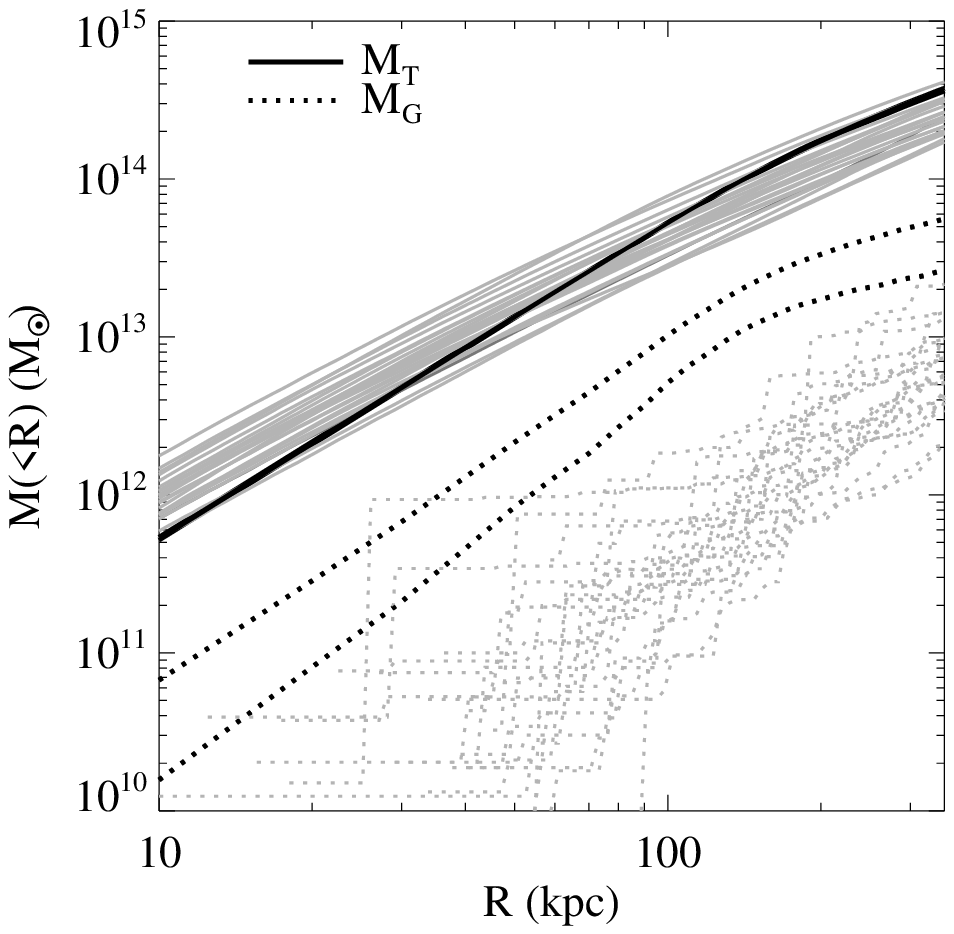}
\caption{Cumulative projected mass, $M(<R)$, profiles from the best-fitting strong lensing model of MACS~0416 (in black) and from dark-matter-only cosmological simulations of galaxy clusters (in gray). The solid and dotted lines represent, respectively, the total and cluster member profiles, at 1$\sigma$ confidence level for MACS~0416 and as different clusters for the simulations.}
\label{fi17}
\end{figure}

To compare with the measured cumulative projected mass profile of MACS
0416 in Section \ref{sec:lensmod:results}, we consider all the
simulated halos above $5 \times 10^{14} M_\odot$ at a redshift of 0.46, the
closest redshift to that of MACS~0416, for a total of 24 systems.  For
each halo we choose a random preferential axis and compute the total
mass within 
projected (two-dimensional) radii, as shown in Figure \ref{fi17}. We also show the mass
profile associated with the identified subhalos with masses $>2 \times
10^9 M_\odot \, h^{-1}$ and circular velocities $>50$\,km\,s$^{-1}$.
For the galaxy members in MACS~0416, we also include only those (165
out of 175) that have circular velocities\footnote{the circular
  velocities of the cluster galaxies can be computed through the
  resulting $\vartheta_{{\rm E}, i}$ of each galaxy, Equations
  (\ref{eq:vdisp_thE}) and (\ref{eq:vcirc_vdisp})}
$>50$\,km\,s$^{-1}$.

Comparing to the simulated $M(<R)$ in gray in Figure \ref{fi17}, the
observed cumulative projected mass profile of MACS~0416 (black curve)
traces the range spanned by the simulated curves given our selection
of simulated halos with masses similar to MACS~0416.  In detail, the
observed $M_{\mathrm{T}}(<R)$ increases more quickly as a function of
radius away from the cluster center,
implying that the observed total mass of the cluster MACS~0416 has a
shallower projected core {\it density} profile than the simulated
ones.  This could perhaps be explained by the apparent merging state
of MACS~0416 in the plane of the sky, although we note that several of
the simulated clusters are in similar dynamical states.

Interestingly, despite the similar total masses of MACS~0416 and of
the simulated clusters, the cumulative projected mass of the cluster
galaxies, $M_{\mathrm{G}}(<R)$, in MACS~0416 is in general larger than
that of the subhalos in the simulated clusters, especially at radius
$R\gtrsim60$\,kpc.  This offset in mass could be due to smaller masses of
subhalos, or fewer numbers of subhalos in the simulations, and we
explore this difference in more detail next.

\subsection{Distribution of substructures}
\label{sec:nbody:substruct}

\begin{figure}
\centering
\includegraphics[width=0.48\textwidth]{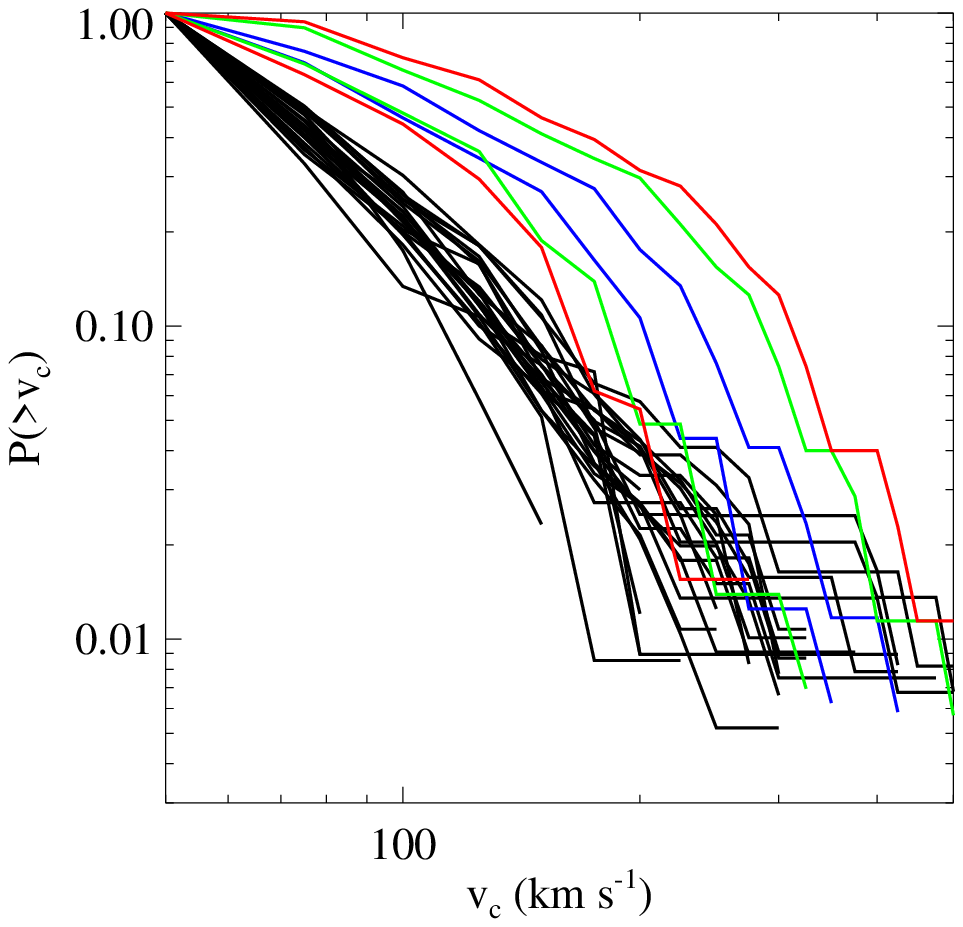}
\caption{Subhalo velocity functions, i.e., probability distribution functions of subhalos with circular velocities larger than fixed values, of massive galaxy clusters estimated from cosmological simulations and our strong lensing modeling. The simulated (in black) and observed (in color) functions, measured within a two-dimensional aperture of 420 kpc, are normalized to their corresponding values at 50 km s$^{-1}$. The 1$\sigma$, 2$\sigma$, and 3$\sigma$ confidence levels, according to the best-fitting 2PIEMD + 175(+1)dPIE ($M_{\mathrm{T}}L^{-1}\sim L^{0.2}$) model of MACS~0416, are shown in blue, green, and red, respectively.}
\label{fi21new}
\end{figure}

In Figure \ref{fi21new}, we plot the velocity function of the cluster
galaxies in MACS~0416 (color curves) and of the subhalos in the
simulated clusters (black curves) within 420\,kpc (corresponding
approximately to the \HST/WFC3 FoV at $z = 0.396$).  
As before, these substructures have circular velocities larger than
50\,km\,s$^{-1}$.  Overall, the observed velocity function is higher and has a
different shape from the power-law like velocity function from
simulations.

\begin{figure*}
\centering
\includegraphics[width=0.48\textwidth]{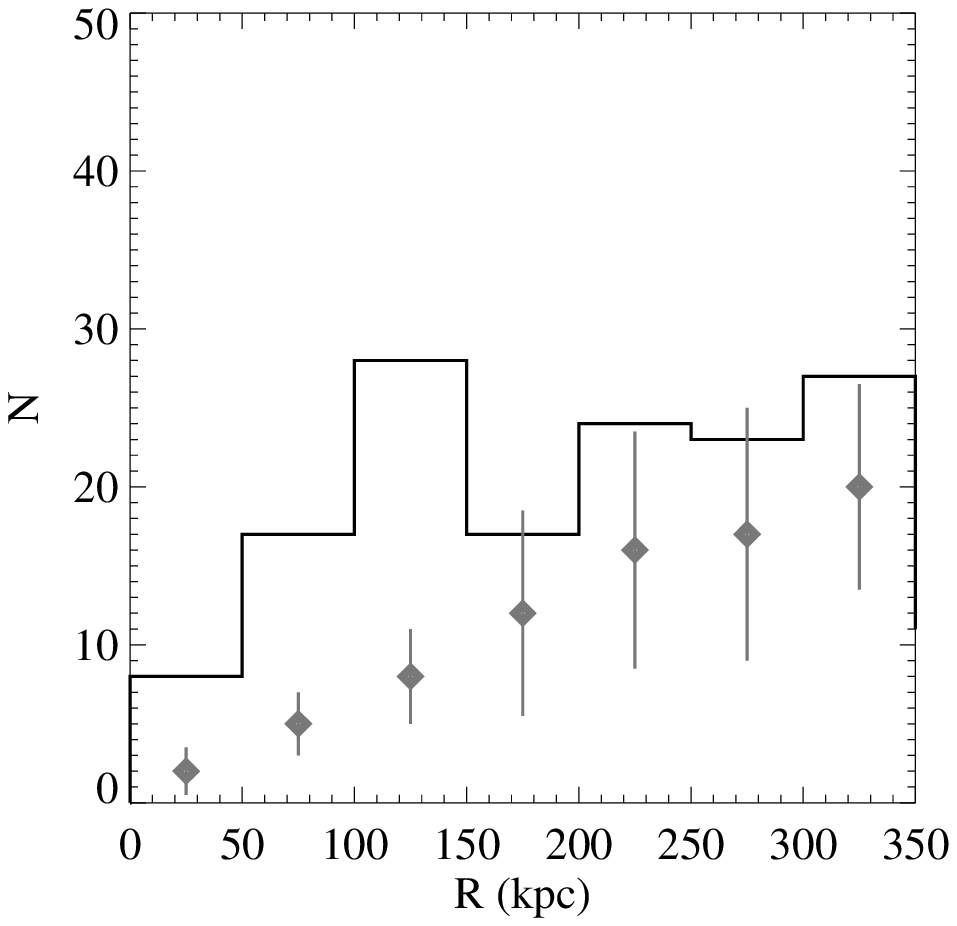}
\includegraphics[width=0.48\textwidth]{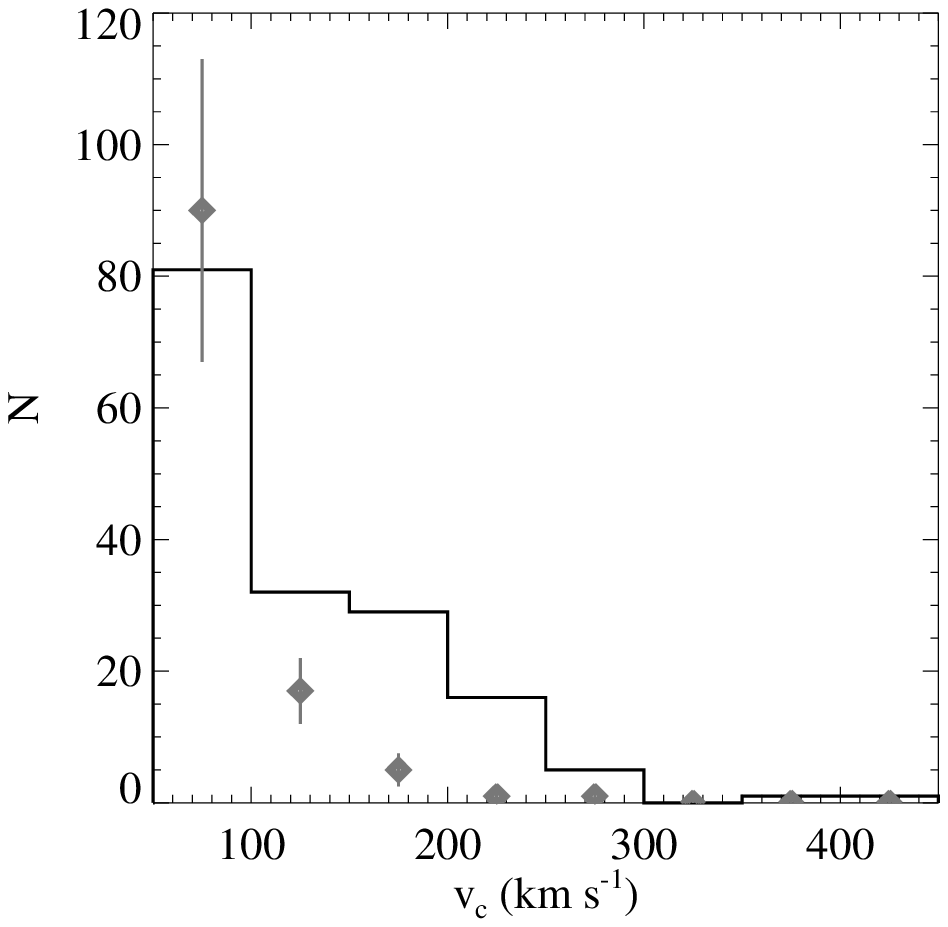}
\caption{Number of subhalos as a function of the projected distance from the galaxy cluster center (\emph{on the left}) and, within a two-dimensional aperture of 420 kpc, of their circular velocity value (\emph{on the right}). The black histograms represent the values derived in our analysis of MACS~0416 and the gray diamonds and bars show, correspondingly, the median values and the 1$\sigma$ uncertainties obtained from cosmological simulations.}
\label{fi19}
\end{figure*}

To probe further the discrepancy between observations and
simulations, we plot the numbers of cluster galaxies and simulated
subhalos in Figure \ref{fi19}.  In both panels, the histograms show the
distributions of the observed number of cluster galaxies in MACS~0416,
whereas the diamond points are the medians with 1$\sigma$ uncertainties
from the simulated clusters.  The simulated halos consistently
underpredict the number of subhalos on all radial scales, as shown in
the left-hand panel.  The underprediction is more severe in the
inner $\sim150$\,kpc of galaxy clusters.  In the right-hand panel, we
categorize the substructures in terms of their 
circular velocities (i.e., masses) regardless of their locations
within the clusters.  We find that the observed number of low-mass cluster
members with circular velocities $\lesssim100$\,km\,s$^{-1}$ is in good
agreement with the predicted number from the $N$-body simulations,
whereas the simulated clusters have fewer substructures with
circular velocities between $\sim100$\,km\,s$^{-1}$ and
$\sim300$\,km\,s$^{-1}$. We remark that within this last circular
velocity range the results of our observations are robust, since our
sample of candidate cluster members can only be marginally
contaminated by foreground/background objects at the
corresponding near-IR luminosities (see Section
\ref{sec:lensmod:masscomp:gal}). In fact, we have checked that by varying from 0.5 to 0.9 the value of the probability threshold of our method used to obtain the cluster members for the strong lensing analysis, i.e., moving from a more complete to a purer sample of cluster members, the number of selected bright (and massive) galaxies, with $F160W < 21$ mag (i.e., stellar mass values larger than $10^{9.8}\,M_{\odot}$), changes from 73 to 69. This confirms that the number of candidate bright cluster members, and thus the comparison with cosmological simulations at the corresponding circular velocities, does not depend appreciably on the selection details.

These findings suggest that the massive subhalos are not formed or
accreted into the simulated clusters as quickly as observed, or that
tidal strippings of massive subhalos are more efficient than observed,
or a combination of these effects.  Also, it appears that the tidal
disruptions of galaxies in the inner parts of MACS~0416 might be less
than those in the simulations, given the higher number of cluster
galaxies in MACS~0416.  We note that the simulations do not contain
baryons.  The addition of baryons into subhalos would likely make the
subhalos more tightly bound, which would in turn make tidal stripping
less effective and result in a higher number of subhalos from
simulations.  This could perhaps partly explain the lower number of
subhalos in simulations, although the effect might not affect 
significantly the number of massive subhalos.  We defer the comparison
of substructure distributions in hydro simulations to future work
which will provide insights on the formation of galaxy clusters and
the role of baryons.

\section{Summary}
\label{sec:conclude}

Thanks to the excellent \HST\ panchromatic observations and VLT spectra
of multiply-imaged sources and galaxy cluster members from our CLASH
and CLASH-VLT programs, we have performed a thorough strong lensing
study and comparison with dark-matter only cosmological simulations in
the inner regions of the galaxy cluster MACS~0416. 

We have emphasized that the use of only multiple-image systems with
spectroscopic redshifts and of a pure sample of cluster members,
selected through extensive multi-color and spectroscopic information,
is key to constructing robust mass maps with
high angular resolution. Further insights into the small scale
structure of the cluster dark-matter distribution would require
the measurement of the cluster member internal velocity dispersions,
which we are pursuing. The main results of our analysis can be
summarized as follows:
  
\begin{itemize}

\item[$-$] We reconstruct the observed image positions of 30 multiple
  images from 10 different sources, with spectroscopic redshift values
  between 1.637 and 3.223, with an unprecedented accuracy of
  approximately 0.3\arcsec. 

\item[$-$] The mass model that best fits the lensing observables is
  constituted of 2 cored elliptical pseudo-isothermal components and
  175 dual elliptical pseudo-isothermal components, representing the
  extended cluster dark-matter halos and candidate cluster members,
  respectively. The latter have been selected by using the full
  covariance matrix of the color distribution of the spectroscopic
  members.

\item[$-$] The two cluster dark-matter halos have mass
  centers at significant projected distances from the luminosity
  centers of the two brightest cluster galaxies and core radii larger
  than 50 kpc.

\item[$-$] When estimated within circular apertures, the total surface
  mass density shows a flat inner profile out to more than 100 kpc.
  The cumulative projected total mass is accurate to 5\% (of both
  statistical and systematical uncertainties) out to 350
  kpc, where it notably matches independent weak lensing measurements.

\item[$-$] The galaxy cluster members are best modeled by mass
  profiles that have total mass-to-light ratios increasing with the
  galaxy near-IR (F160W) luminosities. At more than 100 kpc in
  projection from the cluster barycenter, the mass in the form of
  stars represents approximately 8\% of the cluster members' total mass 
  and 1\% of the cluster total mass.

\item[$-$] We find that simulated galaxy clusters with total mass
  values comparable to that of MACS~0416 contain considerably less mass
  in subhalos in their cores relative to MACS~0416. The mismatch
  is more evident within the central 150 kpc and is associated with a
  deficiency in massive substructures 
  with circular velocities $\gtrsim100$\,km\,s$^{-1}$.

\end{itemize}

\

The new high level of accuracy we have reached in reproducing 
the observed multiple image positions of spectroscopically confirmed
sources paves the way for effectively studying the perturbing lensing
effect of mass structures along the line of sight.  It also lays the 
groundwork for measuring the values of cosmological parameters via ratios 
of angular diameter distances of the multiple background sources at 
different redshifts from a sample of massive strong lensing clusters.
Moreover, our
findings in MACS~0416 of the cored inner density profiles of the two cluster
dark-matter halos and their significant offsets from the brightest
cluster galaxies, together with forthcoming deep Chandra 
observations, might reveal important clues about the nature of dark matter 
(e.g., self-interacting or not).  Further
investigations following our pilot juxtaposition of precise observational
results and predictions from $N$-body simulations in the mass
structure of galaxy
clusters will enable tests of the very physical foundations of the
current $\Lambda$CDM model and the impact of baryonic physics on the
mass assembly of cosmological structures. The exceptional CLASH/HFF
imaging data and the spectroscopic follow-up from the ground like
CLASH-VLT, will be key to achieving all these aims.

\acknowledgments
We thank the ESO User Support group, and specifically Vincenzo
Mainieri, for the continuous and excellent support on the
implementation of the Large Programme
186.A-0798. The CLASH Multi-Cycle Treasury Program is based on
observations made with the NASA/ESA {\it Hubble Space Telescope}. The
Space Telescope Science Institute is operated by the Association of
Universities for Research in Astronomy, Inc., under NASA contract NAS
5-26555. ACS was developed under NASA Contract NAS 5-32864. This
research is supported in part by NASA Grant HST-GO-12065.01-A. The Dark
Cosmology Centre is funded by the DNRF. We acknowledge partial support
by the DFG Cluster of Excellence Origin Structure of the Universe. 
G.B.C. is supported by the CAPES-ICRANET program through the grant BEX
13946/13-7.
We acknowledge financial support from MIUR PRIN2010-2011
(J91J12000450001). K.U. acknowledges support from the National Science
Council of Taiwan (grant NSC100-2112-M-001-008-MY3). Support for
A.Z. is provided by NASA through Hubble Fellowship grant
HST-HF-51334.01-A awarded by STScI. V.P. acknowledges a grant from
"Consorzio per la Fisica - Trieste". A.F. acknowledges the support by
INAF PRIN 2010 grant (VIPERS). 


\clearpage




\clearpage

\clearpage

\end{document}